\documentclass[aps,reprint,pre,showpacs,superscriptaddress,amsmath,amssymb]{revtex4-1}
\usepackage{amsmath}
\usepackage{graphicx}  
\usepackage{float}
\usepackage{gensymb}
\usepackage{bm}        
\usepackage{color}        
\begin{document}

\title{Lattice Boltzmann model for weakly compressible flows}
\author{Praveen Kumar Kolluru}
\affiliation{Jawaharlal Nehru Centre for Advanced 
Scientific Research, Jakkur, Bangalore  560064, India} 
\author{Mohammad Atif}
\affiliation{Jawaharlal Nehru Centre for Advanced 
Scientific Research, Jakkur, Bangalore  560064, India} 
\author{Manjusha Namburi}
\affiliation{Jawaharlal Nehru Centre for Advanced 
Scientific Research, Jakkur, Bangalore  560064, India} 
\author{Santosh Ansumali}
\email{ansumali@jncasr.ac.in}
\affiliation{Jawaharlal Nehru Centre for Advanced Scientific Research, Jakkur, Bangalore  560064, India}

\begin{abstract}
We present an energy conserving lattice Boltzmann model based on a crystallographic lattice for simulation of weakly compressible flows.
The theoretical requirements and the methodology to construct such a model are discussed. 
We demonstrate that the model recovers the isentropic sound speed in addition to the effects of viscous heating and heat flux dynamics. 
Several test cases for acoustics, thermal and thermoacoustic flows are simulated to show the accuracy of the proposed model.
\end{abstract}
    

\maketitle

\section{Introduction}

The lattice Boltzmann (LB) method  with its simplified kinetic description of hydrodynamics in terms of a sequence of
collision and free flight restricted on a $D$-dimensional lattice provides a computationally efficient and easily
parallelizable alternative simulation methodology~\citep{succi2001lattice,chen1998lattice,aidun2010lattice}. 
An LB model in its standard formulation describes  a weakly compressible  flow at a reference temperature $T_0$.
The pressure $p$ and the mass density $\rho$ are related via the ideal equation of state
$p = \rho \theta_0$, where $\theta_0={k_B T_0}/{m}$  is the scaled temperature with  $k_{B}$ as the Boltzmann constant and $m$ is the mass of the molecule.  
These models are  quite suitable for isothermal flows where the relevant time scale is a few orders
of magnitude higher than the acoustic time scales and the object of interest is the velocity or vorticity field.
A consequence of kinetic equation with an isothermal dynamics, as revealed by Chapman-Enskog expansion, 
is that  in  the hydrodynamic limit, the stress tensor is not traceless and is  of form \citep{succi2001lattice,ansumali2005consistent}
\begin{equation}
 \sigma_{\alpha \beta} = \eta \left(\partial_\alpha  u_\beta + \partial_\beta u_\alpha \right), 
\end{equation}
which implies that the bulk viscosity $\zeta= 2/3 \eta$, where $\eta$ is the shear viscosity.  
Even though this expression for bulk viscosity is not realistic for any fluid,  this is not of too much  concern as long as one is interested in  the velocity dynamics in the limit of vanishing
Mach number ${\rm Ma}$.  This is due to the fact that  the divergence of the velocity field for low Mach number is $\partial_{\kappa} u_{\kappa}\sim O({\rm Ma^2})$ \citep{dellar2001bulk}.  
Thus, the
method is routinely used for incompressible hydrodynamic simulations both in low Reynolds number creeping flow regime as
well as for high Reynolds number turbulent flow regime~\citep{namburi2016crystallographic,succi2001lattice,chen1998lattice,aidun2010lattice,ladd1994numerical,ladd2001lattice,
ladd1993short}.

An extension of the lattice Boltzmann method (LBM) for acoustics is relatively recent
\citep{buick1998lattice,dellar2001bulk,crouse2006fundamental,marie2009comparison,li2011lattice,viggen1,viggen2}. These works have established the
 capability of LBM to correctly reproduce fundamental acoustic phenomena and highlighted the low dissipative
behaviour of the LBM. 
The starting point for   acoustic modeling in isothermal LB is the fact that the pressure fluctuation $\delta p$  (linearized around no flow condition) for the method follows the wave equation \citep{landau1959course}
\begin{equation}
 \label{eq:densityWaveEquation}
 \frac{\partial^2   \delta p}{\partial t^2} = \overline{c}_s^2   \nabla^2 \delta p,
\end{equation}
 with the isothermal sound speed being
$\overline{c}_s\equiv \sqrt{\partial p/\partial \rho\rvert_T} = \sqrt{\theta_0}$.
We remind that the sound wave is generated by  compression and expansion of air and is not an isothermal but an adiabatic process.  
Thus, it is not surprising that neglecting the rapidly-fluctuating temperature field  in a sound wave leads to an incorrect value of the sound speed.
The real isentropic 
sound speed is $c_s = \sqrt{\partial p/\partial \rho\rvert_s} = \sqrt{\gamma \theta}$, where $\gamma$ is the specific heat ratio. 
However, in practice for simulating acoustic waves, this incorrect sound speed is not of too much concern, as dynamics can be corrected via rescaling of temperature.
 In particular, 
Eq.\eqref{eq:densityWaveEquation} describes the correct acoustic dynamics  at temperature $\theta= \theta_0/\gamma$.


However, a more elaborate description of the sound wave in LB framework must start from
the true evolution equation for pressure fluctuation  obtained by   Navier-Stokes-Fourier (NSF) equation  linearized around no flow condition as  \citep{chaikin2000principles}
\begin{equation}
 \label{eq:acousticGoverningEquation}
 \left[ -\frac{\partial^2 }{\partial t^2} + \frac{\Gamma}{\rho_0}  
 \frac{\partial }{\partial t} \nabla^2  \right] \delta \rho + \nabla^2 \delta p = 0,
\end{equation}
where   $\delta \rho$ is the density fluctuation from the  equilibrium density $\rho_0$ and $\Gamma = 4\eta/3 + \zeta$. The thermodynamic equation of state relates the density and pressure fluctuations with the entropy fluctuation via the relation \citep{callen1998thermodynamics}
\begin{equation}
 \label{eq:densityFluctuation}
 \delta \rho = \frac{\partial \rho}{\partial p}\bigg\rvert_s \delta p + \frac{\partial \rho}{\partial s}\bigg\rvert_p \delta s,
\end{equation}
where, the entropy density fluctuation $\delta s$ satisfies the relation \cite{landau1959course}
\begin{equation}
 \label{entropyFluctuation}
 \frac{\partial \delta s}{\partial t} = -\frac{1}{\rho_0 T_0 } \nabla_{\alpha} \delta q_{\alpha},
\end{equation}
with  the heat flux due to the temperature fluctuation $\delta q_{\alpha}$ obeying the Fourier's law 
i.e., $\delta q_{\alpha} = - \kappa \nabla_\alpha \delta T$ where $\kappa$ is the thermal conductivity \citep{leal2007advanced}.

Thus, the final evolution equation for the pressure fluctuation obtained using Eqs.\eqref{eq:acousticGoverningEquation},\eqref{eq:densityFluctuation} is
\begin{equation}
\label{FinalPEq}
 \left[ -\frac{\partial^2 }{\partial t^2} + \frac{1}{\rho_0} \Gamma
 \frac{\partial }{\partial t} \nabla^2  \right] \left(\frac{\partial \rho}{\partial p}\bigg\rvert_s \delta p + \frac{\partial \rho}{\partial s}\bigg\rvert_p \delta s\right)+ \nabla^2 \delta p = 0,
\end{equation}
where the entropy fluctuation is governed by Eq.\eqref{entropyFluctuation}.
For short time dynamics, i.e., at time scales much less than the momentum diffusion time scales, the dynamics can
be considered to be isentropic and entropy fluctuations as well as the viscous contribution can be ignored.
This reduces Eq.\eqref{FinalPEq} to the wave equation as solved by an isothermal LB model given in  Eq.\eqref{eq:densityWaveEquation}.
However, a faithful simulation of acoustics by LB models requires that for a linearized flow, the pressure fluctuations obey Eq.\eqref{FinalPEq}, which in turn demands accurate heat flux dynamics via Eq.\eqref{entropyFluctuation}. 
Hence, a model suitable for acoustic simulations should have the following features:
\begin{itemize}
 \item  The right form of bulk viscosity and isentropic sound speed.
 \item  Correct heat flux dynamics to accurately capture the dynamics of pressure fluctuations.
 \item  The correct specific heat ratio and the Prandtl number.
\end{itemize}

Many of these defects disappear if one works with an energy conserving LB model \citep{ansumali2005consistent}. 
For example, it was shown in Ref. \citep{ansumali2005consistent} that the unphysical bulk viscosity does not exist in energy conserving models.
It was further noted in subsequent works that the energy conserving LB models recover the isentropic sound speed \cite{singh2013energy}.
Energy conserving models on lower order lattices such as D3Q27 are restricted to very small temperature fluctuations as the 
heat flux shows a significant departure from Fourier's law.
It is known that multispeed models recover the heat flux dynamics to a higher order of accuracy \citep{frapolli2014multispeed,atif2018higher}.
For example, recently a higher order body centered cubic (bcc) lattice model was proposed for thermal dynamics \citep{atif2018higher}.  
In this work, we propose another higher order lattice Boltzmann model for acoustic simulations and  restrict ourselves to the specific heat ratio of a 
monoatomic ideal gas and the Prandtl number to unity.

  In particular, we propose an energy conserving  multispeed LBM with $41$ discrete velocities in three dimensions that 
  is relevant for acoustics and weakly compressible flows.
  We list the conditions on the equilibrium for the discrete velocity models to recover the 
  Navier-Stokes-Fourier equations as the hydrodynamic limit.
  We also discuss the importance of the cubic accuracy in recovering the Navier-Stokes hydrodynamics.
  We develop the LB model based on recently proposed crystallographic lattice Boltzmann framework using a bcc lattice \citep{namburi2016crystallographic}. 
  It was shown that the bcc lattice gives better spatial accuracy in addition to more accuracy in the velocity space, in comparison to a simple cubic (sc) lattice.
 We outline the procedure for the development of the discrete equilibrium using entropic formulation by a series expansion 
 at the reference state with zero velocity and zero temperature variation. First, the equilibrium at a rest state is
 derived for non-zero variation in temperature, and a non-zero velocity equilibrium is derived as a series expansion around the earlier reference state. 
 This expansion is more stable than a direct two variable (Mach number and deviation from reference temperature) expansion.
  Finally,  we show the capability of the new model to perform aeroacoustic, thermal, thermoacoustic and turbulent simulations by presenting a number
  of benchmark simulations.

The paper is structured as follows: In Sec. \ref{section:lbm} we describe the lattice Boltzmann method briefly and define the restrictions on the
discrete velocity model to recover isothermal compressible hydrodynamics. 
This is followed by a derivation of a compressible thermo-hydrodynamic model and constraints for such a model in  Sec. \ref{section:cthd}.
A brief description on recently proposed class of crystallographic lattice Boltzmann models and followed by the construction of
a new LB model, RD3Q41 model, with 41 velocities on a bcc lattice is given in Sec. \ref{section:clbm}.
An energy conserving equilibrium distribution function is derived
for the proposed RD3Q41 model in Sec. \ref{section:entropiceq}. We show the capability of this model for simulating acoustic phenomena in
Sec. \ref{section:acoustics}, followed by a few test cases to demonstrate its ability to simulate thermal flows
in Sec. \ref{section:thermalFlows}. A nontrivial problem involving both thermal and acoustic phenomena is presented in Sec. \ref{section:TA}.
Further, in Sec. \ref{multiphase} we show that the proposed model is also advantageous for multiphase flows on account of it being more accurate
in the velocity space.
Finally, we simulate a few turbulent flows at high Reynolds numbers like the Kida-Peltz flow, flow in a rectangular channel and flow
past a sphere in Sec. \ref{section:turbulent}.  

\section{Lattice Boltzmann Method}\label{section:lbm}
The lattice Boltzmann method (LBM) with its simplified kinetic picture on a $D$-dimensional lattice  provides a 
computationally efficient  description of hydrodynamics~\citep{chen1998lattice,succi2001lattice}. In this section, we  briefly review the method and the basic 
motivation behind its use for simulating hydrodynamics. In LBM, the  velocity space is discretized to a finite set 
$\mathcal{C} = \{ {\bf c}_i, i=1..N_d \}$  and one associates a discrete population $f_i\in {\bf f}$ with each ${\bf 
c}_i$. The discrete populations $f({\bf c}_i, {\bf x},t)\equiv f_i({\bf x},t)$ are considered a  function of  location  
${\bf x}$ and time $t$. The set $\cal{C}$ is chosen to satisfy an appropriate set of symmetries needed to recover 
hydrodynamics from the evolution equation of $f_i$ in the long time limit~\citep{Wahyu2010}.  
The hydrodynamic variables such as the mass density $\rho$, velocity $\mathbf{u}$, and the energy $E$ are defined as
\begin{equation}
 \rho=\left< f, 1\right>, \quad {\mathbf j}\equiv  \rho \mathbf{u}= \left< f,  \mathbf{c}\right>, \quad 
  E = \left< f,  \frac{\mathbf{c}^2}{2}\right>,
\end{equation} 
where the inner product between two functions of discrete velocities $\phi$ and $\psi$ is defined as
\begin{equation}  
 \left< \phi,
 \psi\right> =\sum_{i=1}^{N_d} \phi_i \psi_i.
\end{equation}
In a three dimensional space ($D=3$), the energy is that of the ideal gas i.e. $E = (\rho u^2 + 3 \rho \theta)/2$.

A few higher order moments which are relevant to the hydrodynamic description are momentum flux as $P_{\alpha \beta}= \left< f,  \mathbf{c_\alpha}  \mathbf{c_{\beta}}\right> $, flux of momentum flux as $Q_{\alpha \beta \gamma} = \left< f,  \mathbf{c_\alpha}  \mathbf{c_{\beta}} \mathbf{c_{\gamma}}\right> $. The fluctuating velocity $\boldsymbol \xi$ is defined as  $\mathbf{c} - \mathbf{u}$  and the heat flux as $q_{\alpha } = \left< f,  \mathbf{{\xi}_\alpha} {\xi}^2/2\right>$ and the  flux of heat flux as $R_{\alpha \beta} = \left< f, {\xi^2} {\xi_\alpha}  {\xi_\beta}\right>  $.

Typically, one works with the  discrete in space and time version of the kinetic evolution equation in the Boltzmann Bhatnagar-Gross-Krook (BGK) 
form~\citep{bhatnagar1954model} as
\begin{equation}
f_i(\mathbf{x}+\mathbf{c}_i\Delta t,t+\Delta t)=f_i(\mathbf{x},t)+2\beta[f_i^{\rm{eq}}(\rho ({\bf f}), {\bf u} ({\bf 
f})) -f_i(\mathbf{x},t)],
\end{equation} 
where  $f_i^{\rm eq}$ represent discrete form of the Maxwell-Boltzmann equilibrium, $\Delta t$ is the chosen  time step 
and $\beta={\Delta t }/(2\tau+\Delta t)$ with  $\tau$ as the mean free time. The dimensionless parameter $\beta$,  bounded in
the interval $0<\beta<1$ with $\beta=1$ as the dissipation-less state,  physically  represents  the 
discrete dimensionless relaxation  towards the equilibrium. Here, the choice of the discrete version of equilibrium 
distribution $f_i^{\rm eq}$  is crucial for recovering the correct hydrodynamic limit  and different  formulations of 
lattice Boltzmann models differ  mainly in the choice of this discrete equilibrium~\citep{ansumali2003minimal,wagner1998h}. A common 
choice is to project the Maxwell-Boltzmann distribution on the  Hermite basis to get a computationally attractive 
polynomial expression of the equilibrium  as~\citep{chen1992recovery,qian1992lattice,benzi1992lattice, 
shan1998discretization}
\begin{align}
\label{feqPoly}
\begin{split}
f_i^{\rm eq}(\rho, {\bf u}) &= w_i\rho\Biggl[1 +
\frac{ u_{\alpha}c_{i\alpha}}{\, \theta_0}   
+ \frac{u_{\alpha} u_{\beta}}{2\, \theta_0^2}  \left(   c_{i\alpha}c_{i\beta} -  \theta_0 \delta_{\alpha \beta}   
\right)  \Biggr],
\end{split}
\end{align} 
where $\theta_0$ is some reference temperature associated with the underlying lattice and $w_i$ are the weights chosen in 
such a way that the mass and momentum  constraints are ensured, i.e.,
\begin{align}
\label{constPol}
 \begin{split}
  \sum _{i=1}^{N_d} f_i^{\rm eq} =  \rho, \qquad
  \sum _{i=1}^{N_d} f_i^{\rm eq} c_{i\alpha}=  j_{\alpha}.
 \end{split}
\end{align}
Furthermore, to get correct hydrodynamic limit for isothermal low Mach number dynamics, it is essential to 
ensure that the second moment of the discrete equilibrium is the same as that obtained from the Maxwell-Boltzmann 
distribution, i.e.,
\begin{equation} \label{eq:stresseq}
  P^{\rm eq}_{\alpha \beta} = \sum _{i=1}^{N_d} f_i^{\rm eq} c_{i\alpha} c_{i \beta} =  \rho u_{\alpha} u_{\beta} + \rho \theta_0 \delta_{\alpha 
\beta}.
\end{equation}

The rationale for adding Eq.~\eqref{eq:stresseq} can be understood by writing first the kinetic equation in its partial differential form (PDE) form (for $\Delta t\to 0$)~\citep{frapolli2014multispeed,yudistiawan2010higher,chikatamarla2009lattices}
as 
\begin{equation} 
\label{eq:kinEqLBM}
\partial_t f_i + c_{i \alpha} \partial_{\alpha} f_i =- \frac{1}{\tau}\left[f_i  - f_i ^{\rm eq}\right],
\end{equation}
from which one can write the mass and momentum conservation laws as
\begin{align}
 \begin{split}
 \partial_t\rho+\partial_{\alpha} j_{\alpha}&=0,\\
 \partial_t j_{\alpha} +\partial_{\beta}P_{\alpha \beta}&=0.\\
 \end{split}
\end{align}
The evolution equation for the pressure tensor $P_{\alpha \beta}$ using Eq.\eqref{eq:stresseq} and Eq.\eqref{eq:kinEqLBM} as
\begin{align}
 \begin{split}
 \partial_t P_{\alpha \beta} +\partial_{\gamma} Q_{\alpha \beta  \gamma}  
&=\frac{1}{\tau}\left(
  \rho u_{\alpha} u_{\beta} + \rho \theta_0 \delta_{\alpha \beta}-P_{\alpha \beta} 
 \right),
 \end{split}
\end{align}
from which it is evident that in the limit of $\tau\to 0$, $P_{\alpha \beta} \to \rho u_\alpha u_\beta + \rho \theta_0 \delta_{\alpha \beta} $ and thus
the zeroth order hydrodynamic equation describes the inviscid hydrodynamics as given by the Euler equation. 
The Navier-Stokes  hydrodynamics is recovered provided  
~\citep{qian1992lattice,yudistiawan2008hydrodynamics,PhysRevLett.97.190601} 
 \begin{equation}
 \label{moment3LB}
  \sum _i^{N_d} f_i^{\rm eq} c_{i\alpha} c_{i \beta}c_{i \gamma}= \rho u_{\alpha} u_{\beta} u_{\gamma} +\rho\theta_0 
\left(u_{\alpha}\delta_{\beta \gamma}+
  u_{\beta} \delta_{\alpha \gamma}+u_{\gamma} \delta_{\alpha \beta}\right). 
 \end{equation}

In most of the widely used lower order lattice Boltzmann models the above condition is 
satisfied only only up to linear order in velocity due to the absence of the cubic term in the equilibrium represented by Eq.\eqref{feqPoly}
~\citep{qian1998complete}.


The Eq.\eqref{feqPoly} along with Eqs.\eqref{constPol},\eqref{eq:stresseq} imply that  
the discrete velocity set and associated weights should satisfy
\begin{align}\label{eq:weights2nd}
 \begin{split}
\sum _{i=1}^{N_d}  w_i=1,\quad \sum_{i=1}^{N_d} w_i c_{i \alpha} c_{i \beta} = \theta_0 \delta_{\alpha \beta},\\
\sum_{i=1}^{N_d} w_i c_{i \alpha} c_{i \beta} c_{i \gamma} c_{i \zeta} = \theta_0^2 \Delta_{\alpha \beta \gamma \zeta},
 \end{split}
 \end{align}
where $\Delta_{\alpha\beta\gamma\zeta} =\delta_{\alpha \beta}\delta_{\gamma\zeta} + \delta_{\alpha 
\gamma}\delta_{\beta\zeta} + \delta_{\alpha\zeta }\delta_{\beta\gamma}$ is the fourth-order isotropic tensor and all odd 
order moments, such as
\begin{align}
 \begin{split} 
  \sum_{i=1}^{N_d} w_i c_{i \alpha} =0, \quad  \sum_{i=1}^{N_d} w_i c_{i \alpha} c_{i \beta}c_{i \gamma}=0, \\
  \sum_{i=1}^{N_d} w_i c_{i \alpha} c_{i \beta}c_{i \gamma}c_{i \kappa} c_{i \zeta}=0,
 \end{split}
\end{align}
are zero.
These conditions are central to models for isothermal incompressible hydrodynamics, and the procedure to construct them is well understood~\citep{succi2001lattice}.
It should be noted here that in case of compressible hydrodynamics, moment chain suggests that one needs to add $\mathcal{O}(u^3)$ contribution in the equilibrium distribution so that Eq.\eqref{moment3LB} is satisfied.
This condition on the third moment can be fulfilled only 
if the discrete equilibrium distribution is of the form~\citep{chikatamarla2009lattices},
\begin{align}
 \begin{split} 
 f_i^{\rm eq} = w_i \left(1 + \frac{ u_{\alpha}c_{i\alpha}}{\, \theta_0}  + \frac{u_{\alpha} u_{\beta}}{2\, \theta_0^2}  
\left(   c_{i\alpha}c_{i\beta} -  
\theta_0 \delta_{\alpha \beta}   \right)  \right. \\
 \left. +  \frac{u_{\alpha} u_{\beta} u_{\gamma}c_{i \gamma}}{6\, \theta_0^3} \left( c_{i \alpha}c_{i \beta}-3\theta_0 
\delta_{\alpha \beta}\right) \right).  
 \end{split}
\end{align}
This adds further restriction on the weights as
\begin{equation} \label{eq:weights6th}
\sum_{i=1}^{N_d} w_i c_{i \alpha} c_{i \beta} c_{i \gamma} c_{i \kappa} c_{i \zeta} c_{i \eta}= \theta_0^3 
\Delta_{\alpha \beta \gamma \kappa \zeta \eta},
 \end{equation}
where $\Delta_{\alpha \beta \gamma \kappa \zeta \eta}$ is the sixth-order isotropic 
tensor~\citep{yudistiawan2010higher}. However, only very high-order on-lattice models are known to satisfy such 
constraint in 3-dimensions~\citep{shan2006kinetic,chikatamarla2009lattices}. In practice, it is easier to satisfy the contracted version 
of Eq. \eqref{moment3LB}
\begin{equation} \label{eq:contractedForm}
  \sum _{i=1}^{N_d} f_i^{\rm eq} c_{i}^2 c_{i \alpha} =  \rho u_{\alpha} u^2 + (D+2)\, \rho \theta_0\,  u_{\alpha},
\end{equation}
which implies
\begin{equation} \label{eq:weights4th}
 \sum_{i=1}^{N_d} w_i c_i^2 \, c_{i \alpha} c_{i \beta} c_{i \gamma} c_{i \kappa} = 7\theta_0^3 \Delta_{\alpha \beta 
\gamma \kappa}.
\end{equation}
The above condition only ensures that the evolution equation for the energy is correct to the leading order.
In this work, we will consider Eq.\eqref{eq:contractedForm} and Eq.\eqref{eq:weights4th} as a requirement on higher-order LBM. 

\section{Compressible Thermo-hydrodynamics}\label{section:cthd}
In the previous section, the requirements for  constructing a model for isothermal compressible hydrodynamics 
and the importance of the cubic accuracy in recovering the Navier-Stokes hydrodynamics has been described.
It is known that on higher-order lattices such cubic accuracy can be imposed~\citep{Wahyu2010,chikatamarla2010lattice}.
However, typically, this has been done in an isothermal setting. A drawback of such models, though, is that the isentropic
speed of sound is not recovered and they also lack the coupling between thermal and acoustic modes~\citep{singh2013energy}. 
These aspects have not got enough attention in the development of new LB models.

The coupling between thermal and acoustic modes requires an accurate description of heat flux~\citep{reichl1999modern,liboff2003kinetic}.
In particular, in the limit of low Knudsen number, one most recover Fourier dynamics.
Hence, Navier-Stokes-Fourier dynamics should be recovered as the first order hydrodynamic description in the compressible models with energy conservation.

The heat flux is given by Fourier law via Chapman-Enskog analysis requires that the 
evolution for the energy flux 
\begin{equation}
 \partial_t q_{\alpha}+\partial_{\beta} R_{\alpha 
\beta}=\frac{1}{\tau}\left(q^{\rm eq}_{\alpha}- q_{\alpha} \right),
\end{equation}
is correct at the leading order
\begin{equation}
 \partial_t^{(0)} q_{\alpha}^{\rm eq} +\partial_{\beta} R_{\alpha \beta}^{\rm 
eq}=\frac{1}{\tau}\left(- q_{\alpha}^{\rm neq} \right).
\end{equation} 

Since only the trace of the equilibrium  of the fourth-order moment in the Maxwell-Boltzmann form appears  in the balance equation of heat flux we only need to ensure 
\begin{align}
 \label{Ralphabeta}
 \begin{split}
   R_{\alpha \beta}^{\rm eq}\left({\bf u}=0 \right)&=   
  5 \rho \theta^2\delta_{\alpha\beta},
 \end{split}
\end{align}
to get the evolution of heat flux at least to be quadratically correct in terms of $\Delta \theta = \theta/\theta_0-1$, the dimensionless departure of $\theta$ from $\theta_0$. 

In order to ensure that $R_{\alpha \beta}$ satisfy Eq.\eqref{Ralphabeta}, we write the equilibrium at $u=0$, as
 \begin{align}
\begin{split}
  \tilde{f}_i  &= w_i \rho\Biggl[ 1+  \frac{\Delta \theta}{2} \left( 
\frac{c_i^2}{\theta_0}-3  \right) 
  +\frac{\Delta^2 \theta}{8} \left( \frac{c_i^4}{\theta_0^2}-10 
\frac{c_i^2}{\theta_0}+15 \right) 
\Biggr].
\end{split}
 \end{align}
The expected value of $ R_{\alpha \beta}^{\rm eq}$ computed with this equilibrium matches the required value
 \begin{align}
 \begin{split}
   R_{\alpha \beta}^{\rm eq}\left({\bf u}=0 \right)&= 5 \rho \theta_0^2 
\delta_{\alpha\beta}\left(  1    +  2 \Delta \theta +     \Delta\theta^2  \right)=5 \rho \theta^2 \delta_{\alpha\beta},
 \end{split}
\end{align}
provided the model satisfies
\begin{equation}
   \sum_{i=1}^{N_d} w_i c_i^8 = 945 \theta_0^4.
   \label{eq:wci8}
  \end{equation}
This condition is not considered by some extensions of LBM where the cubic moment is imposed ~\citep{chikatamarla2009lattices,chikatamarla2015entropic}, where as it is considered in Ref. ~\citep{atif2018higher}.

Thus, we look for LB models which satisfy Eq.\eqref{eq:weights2nd}, Eq.\eqref{eq:weights4th} and Eq.\eqref{eq:wci8}. These tensorial set of equations can be simplified further for the class of discrete velocity models with the discrete velocity set $\mathbf{c}$ which ensure isotropy and avoid preference to any direction in particular. These discrete velocity sets should have the properties of closure under inversion (if $c_i \in \mathbf{c}$  then $-c_i \in \mathbf{c}$)  and closure under reflection (if $c_i(c_{ix},c_{iy},c_{iz}) \in \mathbf{c}$ then all possible reflections of $c_i \in \mathbf{c}$) to ensure that any vector $\psi(\mathbf{c}^2)$ in the discrete case we have
\begin{align}
\begin{split}
&\left<\psi, c_{x}^{2n} \right> = \left<\psi, c_{y}^{2n} \right> = \left<\psi, 
c_{z}^{2n} \right>, \\
&\left<\psi, c_{x}^{2n}c_{y}^{2m} \right> = \left<\psi, c_{y}^{2n}c_{z}^{2m} \right> 
= \left<\psi, c_{z}^{2n}c_{x}^{2m} \right>,\\
&\left<\psi, c_x^{2n+1} \right> = \left<\psi, c_y^{2n+1} \right> = \left<\psi, c_z^{2n+1} \right> =0,
\label{dvmcond}
\end{split}
\end{align} 
where $n$ and $m$ are the natural numbers~\citep{Wahyu2010}.

The above constraints on the discrete velocity set along with Eq.\eqref{eq:weights2nd}, Eq.\eqref{eq:weights4th} and Eq.\eqref{eq:wci8} give us the following  set of seven equations as the constraints on the weights for the proposed model as
 \begin{align}
 \begin{split}
 &\sum w_i  = 1 \quad  \sum w_i c_{ix}^2 = \theta_0 \quad \sum w_i c_{ix}^4 = 3\theta_0^2 \\
 &\sum w_i c_{ix}^2 c_{iy}^2 = \theta_0^2  \quad  \sum w_i c_{ix}^4 c_{i}^2 = 21\theta_0^3 \\
 &\sum w_i c_{ix}^2 c_{i}^4 = 35\theta_0^3 \quad \sum w_i c_{i}^8 = 945\theta_0^4.
 \label{allWeights}
 \end{split}
 \end{align}
 
As reference temperature $\theta_0$ is not specified and is an unknown, we have six more additional degrees of freedom. Hence, to obtain an on-lattice model we need six energy shells each with a weight $w_i$ to solve the system of equations exactly. Furthermore, these weights should be positive. 

\section{Crystallographic lattice Boltzmann model}\label{section:clbm}

Historically, the lattice chosen for the LBM has been the simple cubic (sc) lattice which demands that the grid is refined near the solid body or in zones of extreme flow variations. A recently proposed class of LBM models known as crystallographic LBM show an important connection between crystallography, optimal packing problem in the efficient discretization of PDE~\citep{namburi2016crystallographic}.

Based on this connection, it was argued that the optimal spatial discretization is provided by a body-centered cubic (bcc) arrangement of grid points and not by a simple cubic arrangement of grid points as used by conventional structured grid-based methods~\citep{atif2018higher,murthy2018lattice,namburi2016crystallographic}.

\begin{figure}
\centering
  \includegraphics[scale=1.0]{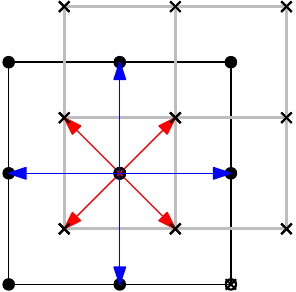}  
  \caption{Building block of a crystallographic lattice in two dimensions, 
simple cubic links (blue) and body-centered links (red) are depicted here. }
  \label{replica_shells2d}
\end{figure}

This lattice comprises of two simple cubic lattices displaced by a distance of $0.5\Delta x$ in each direction. 
Figure \ref{replica_shells2d} depicts the building blocks and the links of a bcc lattice in two dimensions for illustration purpose. 

\begin{figure}
  \includegraphics[scale=0.25]{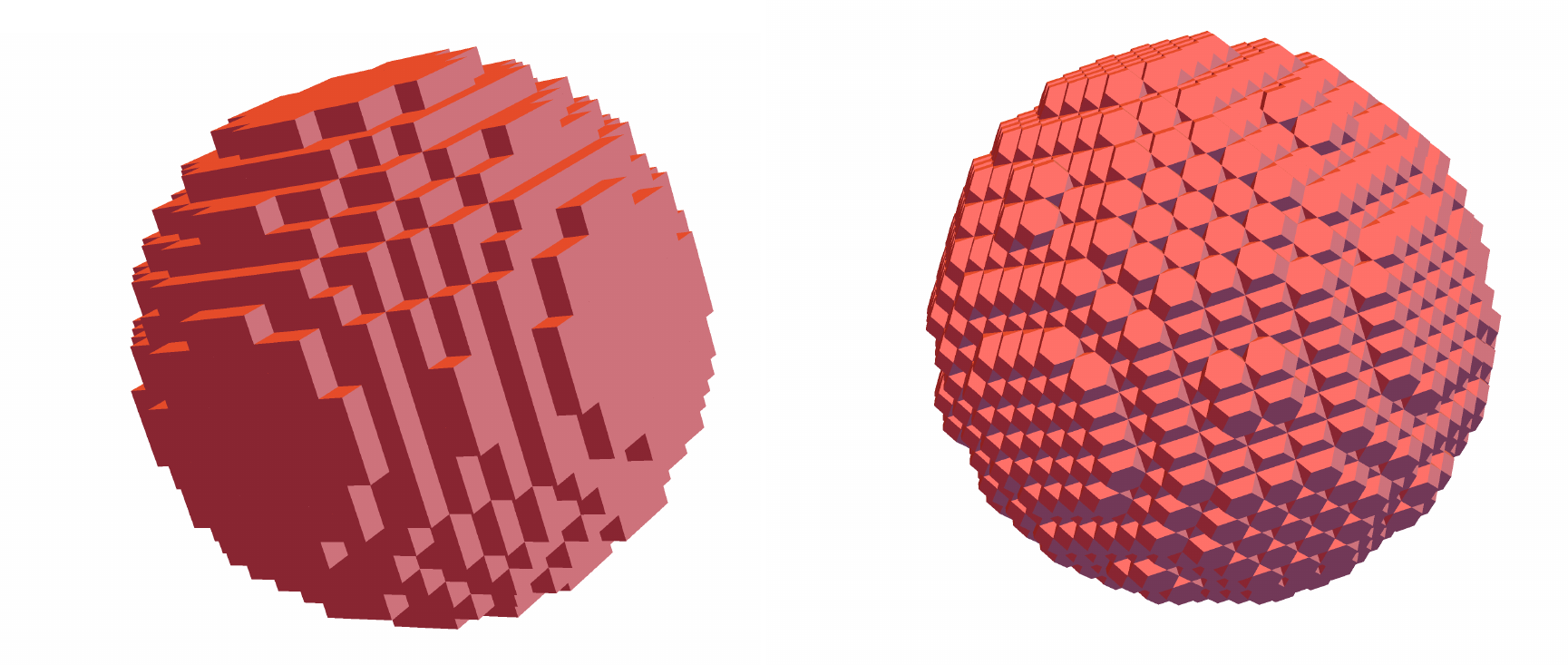}  
  \includegraphics[scale=0.25]{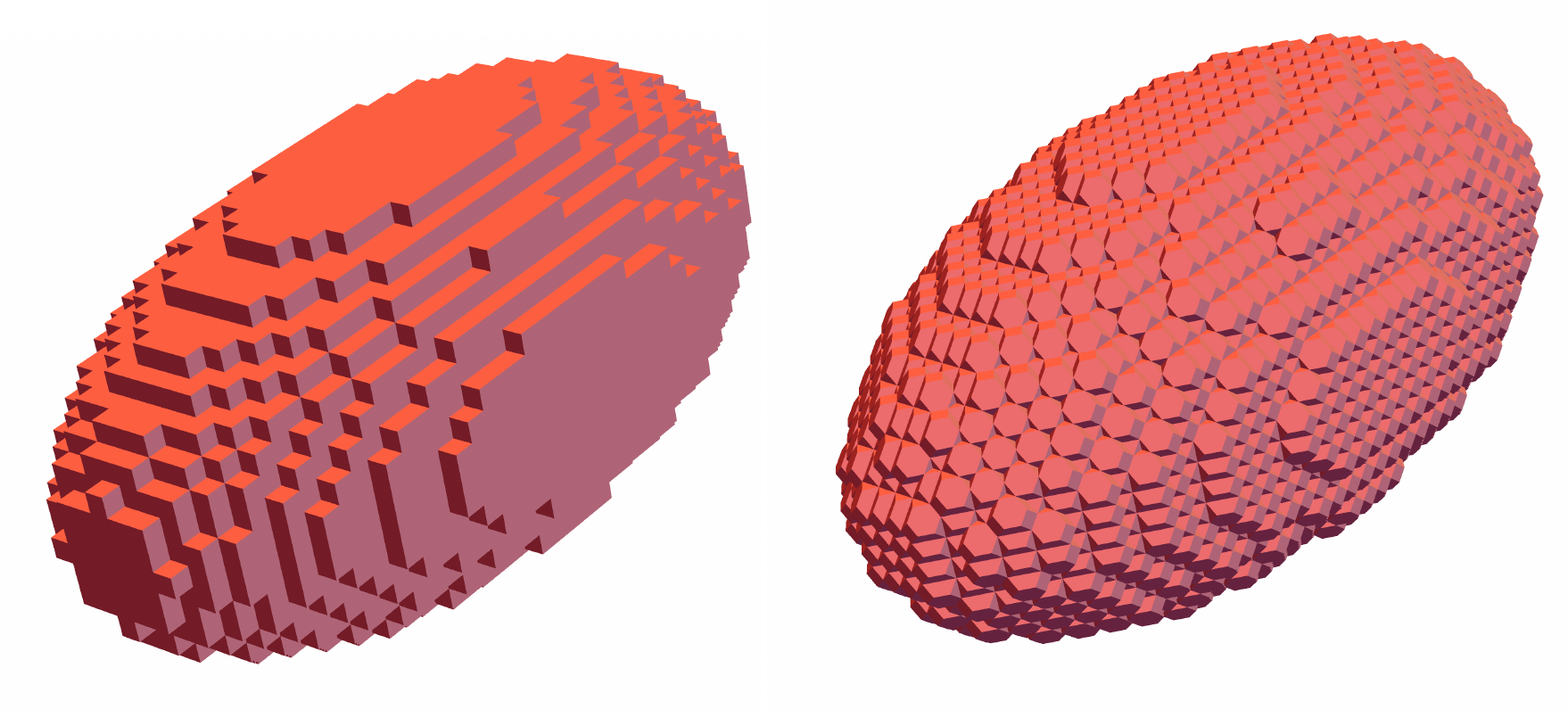}
  \caption{A representation of sphere and an ellipsoid on sc(left) and bcc(right) lattices.}
  \label{fig::BCC_SC}
\end{figure}

Another well-known fact in the computer graphics literature is that the volume representation (or rendering) is better on the bcc lattice~\citep{alim2009}.  
As the bcc grid has more points at the boundaries, it was also found to represent the boundaries well. To illustrate the difference between sc and bcc lattices, we show a depiction of a sphere and an ellipsoid on both sc and bcc lattices in Fig. \ref{fig::BCC_SC}.

Additionally, this class of bcc lattices removes an important artefact of sc lattices with velocity component unity (D3Q15, D3Q27) of imposing 
an artificial closure on the third-order moments where
\begin{equation}
\left<f,c_{\alpha}^3 \right> = c^2 \left<f, c_{\alpha} \right>.
\end{equation}
This effect plays an important role in regimes where the Knudsen boundary layer is important~\citep{ansumali2007hydrodynamics}. 

Like the traditional sc grids, the bcc grid also preserves the ease of streaming along the links while increasing the local accuracy.
  
\begin{figure*}
 \includegraphics[scale=0.5] {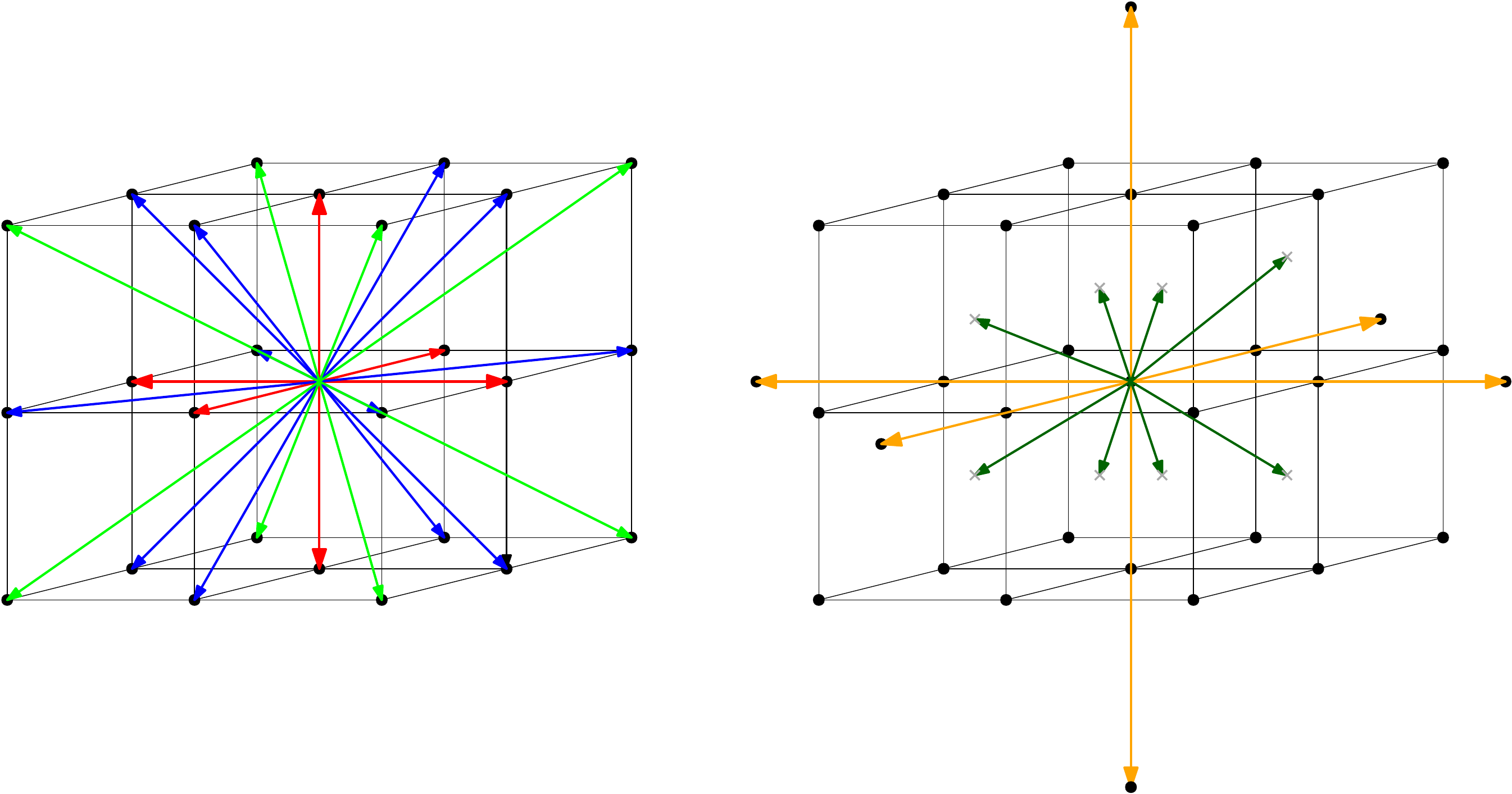}
\caption{Energy shells of the RD3Q41 model - sc-1(red), fcc-1(blue), bcc-1(light green) shown on a regular lattice and sc-2(orange), bcc-$\frac{1}{2}$(dark green). }
\label{fig:replica_shells3d}  
\end{figure*} 

The proposed model has 41 velocities with five different energy shells and is referred to as RD3Q41 model hereafter.
Figure \ref{fig:replica_shells3d} shows the building blocks of the RD3Q41 model. 
The weights are derived by imposing the constraints from Eq.\eqref{allWeights}. It has two sc shells, one face-centered cubic (fcc) shell and two bcc energy shells.
The energy shells and corresponding velocity sets with weights for this model are given in the Table \ref{tab:weights}.
\begin{table}
\resizebox{0.5\textwidth}{!}{
\begin{tabular}{cccc} 
\hline \hline
Shells              &  Discrete Velocities($c_i$)                                                                      & Weight($w_i$) \\ \hline
sc-1                & $\left(\pm 1, 0, 0  \right),\left(0, \pm 1, 0  \right),\left( 0, 0, \pm 1  \right) $             &   0.04743040745116578 \\
sc-2                & $\left(\pm 2, 0, 0  \right),\left(0, \pm 2, 0  \right),\left( 0, 0, \pm 2  \right) $             &  0.00165687664501576 \\
fcc-1               & $\left(\pm 1, \pm 1, 0  \right),\left(\pm 1, 0, \pm 1  \right),  \left( 0,\pm 1, \pm 1  \right)$ &  0.00651175327832464 \\
bcc -1              & $\left(\pm 1, \pm 1,  \pm 1  \right)$                                                            &  0.00454087801154440 \\
bcc -$\frac{1}{2}$  & $\left(\pm 0.5, \pm 0.5,  \pm 0.5  \right)$                                                      &  0.04917980624482672 \\ \hline \hline
\end{tabular}}
\caption{Energy shells and their corresponding velocities with weights for RD3Q41.}
\label{tab:weights}
\end{table}

\section{Discrete entropic equilibrium} \label{section:entropiceq}
The entropic formulation of the LBM restores the  thermodynamic consistency embedded in the Boltzmann description.
In this method, one starts with a discrete  $H$ function typically  in the Boltzmann form 
 \begin{equation}
 \label{entropy}
   H=  \sum_{i=1}^{N_d} f_i\left(\log{\frac{f_i}{w_i}}-1\right),
 \end{equation}
 and construct equilibrium as its minimizer under the constraints of local conservation laws
~\citep{karlin1998maximum,boghosian2003galilean,wagner1998h,chen2000h,karlin1999perfect,succi2002colloquium,ansumali2003minimal,ansumali2005consistent}. 
 Typically, the equilibrium is constructed in an isothermal setting which lacks energy conservation. 
 In higher-order LBM energy conservation is included in deriving the thermal entropic equilibrium distribution.

In this section, we briefly derive the energy conserving 
equilibrium for the case of small temperature variation around reference temperature $\theta_0$ using the entropic LBM.  It should be reminded that in the case of entropic lattice Boltzmann model, for every discrete velocity model, one finds the Lagrange multiplier with a high degree of accuracy to preserve the positive form of equilibrium~\citep{chikatamarla2006entropic}. This is typically done numerically. To analyze the hydrodynamic limit of LBM, we derive the series form of the equilibrium. This allows us to calculate the closed-form expression for the moments. These expressions for the moments are relevant for analyzing the errors in the hydrodynamic limit.

Following Refs.~\citep{ansumali2005consistent,atif2018higher}, we consider energy conserving equilibrium and   
include energy in the  set of constraints to obtain the equilibrium distribution which minimizes entropy (Eq.\eqref{entropy}) as
\begin{equation}
f_i^{\rm eq} = w_i \rho \exp(-\alpha - \beta_\kappa c_{i\kappa} -\gamma c_i^2).
\label{discreteentropiceq}
\end{equation}
where  $\alpha, \beta_\kappa, \gamma$  are the  Lagrange multipliers associated with mass, momentum and energy 
respectively~\citep{ansumali2005consistent,ansumali2003minimal}. The  explicit expression can be obtained by inverting the 
following relations
\begin{align}
\begin{split}
\left< f_i^{\rm eq} ,\{1,c_{i\alpha},c^2\} \right> &=\{\rho,\rho u_\alpha,\rho u^2 + 3\rho\theta\}.
\label{fivelagrangemultipliersystem}
\end{split}
\end{align} 

However, other than a few special cases such as the $D1Q3$ model and its higher dimension extensions $D2Q9$ and 
$D3Q27$, the explicit solutions are not known \citep{ansumali2005consistent,ansumali2003minimal}. The system of equations in 
Eq.\eqref{fivelagrangemultipliersystem} are not explicitly invertible, therefore we choose a reference state with mean 
velocity $u_\alpha=0$ where $\beta_\kappa=0$. Thus, the system of equations simplifies as
\begin{align}
 \begin{split}
 \exp(-\alpha^{0} ) \sum_{i=1}^{N_d}  w_i  \exp(-\gamma^{0}  c_i^2) \{1,  c_i^2 \}  &\\ 
                    = \{1,  3\, \theta \}                  \equiv \{1,3\, \theta_0\left(1+\eta\right) \},
\end{split}
\end{align} 
where, $\alpha^{0}$ and $\gamma^{0}$ are Lagrange multipliers corresponding to the state ${\bf u}={\bf 0}$. However, even 
for ${\bf u}={\bf 0}$, explicit solutions for other Lagrange multipliers are not known for most of the 
models\citep{ansumali2005consistent}. Therefore, at  $u_\alpha=0$ itself we chose another reference state  
$\theta=\theta_0$  for which it is trivial to check that the solution is $\alpha^{0}(\theta=\theta_0)=0$ and 
$\gamma^{0}(\theta=\theta_0)=0$.

Following the procedure of Ref.~\citep{atif2018higher},  a perturbative series around this reference state can be built by 
expanding the Lagrange multipliers around $\alpha^{\left(0\right)}$ and $\gamma^{\left(0\right)}$ in powers of  the 
smallness parameter $\eta= \theta/\theta_0-1$ (denoting smallness of the temperature deviation).
The explicit solution is evaluated up to  ${\cal O}(\eta^3)$ as

\begin{align*}
 \begin{split}
  \tilde{f}_i  &= w_i \rho\Biggl[ 1+  \frac{\eta}{2} \left( \frac{c_i^2}{\theta_0}-3  \right)
 +\frac{\eta^2}{8} \left( \frac{c_i^4}{\theta_0^2}-10 \frac{c_i^2}{\theta_0}+15 \right) \\
 &+ \frac{\eta^3}{48}\biggl(  \frac{c_i^6}{ \theta_0^3} -21 \frac{c_i^4}{\theta_0^2}   + 105 \frac{ c_i^2}{\theta_0}  
 - 105 \biggr) \Biggr].
 \end{split}
\end{align*}

The  equilibrium distribution at non-zero velocity is derived by expanding the Lagrange multipliers in 
$\epsilon$ (representing smallness of the velocity scale) and the expression for discrete 
equilibrium accurate up to ${\cal O}(\epsilon^3)$ is obtained to be
 \begin{align}
\begin{split}\label{eq:feqGenU3}
 f_i^{\rm eq} =& \tilde{f}_i \Biggl\{ 1+ \frac{u_{\alpha} c_{i \alpha}}{\theta} - \frac{u^2}{2\theta} (1-A_1)  + \frac{1}{2} \left( \frac{u_{\alpha} c_{i \alpha}}{\theta} \right)^2 \\
&  + \frac{1}{6} \left( \frac{u_{\alpha} c_{i \alpha}}{\theta}\right)^3 - \frac{u^2c_i^2}{6\theta^2} A_1  \\ 
&  + \frac{u^2 u_\alpha c_{i\alpha}}{6 \theta^2 } A_2  -\frac{u^2 c_i^2 u_\alpha c_{i\alpha}}{6 \theta^3 } A_1
 \Biggr\},\end{split}
 \end{align}
 where $M_2$ and $M_2^{\prime}$, the error terms with respect to zero velocity equilibrium moments, and $A_1$ and $A_2$ are
 \begin{align}
\begin{split}
M_2 &= - 0.807953 \Delta^3 \theta \left( \frac{\theta_0}{\theta} \right)^2, \\
M_2^{\prime}& = -\left[ 0.520459 \Delta^2 \theta + 1.41017 \Delta^3 \theta \right] \left( \frac{\theta_0}{\theta} \right)^2, \\
A_1 &= \frac{5/2 M_2}{1+ 5/2 M_2},\\
A_2 &= 5 A_1  \left(1-M_2 \right) -\left(3+M_2^{\prime} - 5 M_2 \right),
\end{split}
 \end{align}
with $\Delta \theta = \theta - \theta_0$.
The equilibrium moments for this model are,
\begin{align}
 \begin{split}
  P^{\rm eq}_{\alpha \beta} &= \rho \theta \delta_{\alpha \beta} + \rho u_{\alpha} u_{\beta} \left(1-M_2^{\prime} \right) \\
  & + \frac{u^2}{2}\delta_{\alpha \beta} \left[ 1+ 4 M_2 ^{\prime} - 5 M_2 - \frac{\left(1+ 
\frac{25}{6}M_2 - \frac{25}{6}M_2^2  \right)}{1+\frac{5}{2}M_2}   \right], \\
q^{\rm eq}_{\alpha} &= 5 \theta u_{\alpha} \left( 1-M_2 \right) +\frac{1}{6} u^2 u_{\alpha} \Biggl[ 5 A_2 \left( 1-M_2 \right) \\
&- 7 A_1 \left( 1-M_3 \right) 
+ 7 \left( 3 +2 M_3^{\prime} - 5M_3 \right) \Biggr],\\
 \end{split}
\end{align}
where
\begin{align}
 \begin{split}
M_3           &= -\left[0.69253  \Delta^2 \theta - 2.49925  \Delta^2 \theta \right]\left( \frac{\theta_0}{\theta}\right)^3, \\
M_3 ^{\prime} &= -3.18168 \left( \frac{\theta_0}{\theta}\right)^3 \times \\
              & \quad \quad \quad \left(0.0934743  \Delta \theta  + 0.636954  \Delta^2 \theta + \Delta^3 \theta \right).
 \end{split}
\end{align}

The $P_{\alpha \beta}$ and $q_{\alpha}$ moments of the equilibrium distribution have errors of the order ${\cal O}(u^2 \eta^3)$ and ${\cal O}(u^3 \eta^3)$ respectively and match with the moments of the Maxwell-Boltzmann distribution up to high accuracy. 
Thus, model accurately recovers linearized hydrodynamics.

In the upcoming sections, we validate the accuracy and robustness of the proposed RD3Q41 model by simulating
various canonical flows related to acoustics, compressible, turbulent, multiphase and thermal flows.

\section{Acoustics}\label{section:acoustics}
In the previous sections, the RD3Q41 model has been shown to recover the full Navier-Stokes-Fourier equations as its macroscopic limit.
In this section, we select a few well-studied benchmarking problems related to the propagation of an acoustic pulse.

As a first example, we present an isothermal simulation of propagation of a 2D acoustic pulse.
The third direction has two lattice points, and periodic boundary conditions are imposed. 
An axisymmetric density pulse is initialized at the center of a uniform fluid of size $[-1,1]$ in both $x$ and $y$ directions as 
\begin{equation}  
\rho(x,y,t=0) = \rho_0 [1.0 + \rho'(x,y,t=0)], 
\end{equation}
where
\begin{align}
 \begin{split}
 \rho'(x,y,t=0) = \epsilon e^{-\alpha r^2},\quad \epsilon = 0.001, \quad \alpha = \frac{\ln(2)}{b^2}, \\
 b = 0.1 ,\quad \quad \quad  r=\sqrt{x^2 + y^2}.\quad \quad 
 \end{split}
\end{align}
A small value of $\epsilon$ is chosen to keep the amplitude of the acoustic perturbation small.
For low amplitudes of density fluctuations and low viscosity, the exact solution of the density fluctuation is given by
the analytical solution of the linearized Euler equations~\citep{tam1993dispersion,gendre2017grid}  as
\begin{equation}
 \rho'(x,y,t) = \rho_0 \times \frac{\epsilon}{2 \alpha} \int_0^{\infty} \exp\left(\frac{-\xi^2}{4 \alpha}\right) \cos(c_s \xi t) J_0(\xi r) 
\xi\ d\xi,
\end{equation}
where $J_0$ is the Bessel function of the first kind of zero-order \cite{abramowitz1965handbook}.

\begin{figure}
\centering
 \includegraphics[scale=0.63]{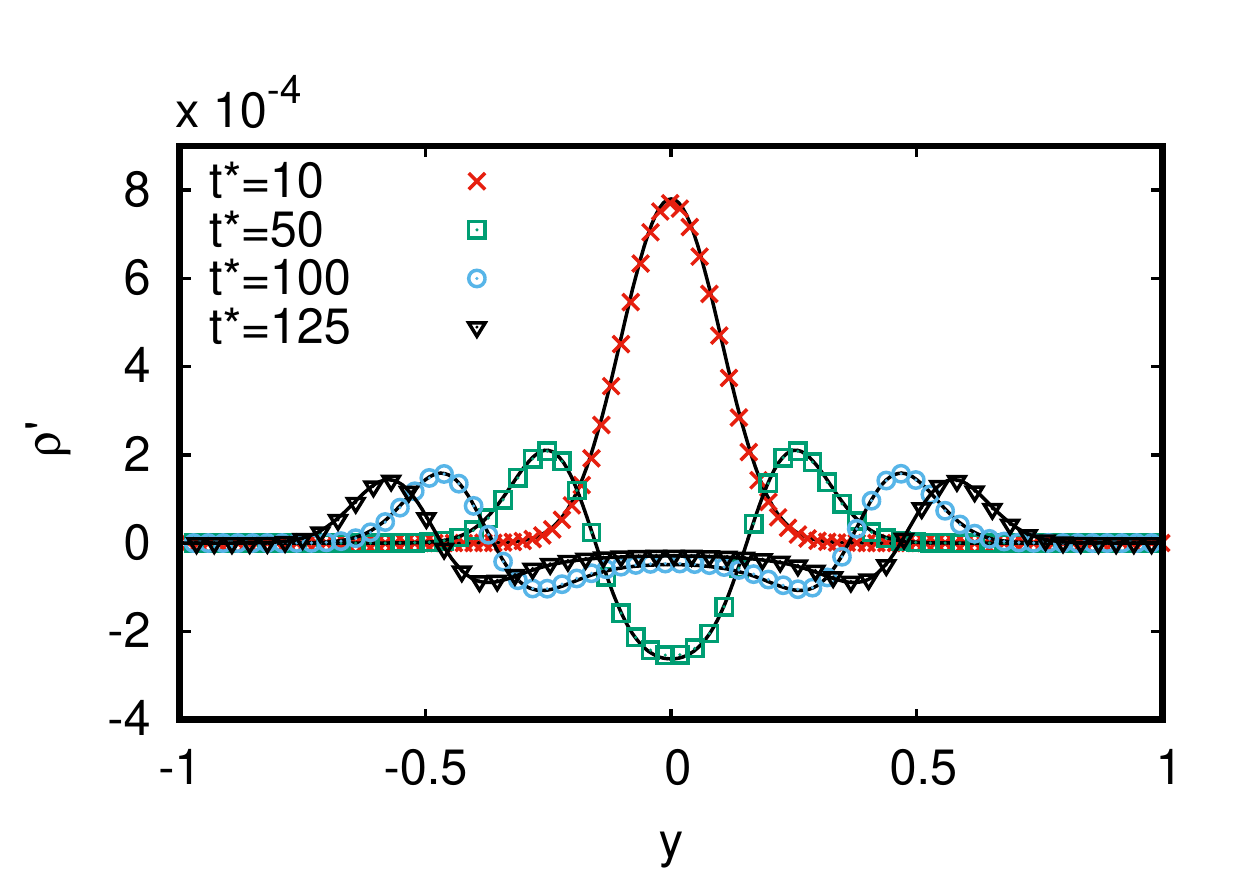}
\caption{Density fluctuations along the centerline in isothermal case (solid line) at different time steps compared with analytical solution (points).}
 \label{fig:soundSpeed_isothermal}
\end{figure}

An excellent match is observed upon comparing density fluctuations along the centerline from 
isothermal simulation using our model and from the analytical solution at various time steps ($t^*$) 
as shown in Fig. \ref{fig:soundSpeed_isothermal}.
This confirms that the linearized acoustics are captured accurately in an isothermal setting.  

\begin{figure}
\centering
 \includegraphics[scale=0.63]{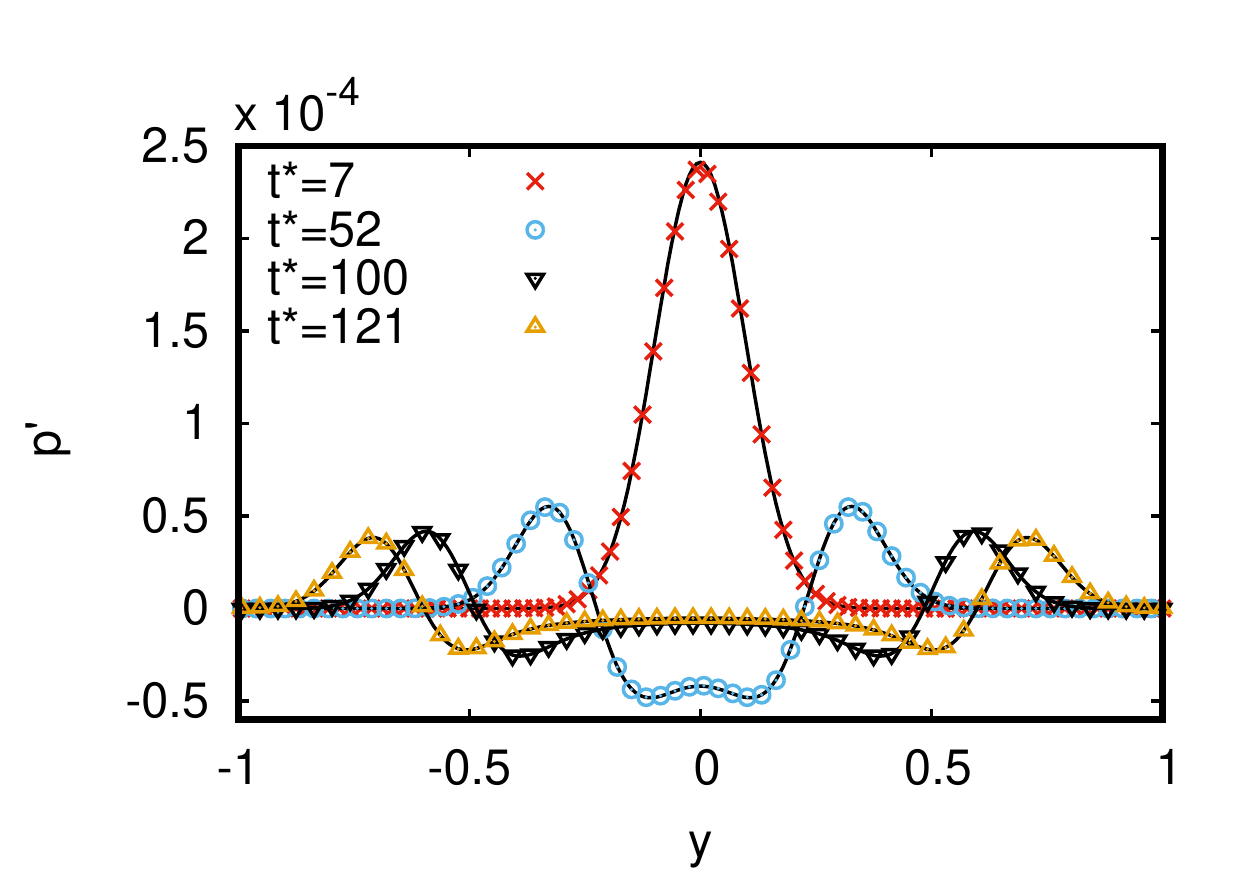}
\caption{Comparison of pressure fluctuations along the centerline at time $t^*$ from a thermal simulation (solid line) 
and isothermal simulation (points) at time $\sqrt{\gamma} \times t^*$
}
 \label{fig:soundSpeed_thermal}
\end{figure}

To show that the ratio of sound speeds in a thermal to isothermal model is $\sqrt{\gamma}$,
we perform thermal simulation of the same setup and compare the pressure fluctuations at times ($t^*$) with that obtained from an 
isothermal simulation at times $t=\sqrt{\gamma} \times t^*$.
The specific heat ratio $\gamma$ for the current model is $5/3$. 
The pressure fluctuation $p'$ is defined as  the deviation of pressure from rest condition as $p'= p - \rho_0 \theta_0$. 
It can be seen from Fig. \ref{fig:soundSpeed_thermal} that the profiles match very well confirming that the ratio of speed of sound is $\sqrt{\gamma}$ and 
energy conserving LB model indeed recovers the correct isentropic sound speed.

\subsection{3D acoustic spherical pulse source}
We demonstrate the utility of the present model in 3D via a simulation of a spherical pulse source. 
An acoustic pulse is initialized in the center of the domain of size $[-1,1]$ in $x, y$ and $z$ directions as
\begin{equation}  
\rho(x,y,z,t=0) = \rho_0 [1.0 + \rho'(x,y,z,t=0)], 
\end{equation}
where
\begin{align}
 \begin{split}
 \rho'(x,y,z,t=0) = \epsilon e^{-\alpha r^2},\quad \epsilon = 0.001, \quad \alpha = \frac{\ln(2)}{b^2}, \\
 b = 0.03 ,\quad \quad \quad  r=\sqrt{x^2 + y^2 + z^2}.\quad \quad 
 \end{split}
\end{align}

The exact solution for the density fluctuation is given as ~\citep{bogey2002three} 
\begin{equation}
 \rho'(x,y,t) = \frac{\epsilon}{2 \alpha \sqrt{\pi \alpha}} \int_0^{\infty} \exp\left(\frac{-\xi^2}{4 \alpha}\right) \frac{\sin(\xi r)}{\xi r} \xi^2 d\xi.
\end{equation}

\begin{figure}
\centering
 \includegraphics[scale=0.7]{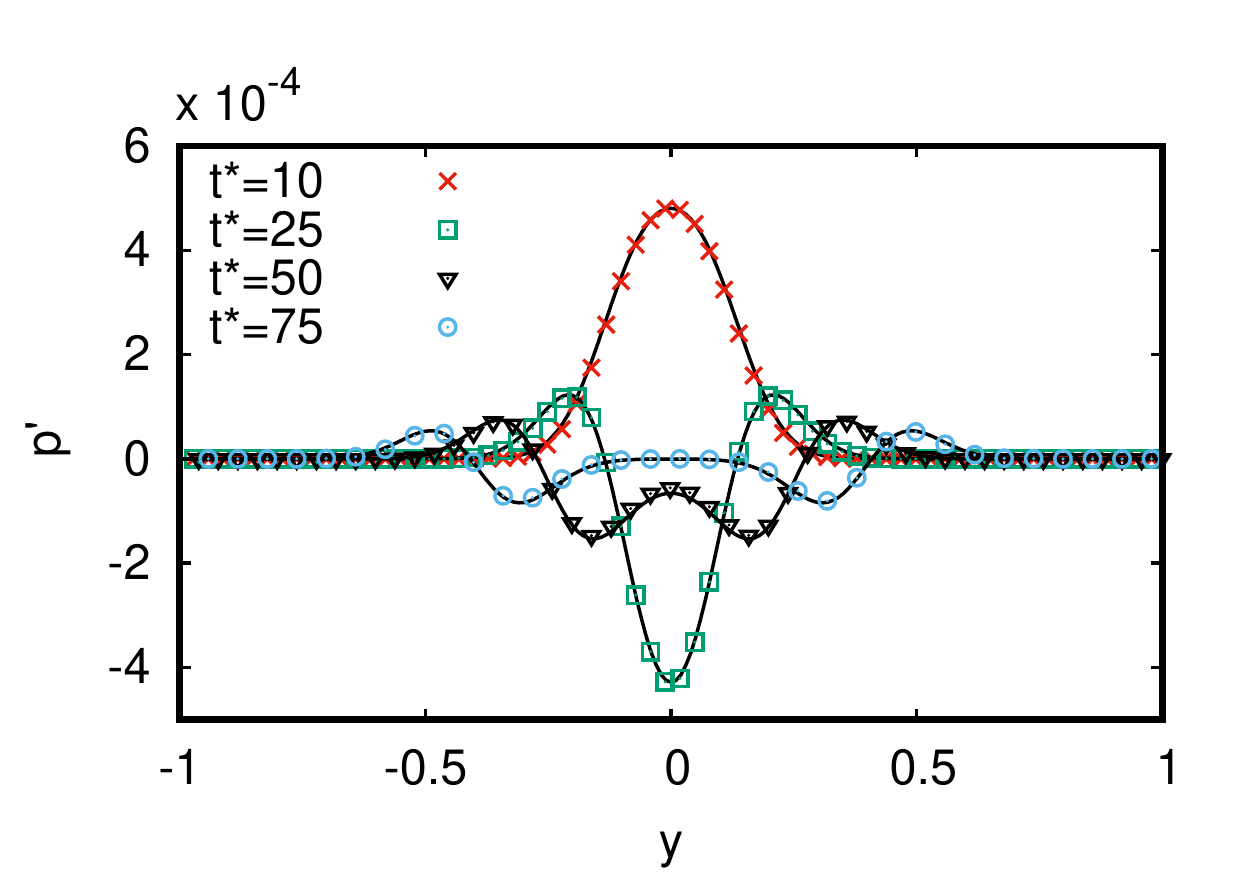}
\caption{Density fluctuations due to a 3D spherical pulse source along the y-axis at $(x,z) = (0.5,0.5)$ from LB simulation (line) and exact solution (points).}
 \label{fig:3d_pulse}
\end{figure}

The density fluctuations  from the LB simulation and the exact solution are plotted at a few time steps ($t^*$) along the $y$-axis at $(x,z) = (0.5,0.5)$ in Fig. \ref{fig:3d_pulse} and they show an excellent agreement. 

\subsection{Acoustic pulse reflecting off a planar wall}
\begin{figure}
\centering
 \includegraphics[scale=0.63]{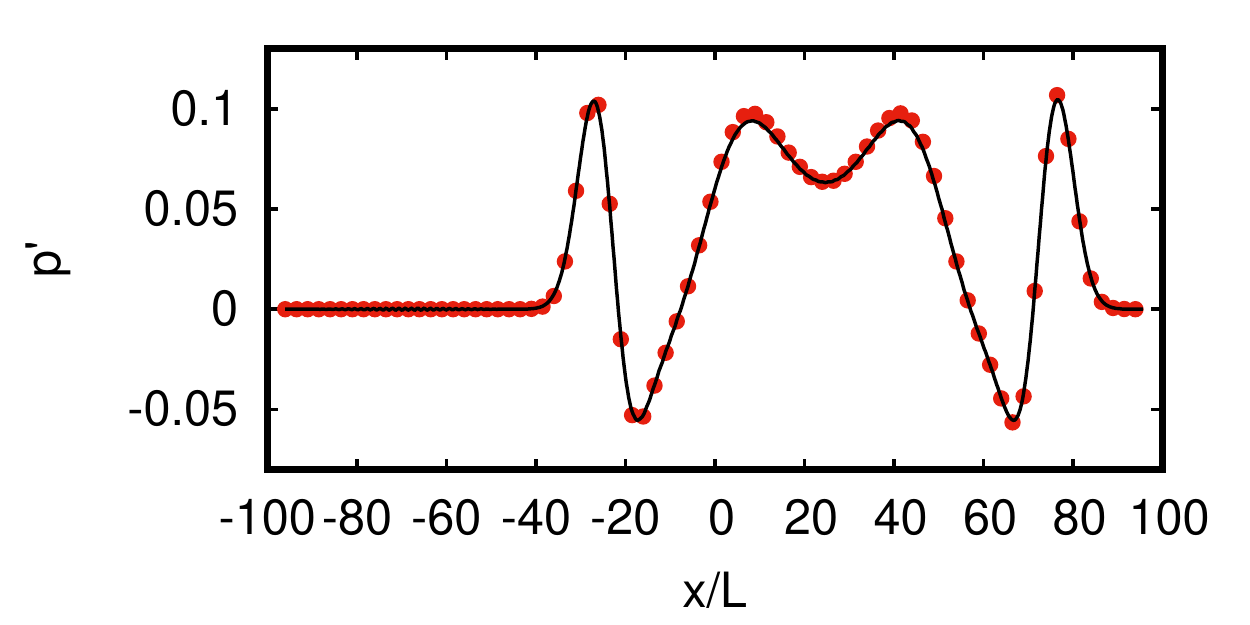}
\caption{Pressure fluctuations normalised with $A$  at $y=24L$ after $50$ convection times (LB-line, Analytical-points).}
 \label{fig:wallReflection_isothermal}
\end{figure}

We now demonstrate the capability of the model to simulate the interaction of acoustic waves with simple boundaries.
The interaction of an acoustic wave in a mean flow of Mach number ${\rm Ma}=0.5$ with an inviscid planar wall is simulated.
A domain of length $[-100L,100L]$ and $[0,200L]$ is chosen along the $x$ and $y$ axes. 
An acoustic pulse is initiated at $t=0$ as
\begin{equation}
p' = A \exp \left( { -\ln(2) \left[ \frac{x^{2}+ (y-25)^{2}}{25} \right] }  \right),
\end{equation}
with $u = 0.5c_s,\ v = 0 $ and $A=10^{-4}$.
This setup is identified as an effective test case to check the wall boundary conditions and 
an analytical solution for the pressure fluctuations is given in Ref.~\citep{hardin1994icase} as
\begin{equation}
p' = \frac{A}{2 \alpha} \int_0^{\infty} \exp\left(\frac{-\xi^2}{4 \alpha}\right) \cos(\xi t) \left[J_0(\xi \eta) + J_0(\xi \zeta) \right] \xi d\xi,
\end{equation}
where $\alpha= \ln(2)/25$, $\eta= \sqrt{(x-{\rm Ma} \times  t)^2+(y-25)^2}$, and $\zeta =\sqrt{(x+{\rm Ma}  \times  t)^2+(y-25)^2}$.
We choose four lattice points per $L$ for this simulation.
Pressure fluctuations normalized with $A$ obtained from the current simulation and the analytical solution after a time of $50t_c$  at $y=24L$ shows an excellent match in Fig. \ref{fig:wallReflection_isothermal}. 
The convection time $t_c$ here is defined as $L/c_s$. 

\begin{figure}
\begin{center}
 \includegraphics[scale=0.63]{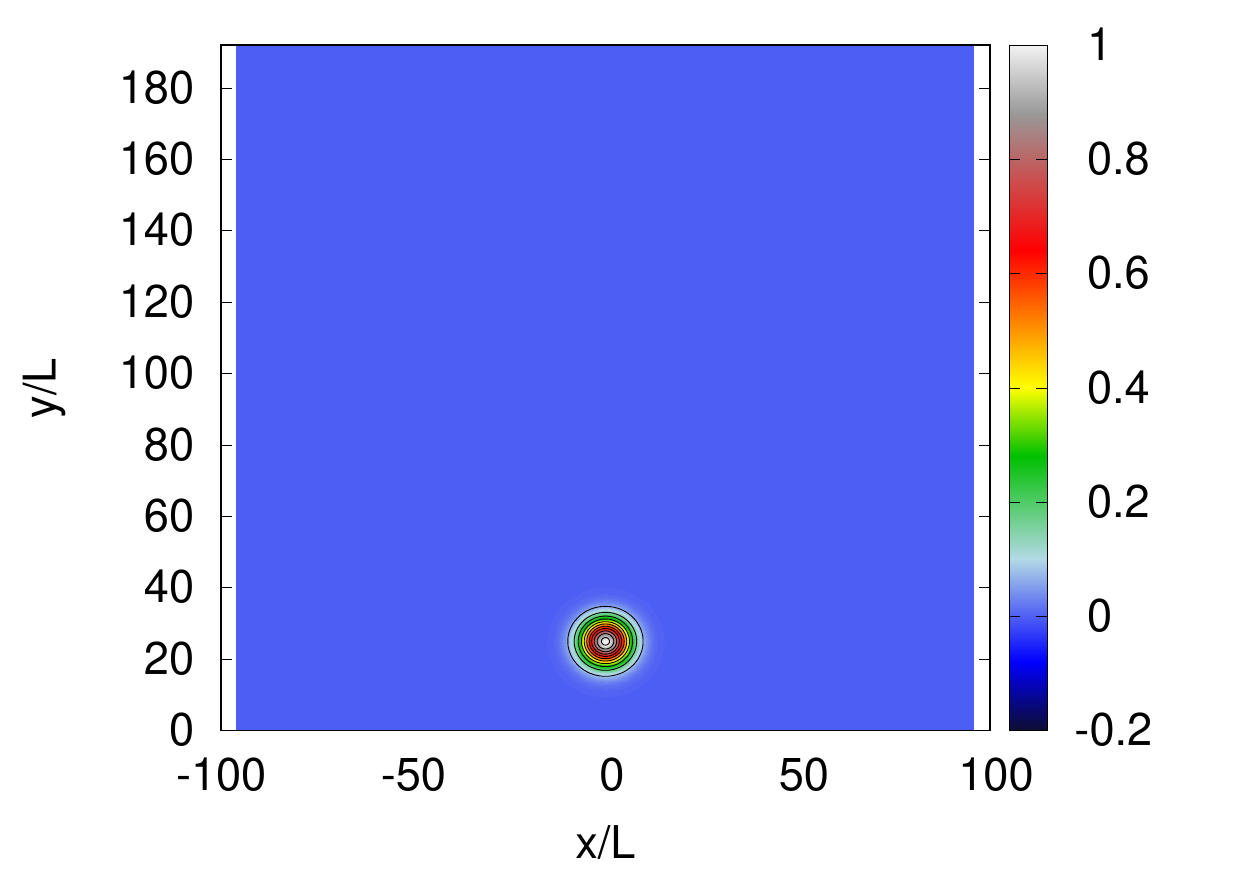}
 \includegraphics[scale=0.63]{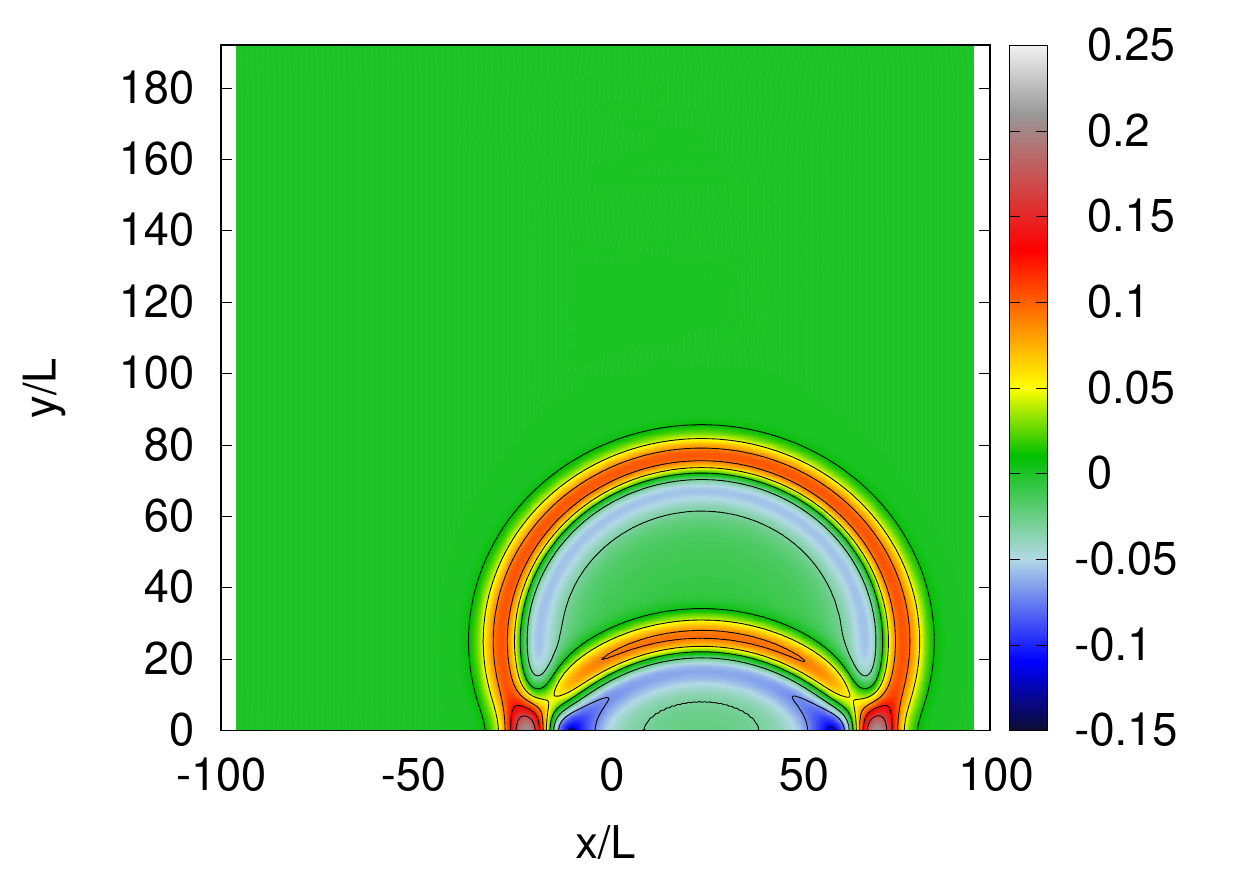}
 \caption{Contours of pressure fluctuations normalised with $A$ at $t=0$ (up) and $t=50$ convection times (bottom).}
 \label{fig:pressureContours_wall}
\end{center}
\end{figure}

\subsection{Acoustic scattering off a rigid cylinder}
The acoustic scattering off a rigid cylinder is one of the benchmark problems identified as a simplified model to find the sound scattered by aircraft fuselage produced by the propeller~\citep{tam1997second}. The fuselage was approximated as a circular cylinder and the source of sound reduced to a point source.
Presence of a curved boundary also makes this a natural extension to the previous test case for validating the wall boundary conditions.

A cylinder of diameter $D_0$ is placed at the center of a domain of length $[-15D_0,15D_0]$ in both $x$ and $y$ directions. 
At a distance of $4D_0$ from the center of the cylinder along the $x$-axis a Gaussian acoustic pulse is initialized at $t=0$ as
\begin{equation}
p' = A \exp \left( { -\ln(2) \left[ \frac{(x-4)^{2}+ y^2}{0.2^2} \right] }  \right),
\end{equation}
where $A = 10^{-4}$.
The analytical solution for this test case is given as~\citep{tam2004optimized} 
\begin{equation}
p' = Re \left \{ \int_0^{\infty} \left( A_i(x,y,\omega) + A_r(x,y,\omega) \right)  \omega e^{- i\omega t} d\omega \right \}.
\end{equation}

The $A_i$ and $A_r$ stand for the amplitudes of incident and reflected waves given by
\begin{equation}
A_i(x,y,\omega) = \frac{1}{2b} e^{-\omega^2/(4b)} J_0(\omega r_s),
\end{equation}
where $r_s = \sqrt{(x-4D_0)^2+y^2}$, $J_0$ is Bessel function of zero order, and
\begin{equation}
A_r(x,y,\omega) = \sum_{k=0}^{\infty} C_k(\omega) H_k^{(1)}(r\omega) \cos(k\omega),
\end{equation}
with 
\begin{align*}
C_k(\omega) &= \frac{1}{2\pi} \exp \left\{-\omega^2/(4b) \right\} \frac{\epsilon_k}{\left[H_k^{(1)}\right]^{'}} \\
            &   \quad   \quad   \quad  \quad \int_{0}^{\pi} J_1(\omega r_{s0}) \frac{r_0 - 4 \cos(\theta)}{r_{s0}} \cos(k\theta) d\theta,
\end{align*}
where $\epsilon_0 = 1$, $\epsilon_k = 2$ for $k\neq 0$, $r_0 = 1/2$ and $r_{s0} = \sqrt{0.25+x_s^2D_0^2 - x_s D_0 \cos(\theta)}$ while 
$H_k^{(1)}$ is the Hankel function of first order \citep{abramowitz1965handbook}.

\begin{figure}
\begin{center}
  \includegraphics[trim={0cm 0.9cm 0.0cm 0.0cm},clip,scale=0.63]{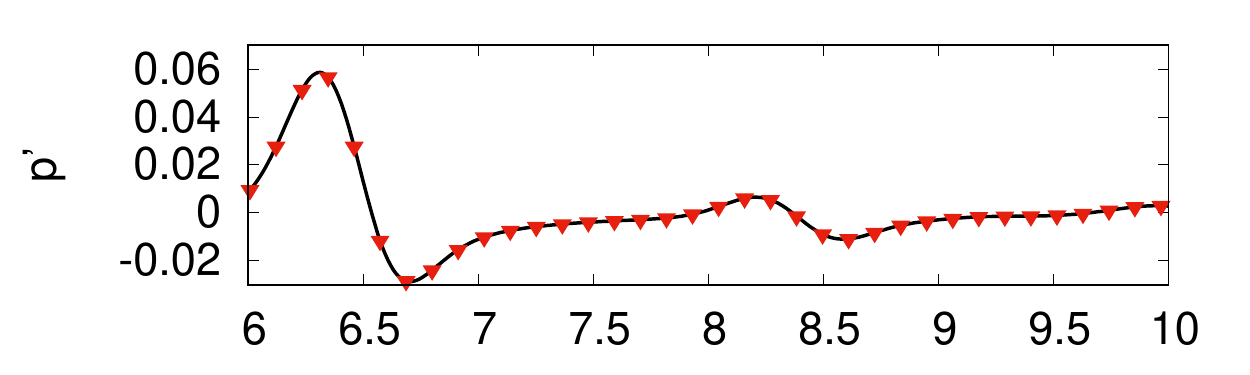}  
  \includegraphics[trim={0cm 0.9cm 0.0cm 0.3cm},clip,scale=0.63]{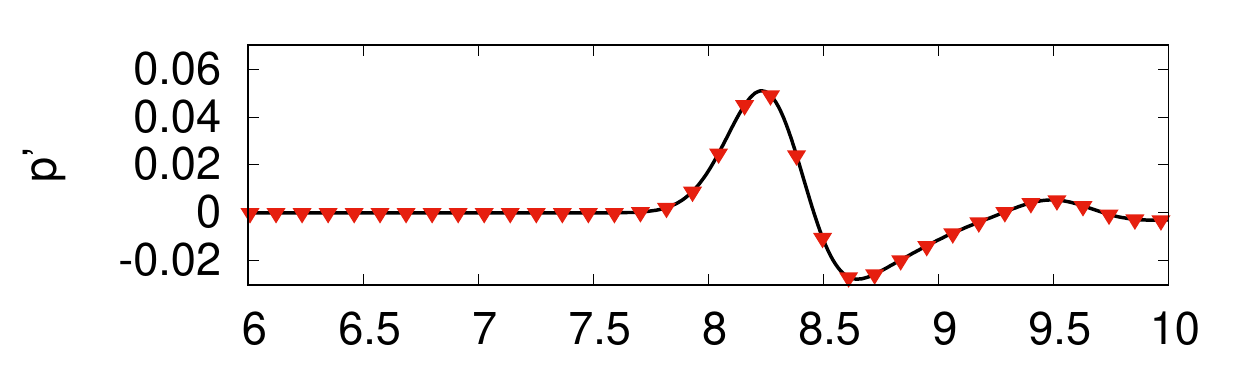}                    
  \includegraphics[trim={0cm 0cm   0.0cm 0.3cm},clip,scale=0.63]{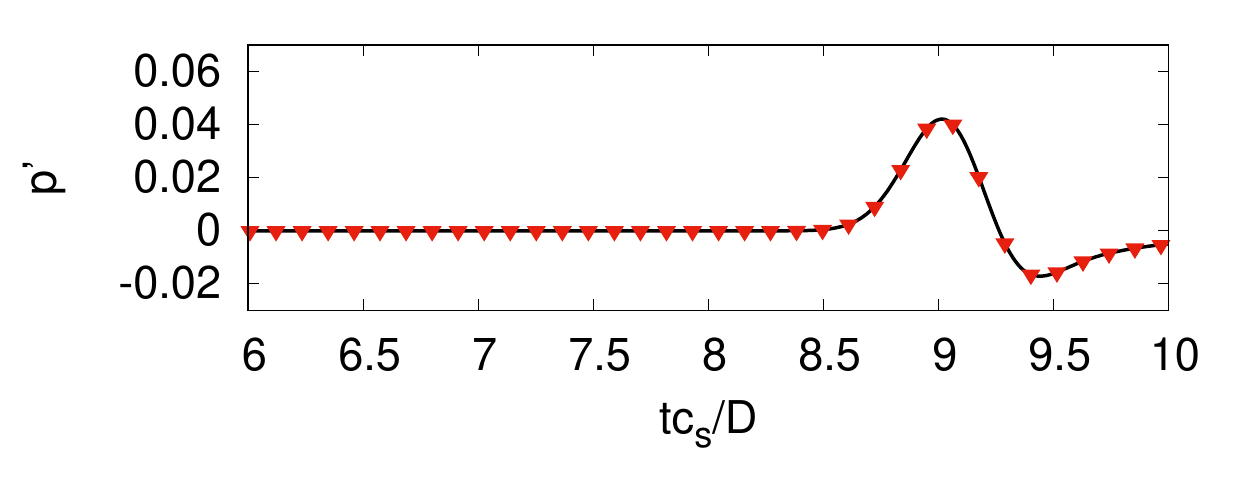}                    %
 \caption{Time evolution of pressure fluctuations normalised by $p_0 A$ at points $r=5D_0$ and $\theta=90\degree, 135\degree$ and
 $180\degree$ from top to bottom (LBM-solid line, Analytical-points).}
 \label{fig:pulse_cylinder}
\end{center}
\end{figure}

Pressure fluctuation profiles normalised with $p_0 A$ from times $t = 6$ to $10$ convection times at three points 
$A(r=5D_0,\theta=90\degree)$, $B(r=5D_0,\theta=135\degree)$ and $C(r=5D_0,\theta=180\degree)$ are shown in Fig. \ref{fig:pulse_cylinder}. 
The convection time scale is defined based on the diameter of the cylinder and speed of sound as $D_0/c_s$.
The points and time interval are chosen such that only the acoustic wave reflected off the cylinder passes through these points.
The pressure fluctuation profiles show a good agreement with the exact solution demonstrating the capability of the current model for
solving computational aeroacoustics problems with non-trivial boundary shapes.

\begin{figure*}
\centering
 \includegraphics[scale=0.63]{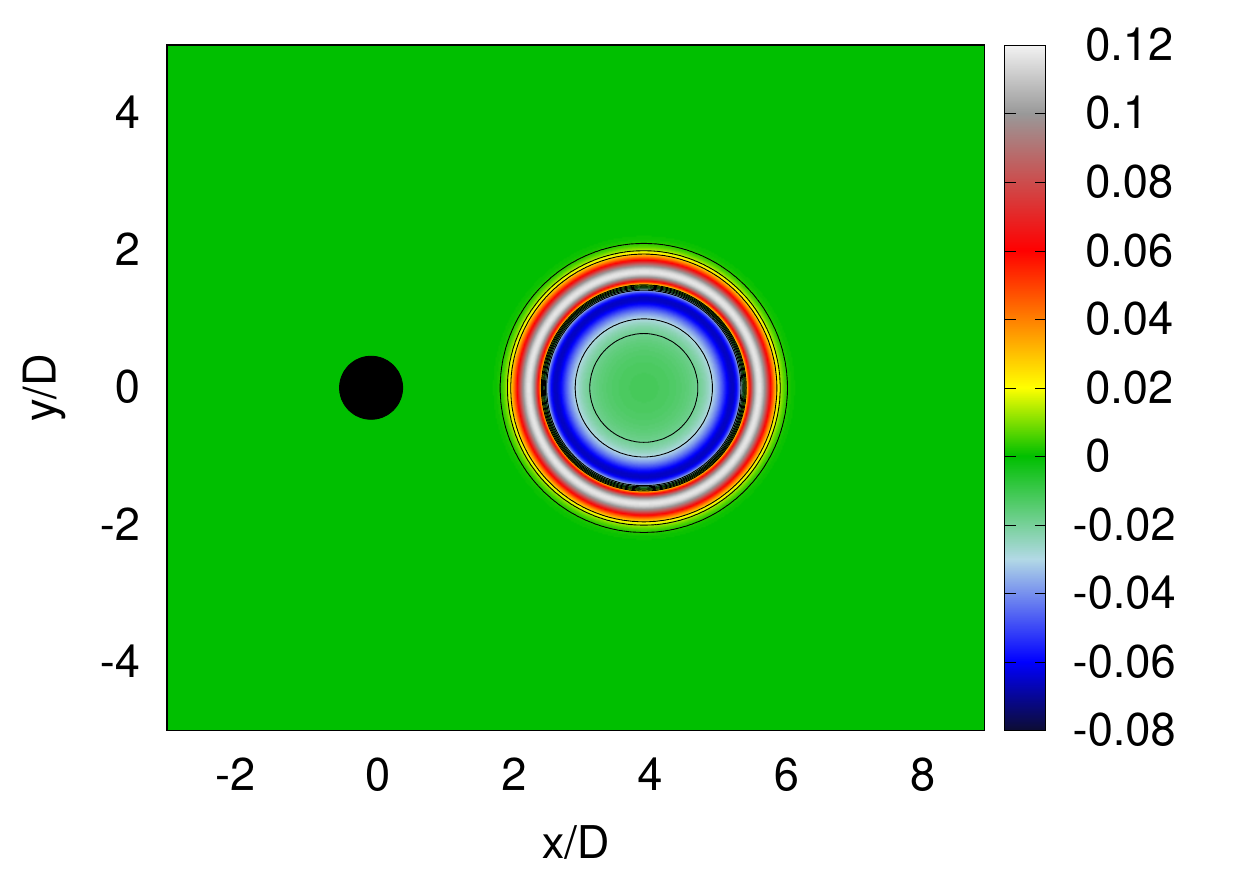}
 \includegraphics[scale=0.63]{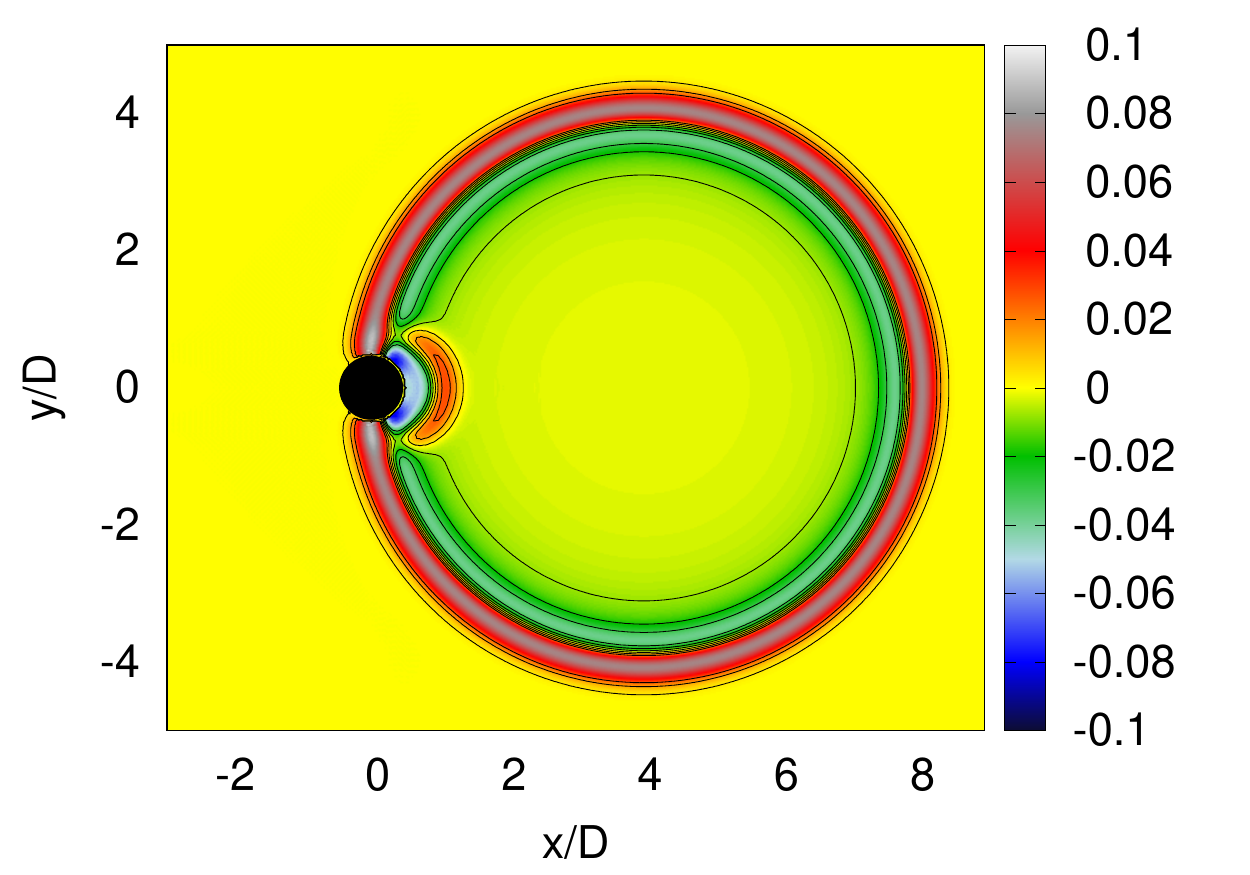}
 \includegraphics[scale=0.63]{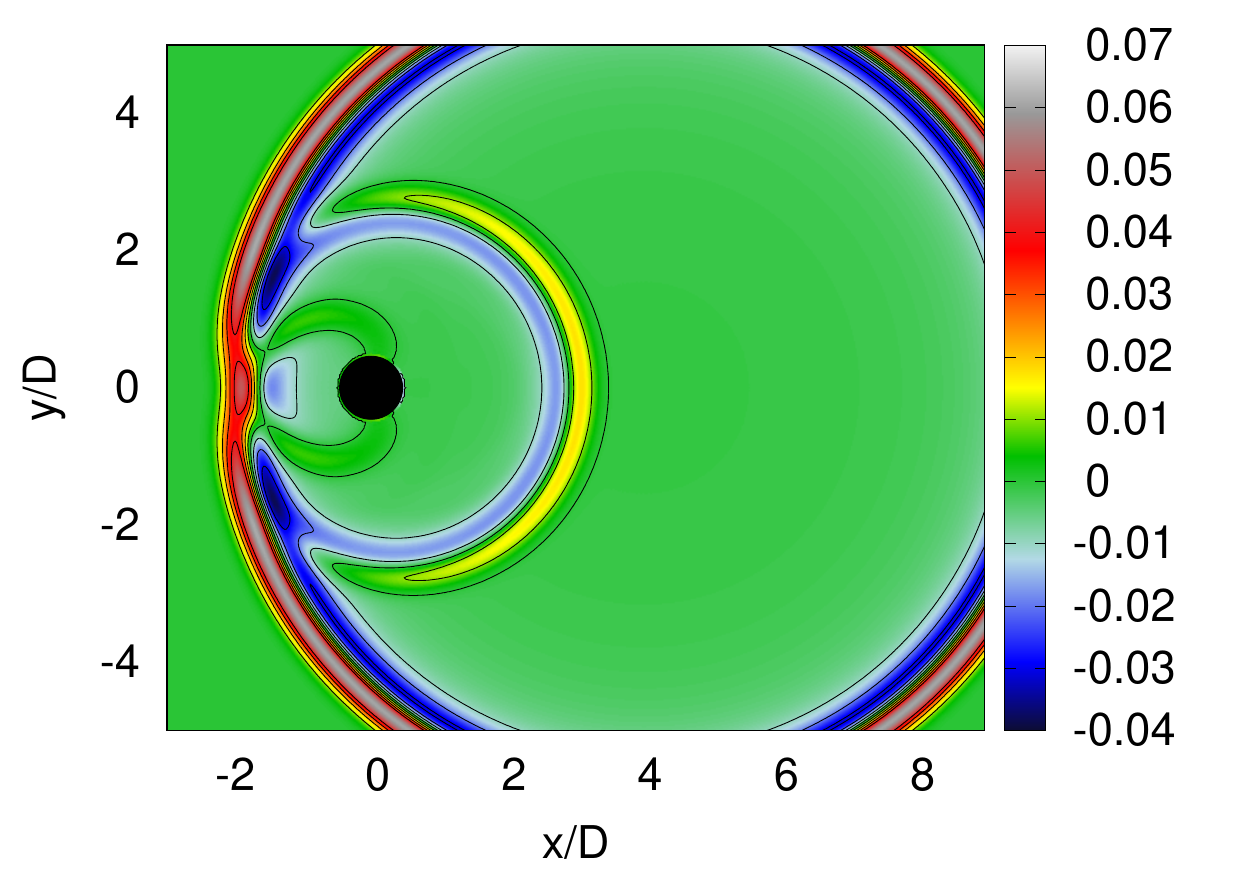}
 \includegraphics[scale=0.63]{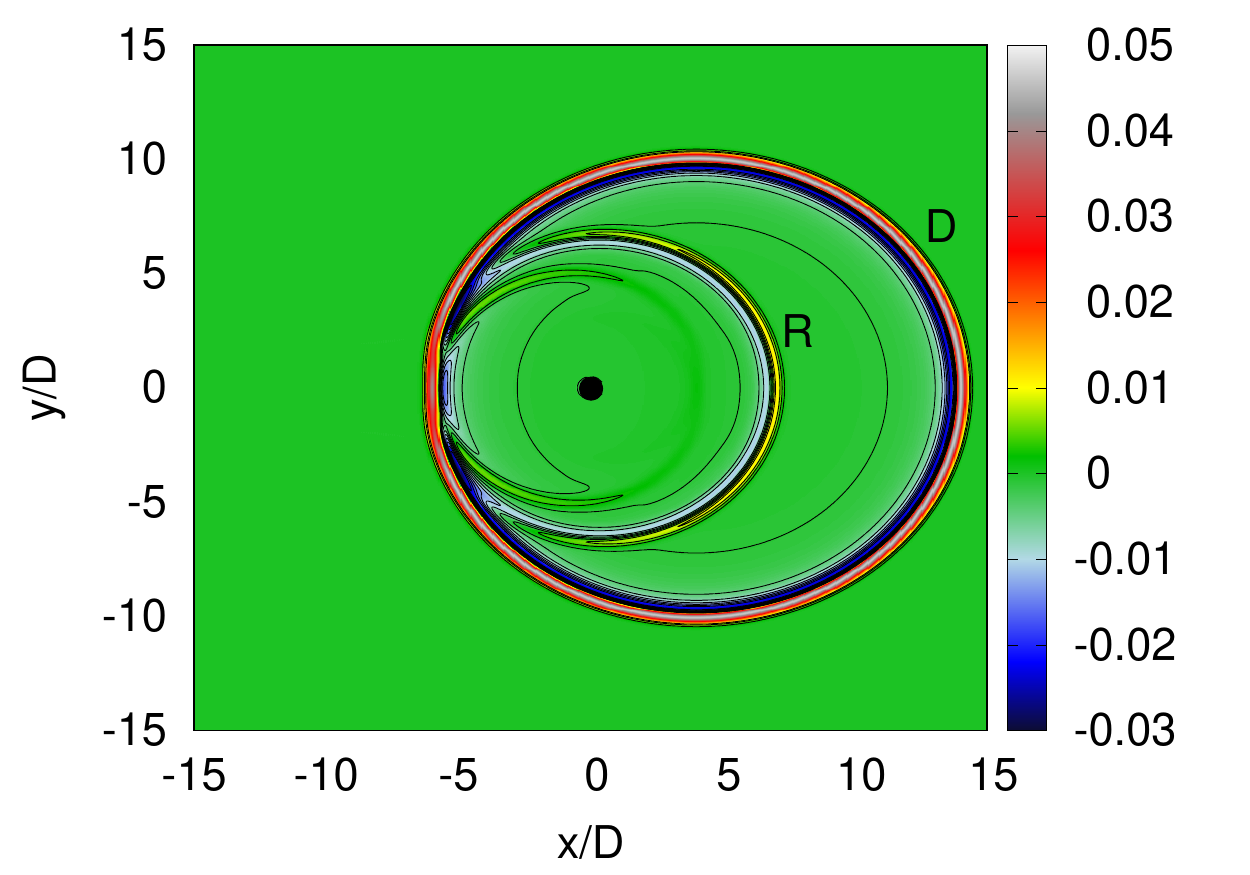}
 \caption{Isocontours of pressure fluctuations normalised by $p_0 A$ at $t = 1.6, 4.0, 6.0 $ and $10.0$ convection times where D is the direct wave 
 and R is the wave reflected off the surface of the cylinder.}
 \label{fig:contours}
\end{figure*}

Isocontours of pressure fluctuations normalized by $p_0 A$ are shown in Fig. \ref{fig:contours} at times $t = 1.6, 4.0, 6.0 $
and $10.0$ convection times. The contours also show the direct wave D and the acoustic waves reflecting off the surface of the cylinder R.

\section{Thermal Flows}\label{section:thermalFlows}
In this section, we demonstrate the efficacy of the RD3Q41 model for simulating thermal
flows by studying two test cases: a Couette flow with a temperature gradient
(where the viscous heat dissipation becomes crucial) and heat conduction in 
a 2D cavity.

\subsection{Viscous heat dissipation}

We consider the steady state of flow induced by a wall at $y=H$ moving 
with a constant horizontal velocity $U_0$ and maintained at a constant elevated temperature $T_1$.
The lower wall at $y=0$ is kept stationary at a constant temperature $T_0$ ($T_1 > T_0$). 

This setup is well suited to validate the effect of viscous heat dissipation. 
Each layer of fluid drags the layer below it due to friction, which results in
the mechanical energy being converted to thermal heating, and therefore, 
the heat produced affects the temperature profile in bulk. 
The analytical solution for the temperature profile for this
setup is~\citep{bird2015introductory}
\begin{equation}
 \frac{T - T_0}{\Delta T} = 
\frac{y}{H} + \frac{\rm Ec}{2} \frac{y}{H} \left(1 - \frac{y}{H} \right),
 \label{eckert}
\end{equation}
where $\Delta T = T_1 - T_0$ is the temperature difference between the 
two walls and ${\rm Ec} = U_0^2/(c_p \Delta T)$ is the Eckert number 
that represents the ratio of viscous dissipation to heat conduction with 
$c_p=5/2$ as the specific heat at constant pressure.

\begin{figure}
\centering
 \includegraphics[scale=0.6]{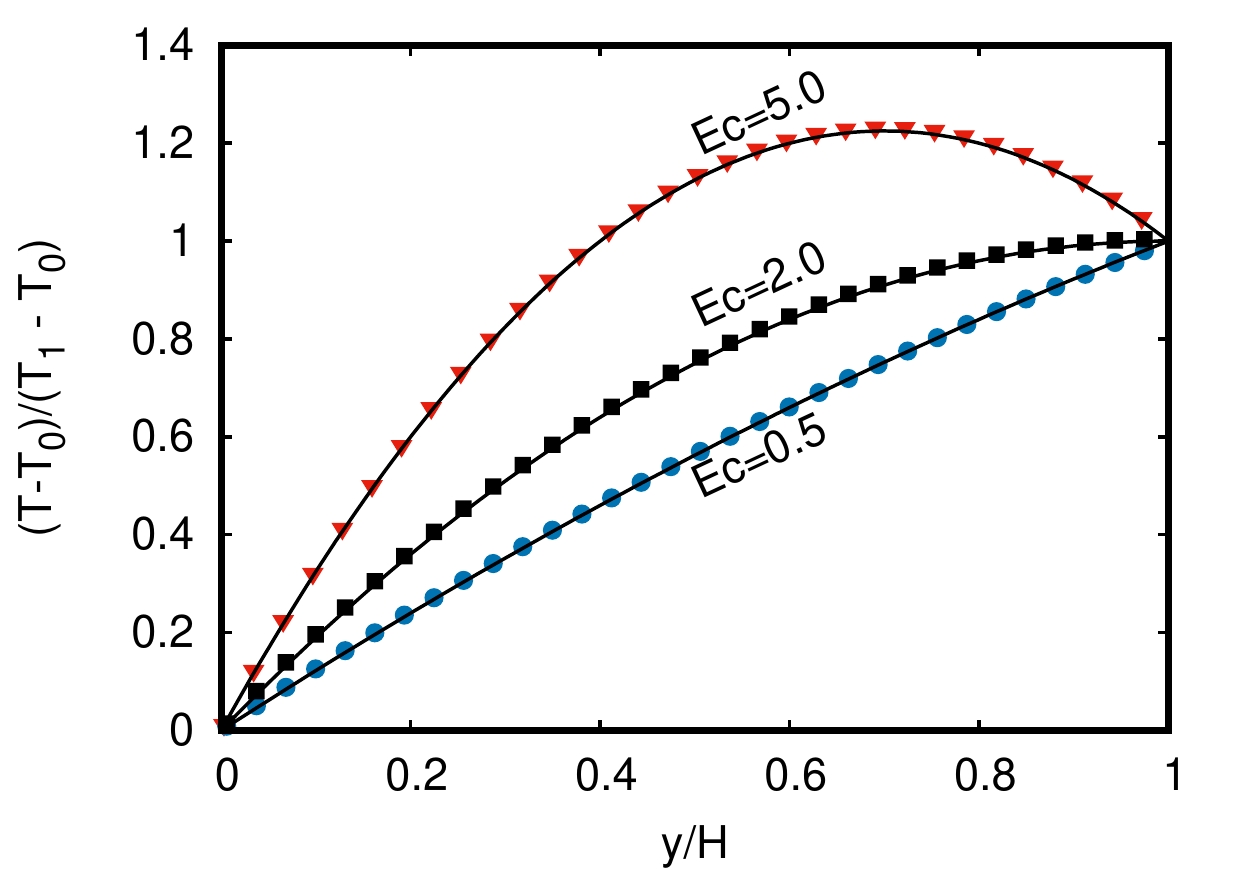}
\caption{Mean planar temperature profiles obtained from $RD3Q41$ at steady 
state (symbols) compared to the analytical solution (lines). }
\label{fig:eckert}
\end{figure}

Simulations were performed for ${\rm Ec} = 0.5, 2.0, 5.0$ with $U_0 = 0.02$ and  $\Delta \theta$ 
calculated according to respective Eckert numbers. The walls were maintained at 
temperatures $\theta_0+0.5\Delta \theta$ and $\theta_0-0.5\Delta \theta$. 
Kinetic boundary conditions as described in Ref.~\citep{ansumali2002kinetic} have been applied at the walls, and periodic boundary conditions are used in the other two directions.
Figure \ref{fig:eckert} compares the temperature profiles obtained analytically and via simulations and they are found to 
agree well.

\subsection{2D cavity heated at the top}

\begin{figure}
 \includegraphics[scale=0.63]{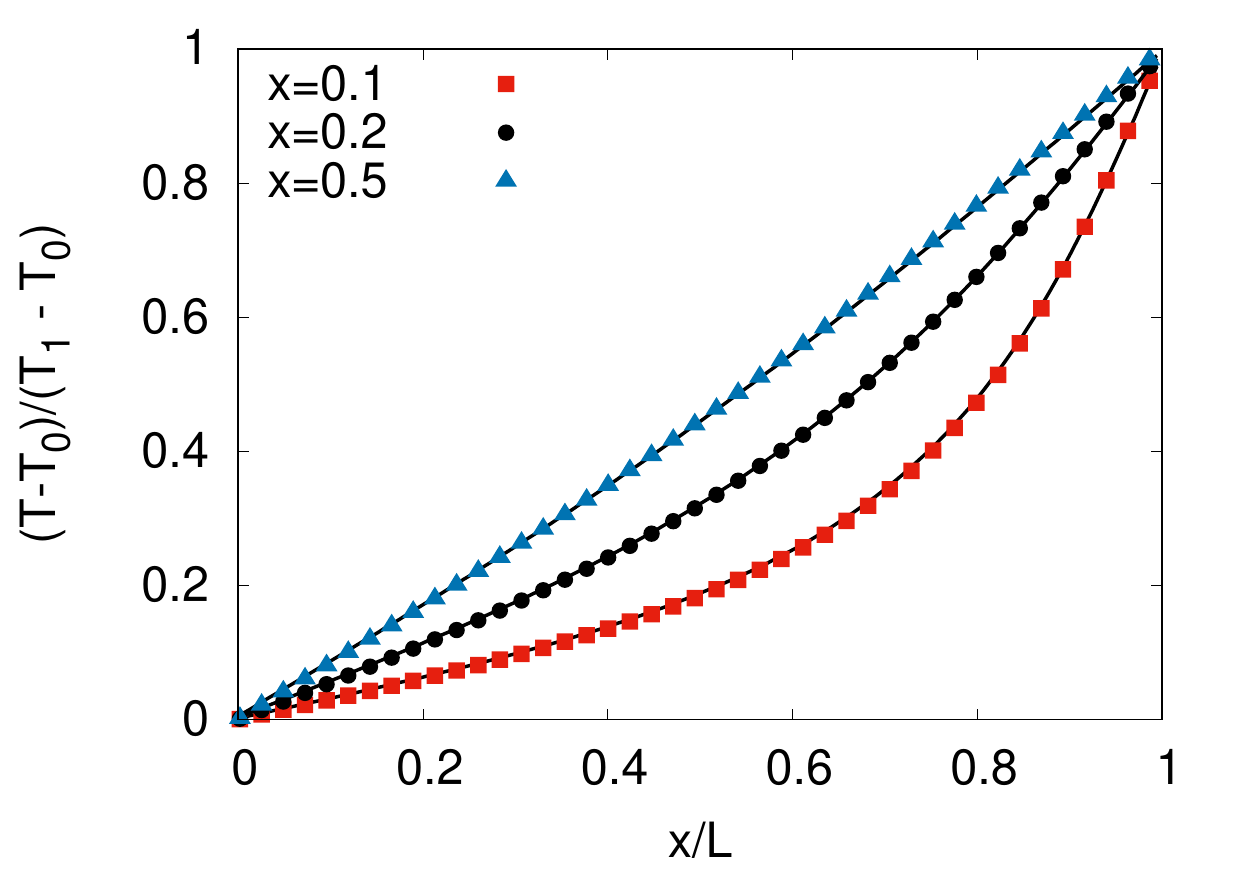}
 \includegraphics[scale=0.63]{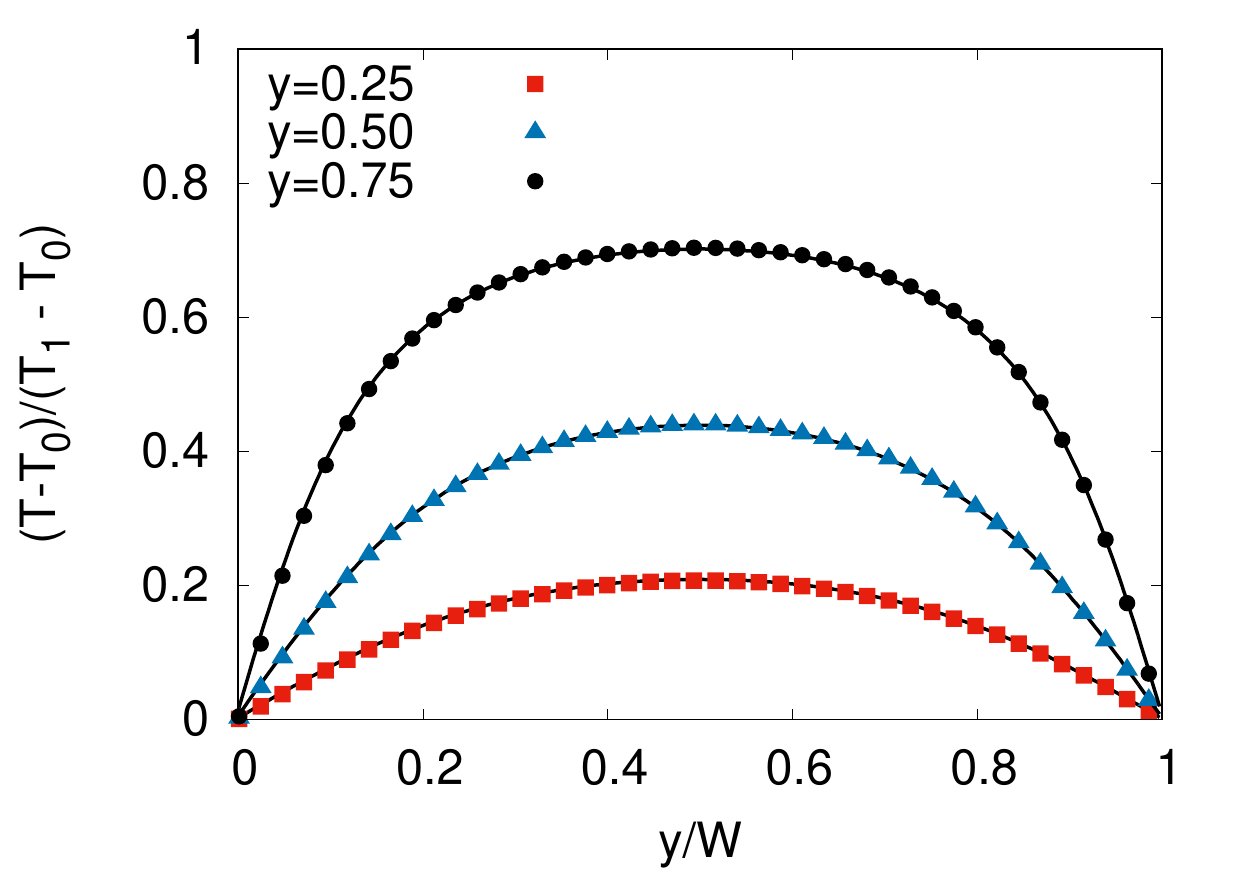}
 \caption{Steady state temperature profiles at sections along X-axis (left) and Y-axis (right). The symbols are the solution form the RD3Q41 model while the lines represent the analytical solution.}
 \label{fig:2DConduction}
\end{figure}

In this setup, the fluid is confined in a rectangular cavity bounded with stationary walls on all four sides. 
The height of the cavity is $W$, and its length is $L$.
The top wall is maintained at temperature ($T_1$), and the other three walls are maintained at temperature $T_0$ ($T_1 >T_0$). 
\begin{figure}
\centering
 \includegraphics[scale=0.4]{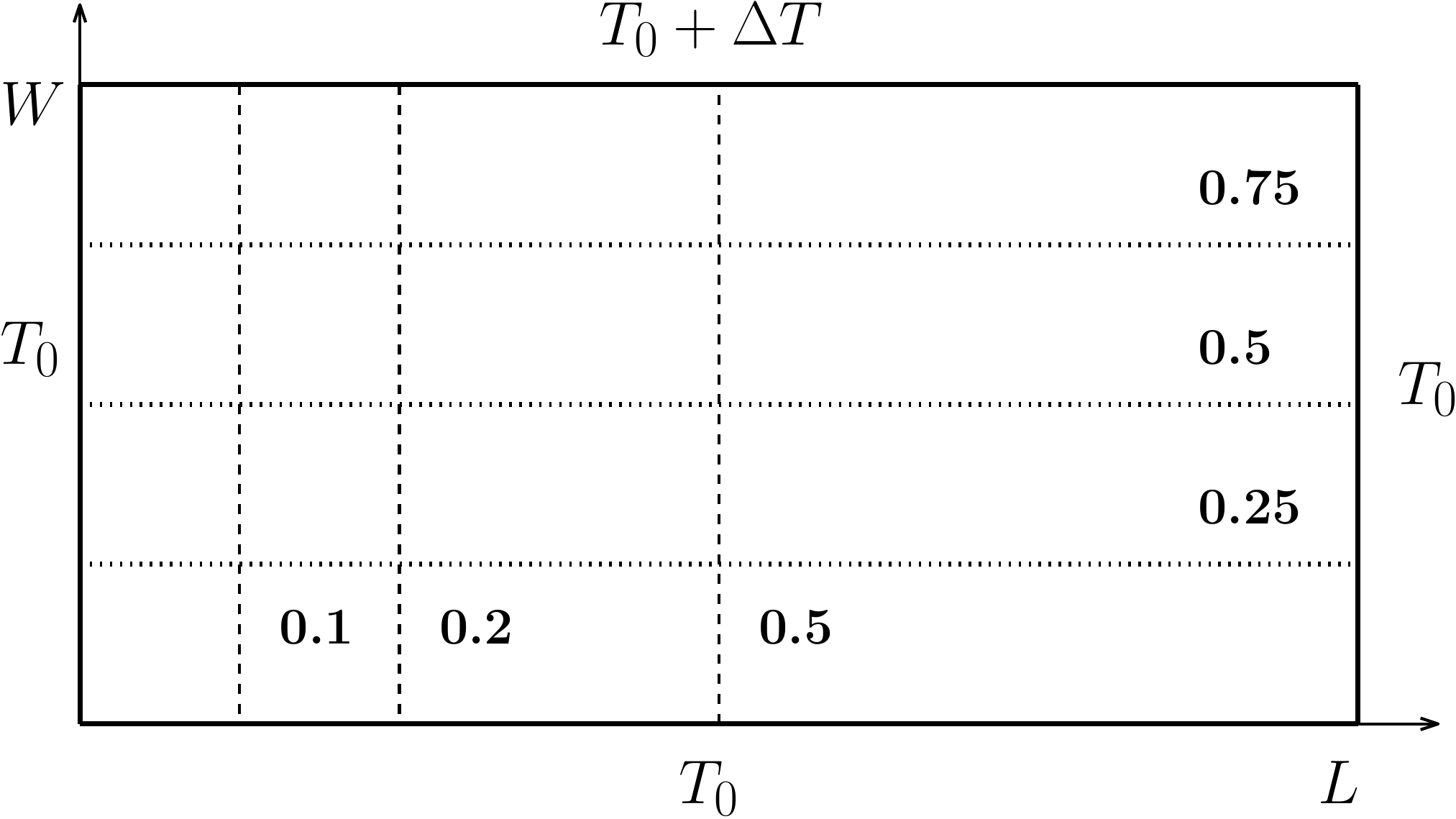}
\caption{Setup for the 2D cavity heated at the top.}
 \label{fig:2DConduction_setup}
\end{figure}
The temperature profile for this setup at the steady state is given by
\begin{equation}
 \frac{T - T_0}{T_1 - T_0} = \frac{2}{\pi} \sum_{n=1}^{\infty} \frac{(-1)^{n+1}+1}{n} \sin\left({n\pi x}\right) 
\frac{\sinh(n\pi y)}{\sinh(n \pi H/L)}.
\end{equation}
Kinetic boundary conditions as described in Ref.~\citep{ansumali2002kinetic} have been applied at the top and bottom walls and periodic boundary conditions are applied in z-direction. 
The temperature profiles along the constant $x=0.1,0.2,0.5$ and constant $y=0.25,0.5,0.75$ are shown in Fig. \ref{fig:2DConduction}
and can be seen to match well with the analytical solution.

These two test cases prove that the thermal transport phenomenon is modeled correctly in the RD3Q41 model.

\section{Thermoacoustics}\label{section:TA}
In the previous sections, we showed that the current model accurately predicts acoustic and thermal phenomena individually.
We now demonstrate the capability of the RD3Q41 model in simulating flows involving both thermal and acoustic phenomena.

\begin{figure}
\centering
 \includegraphics[scale=0.4]{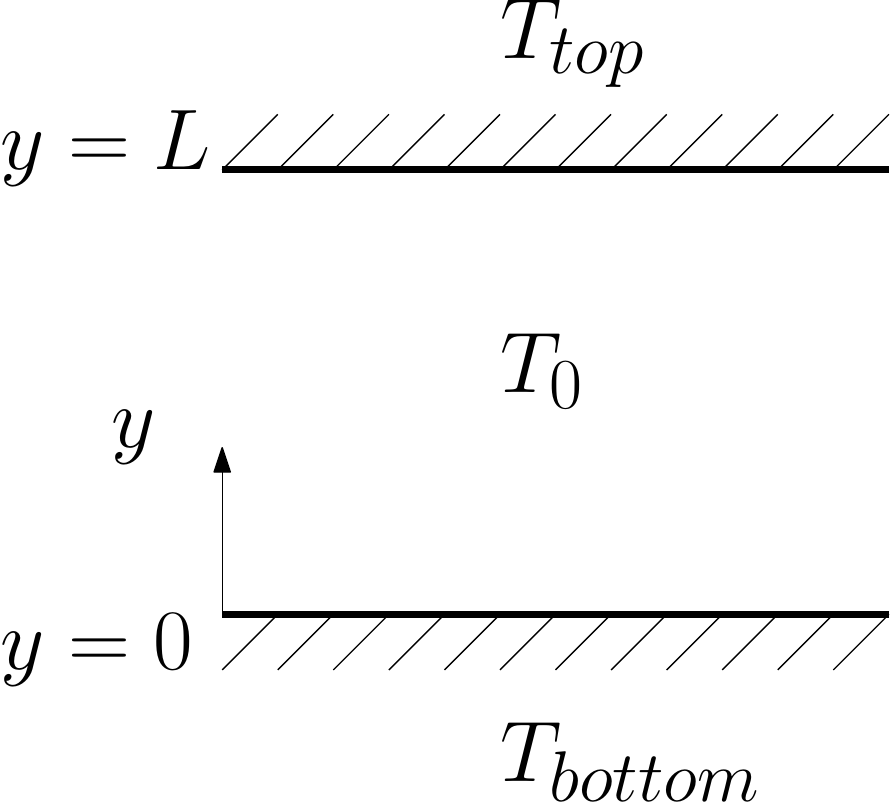}
\caption{A schematic of the setup for thermoacoustic convection}
 \label{fig:SetupThermoacoustics}
\end{figure}

A simple example is when a compressible fluid confined between two walls is heated rapidly on one end, it sets up a convective current   
 \citep{larkin1967heat,parang1984thermoacoustic,HUANG1997407}. 
The heated wall creates a pressure wave which reflects back and forth in the medium until it gets dissipated by viscosity. 
These pressure waves are called thermoacoustic waves because of the acoustic nature of the origin of these waves.
It is understood that this thermally induced motion is known to enhance the heat transfer in the medium by addition of convective mode to the conductive mode of heat transfer~\citep{larkin1967heat}.
The convective mode slowly dissipates due to the dissipation of the pressure wave, and conduction becomes the only mode of heat transfer~\citep{spradley_churchill_1975}. 
The many time and length scales present in the system along with the compressible nature of the fluid makes numerical modeling of thermoacoustic convection a challenging problem.

The setup consists of a fluid column of length $L$ enclosed between two walls. 
It is initially at a uniform temperature of $T_0$, as shown in Fig. \ref{fig:SetupThermoacoustics}. 
At times $t>0$ the bottom wall is maintained at temperature $T_0$ while the top wall is rapidly heated to a temperature $T_0 + \Delta T$.
Kinetic boundary conditions, as described in Ref.~\citep{ansumali2002kinetic}, have been applied at the walls.
The compression and rarefaction of the thermoacoustic waves due to rapid heating create a fluctuating velocity in the domain which results in a 
significant increase in the rate of heat transfer relative to pure conduction.

\begin{figure}
\centering
 \includegraphics[scale=0.63]{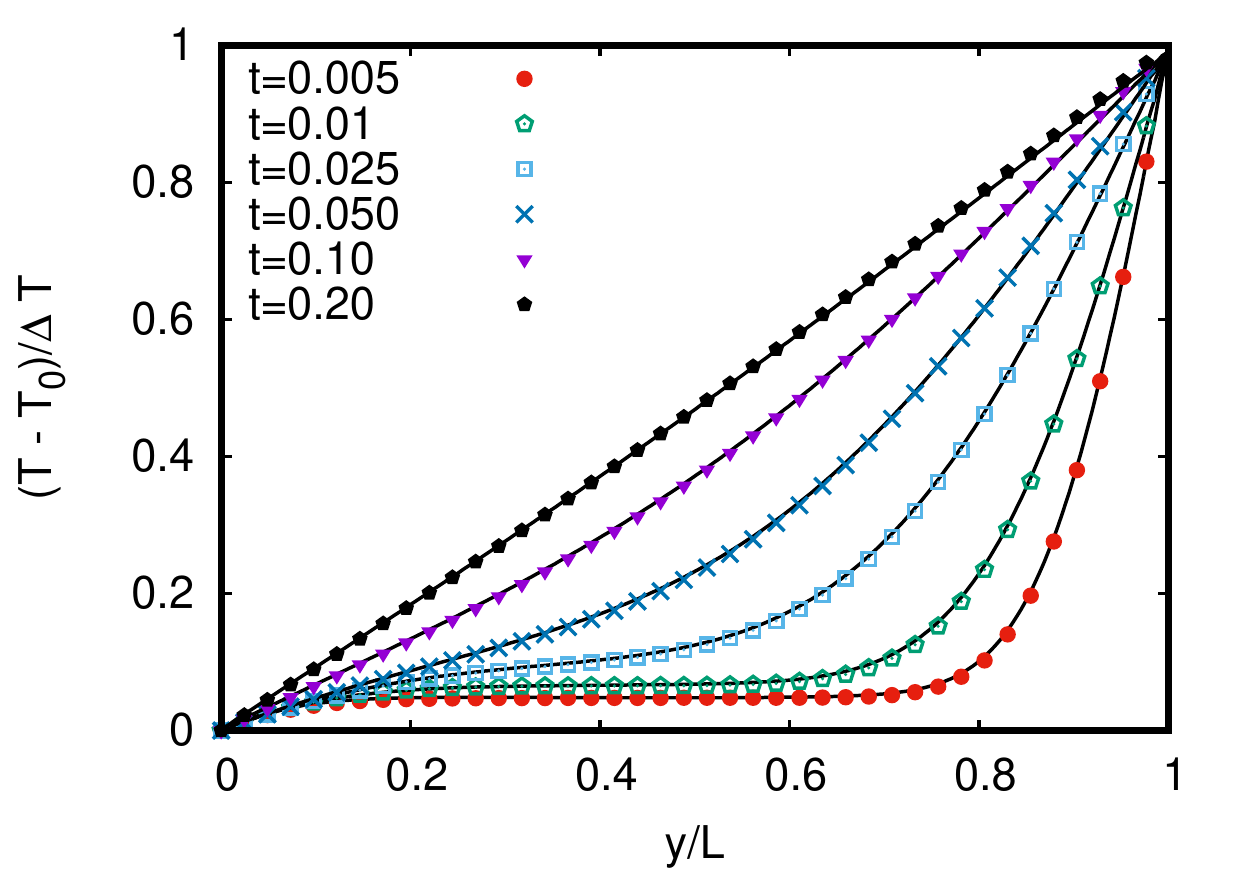}
 \includegraphics[scale=0.63]{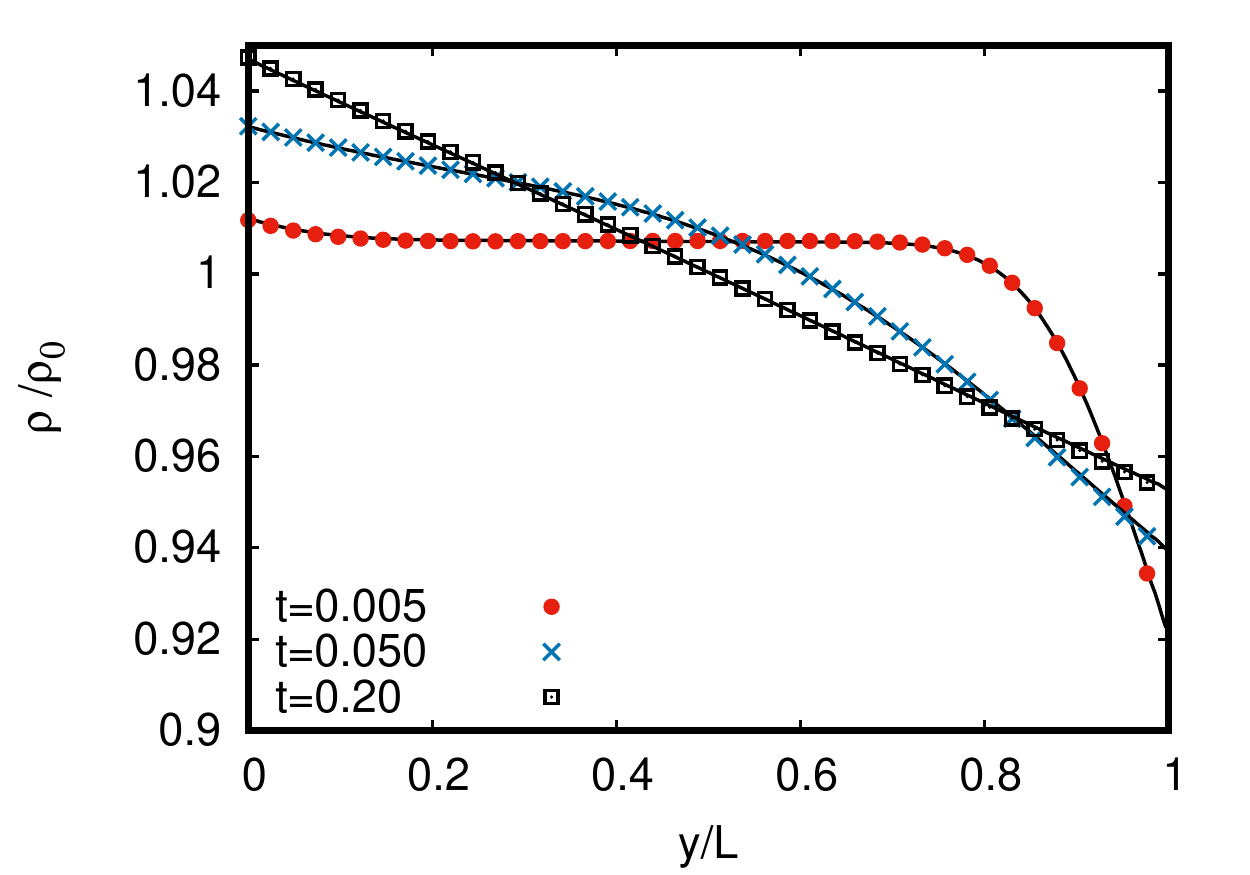}
 \includegraphics[scale=0.63]{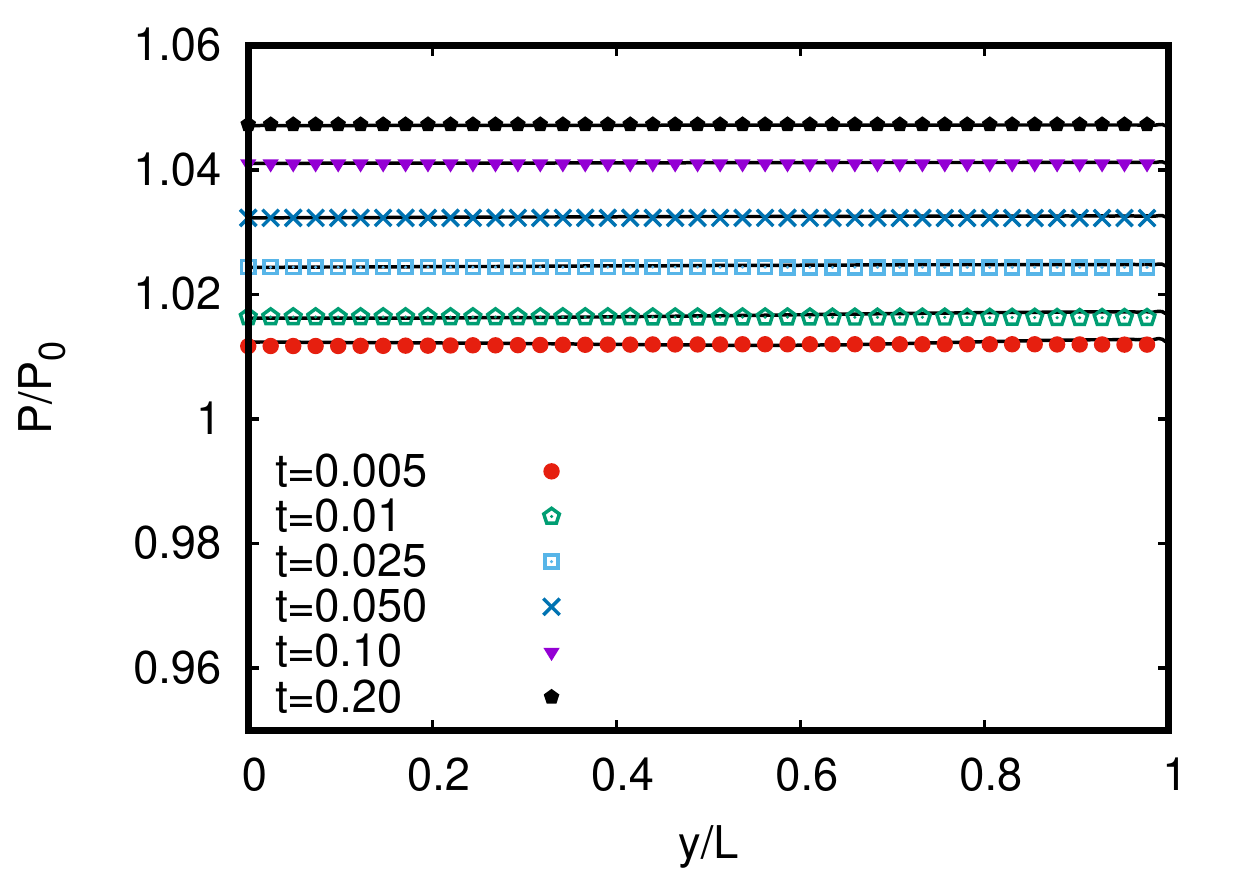}
 \caption{Nondimensional temperature, density and pressure at a few intermediate times scaled by diffusion time.(LBM- points, solution of NSF equations - lines).}
 \label{1DTemperature}
\end{figure}

Navier-Stokes-Fourier equations are solved as described in Ref.~\citep{larkin1967heat} to compare the evolution of
the nondimensional temperature, density, and pressure obtained from the RD3Q41 model at a few intermediate times scaled by diffusion time. 
Diffusion time is defined as $L^2/\nu$, where $\nu$ is kinematic viscosity of the fluid.
Results from both the methods are in good agreement, as shown in Fig. \ref{1DTemperature}.

\section{Nonideal Fluids and two-phase flow}\label{multiphase}
The standard lattice Boltzmann method leads to an ideal gas equation of state. 
Several variations to simulate nonideal fluids have been proposed \citep{colorfluid, shanchen, freeenergy, hechendoolen, ansumalihardsphere}. 
Most of the mentioned approaches model the microscopic physics and the interfacial dynamics at the mesoscopic level at an affordable computational expense. 
The LBM is considered advantageous for multiphase flows as it maintains a stable interface and does not require explicit interface tracking. 
Ref. \cite{shanchen} considered the microscopic interactions between the nearest neighbours to model the collision
operator for which the surface tension could be maintained automatically \citep{chen1998lattice}.
The interaction potential controlled the form of the equation of state of the fluid and gave rise to phase separation,
however, the surface tension could not be freely adjusted \citep{he2002thermodynamic}.
Among the various multiphase models, of particular interest is the free-energy model \citep{freeenergy} as it allows for surface tension to be independent of the viscosity in addition to being thermodynamically consistent. 
Recently, the entropic lattice Boltzmann model was extended for multiphase flows to control the spurious currents at the liquid-vapour interface thereby opening the possibility to simulate large density ratios \citep{chikatamarla2015entropic}.

At the macroscopic scale, the multiphase LBMs can be considered diffuse interface models.
These models smoothen the discontinuity at the interface over a thin but numerically resolvable layer. 
The fluid properties transition over this layer smoothly as opposed to singular interfaces which have a sharp discontinuity.
The surface tension is transformed into a volumetric forcing \citep{leefischer} and is spread over the diffused interfacial region. 
An undesirable feature of the diffuse interface in LBM manifests in the form of the spurious currents that develop in the vicinity of the interface \citep{leefischer,he2002thermodynamic}. 
Several attempts have been made to identify their origin and to alleviate these currents \citep{leefischer,shan2006analysis}. 
The departure of the LBM simulations from the theoretical phase densities and the magnitude of the spurious current depends upon the liquid-vapour density ratio, the equation of state, and the surface tension. Often they have been attributed to the violation of Gibbs-Duhem equality due to the discrete derivative operator \citep{wagner2006}.
It was mentioned in Ref. \citep{wagner2006} that an open problem in this field is the identification of a discrete derivative operator that preserves the Gibbs-Duhem equality at the interface.

The RD3Q41 lattice is expected to offer
an edge over the conventional lattices because of the presence of the bcc grid points that allow the derivatives 
to be calculated more accurately. 
In this section, we propose an alternate way of treating the discrete derivative such that the violation of Gibbs-Duhem relation at the interface is reduced.
We show that this way of discretization leads to accurate liquid vapour phase densities and reduces the spurious currents.
We first review the van der Waals theory of a single component two-phase fluid followed by the equations of hydrodynamics
and the methodology to incorporate nonideal effects in the kinetic theory and lattice Boltzmann model.
It is followed by expressions for evaluating second and fourth-order accurate derivatives on a lattice and the justification of the choice of stencil employed to evaluate derivatives. 
We then propose a discrete derivative operator to reduce the spurious currents. 
To prove the thermodynamic consistency we choose an equation of state and compare densities obtained from a deeply-quenched 
liquid-vapor system with their corresponding theoretical values. Finally, we extend the entropic formulation of 
LBM to two-phase flows and simulate a head-on collision between two droplets.

In this section, we consider an isothermal system of a nonideal fluid. Van der Waals modified the free-energy density by incorporating terms that are large only when the density gradients are significant \citep{van1979thermodynamic,rowlinson}.  Therefore, the underlying free-energy functional $\Psi(x)$ is of the form 
\begin{equation}
\Psi(\bm x) =   \int [\mathcal{F}(\rho(\bm x)) + \mathcal{I}(\nabla \rho(\bm x), \nabla^2 \rho(\bm x), ...)] d\bm x,
\label{totalFE}
\end{equation}
where $\rho$ is the density, $\mathcal{F}$ is the bulk free-energy, and $\mathcal{I}$ is the interfacial free-energy. The interfacial free-energy $\mathcal{I}(\nabla \rho(x), \nabla^2 \rho(x), ...)$ is approximated to the lowest order as \citep{rowlinson,wagner2006}
\begin{equation}
\mathcal{I}(\nabla \rho(x), \nabla^2 \rho(x), ...)= \frac{\kappa(\rho)}{2} |\nabla \rho(x)|^2.
\end{equation} 
For simplicity, $\kappa(\rho)$, related to the surface tension, is taken constant. 
With this definition, the excess free-energy is the surface tension $\sigma$
\begin{equation}
\sigma = \int_{-\infty}^{\infty} \frac{\kappa}{2} |\nabla \rho(x)|^2 dx.
\end{equation}
The above model of free-energy is the simplest model that gives two stable phases provided $\mathcal{F}$ has two minima and was first formulated by van der Waals \citep{van1979thermodynamic}.

The macroscopic mass and momentum conservation equations are given by \citep{LEE200516,suryanarayanan_singh_ansumali_2013}
\begin{equation}
\partial_t \rho + \partial_\alpha (\rho u_\alpha) = 0,
\end{equation}
\begin{equation}
\partial_t (\rho u_\alpha) + \partial_\beta \left[p \delta_{\alpha\beta} + \rho u_\alpha u_\beta + \sigma_{\alpha\beta} + \sigma^{(\kappa)}_{\alpha\beta} \right] = 0,
\end{equation}
where $\sigma_{\alpha\beta}$ is the viscous stress tensor 
\begin{equation}
\sigma_{\alpha\beta} = -\rho \nu \left(\partial_\beta u_\alpha + \partial_\alpha u_\beta \right) + \frac{2}{3}\rho\nu {\partial_\gamma u_\gamma}\delta_{\alpha\beta},
\end{equation}
where $\nu$ is the kinematic viscosity, and $\sigma^{(\kappa)}_{\alpha\beta}$ takes the form
\begin{equation}
\sigma^{(\kappa)}_{\alpha\beta} = \kappa \left[\left( -\frac{1}{2}\partial_\gamma \rho \, \partial_\gamma \rho - \rho  \partial^2 \rho \right)\delta_{\alpha\beta} +  \left\{ \partial_\alpha \rho \, \partial_\beta \rho \right\} \right].
\end{equation}
and accounts for the interfacial stresses. The term within the curly braces in the above expression is known as the van der Waals stress \citep{rowlinson}.

The non-local pressure tensor (also known as Korteweg's stress tensor) derived from the above description is consistent with the definition of the free-energy functional and is written as \citep{korteweg,evans1979,freeenergy} 
\begin{equation}
P_{\alpha\beta} = \left[p - \frac{\kappa}{2}\partial_\gamma \rho \, \partial_\gamma \rho - \kappa\rho \, \partial^2 \rho \right]\delta_{\alpha\beta}  + \kappa \partial_\alpha \rho \, \partial_\beta \rho,
\end{equation}
where $\delta_{\alpha\beta} $ is the Kronecker delta and 
\begin{equation}
p = \rho \mu_0 - \mathcal{F},
\end{equation}
is the equation of state describing the nonideal fluid, with $\mu_0 = {\partial \mathcal{F} } /{ \partial \rho }$ as the bulk chemical potential. 
For a continuous system, the Gibbs-Duhem equality is trivially satisfied. 
However, it gets violated for a discrete system due to the definition of the discrete derivative which leads to thermodynamic inconsistency \citep{wagner2006}.

To incorporate the deviation from the ideal gas the intermolecular attraction, the repulsion between the particles due to their non-vanishing size, and the interface dynamics needs to be modeled. 
In LBM framework, the attractive and repulsive parts are added as a force term.
In order to do so, one begins with the 
 Boltzmann BGK equation for an isothermal nonideal gas \citep{hechendoolen}
\begin{equation}
\frac{\partial f_i}{\partial t} +  {c}_{i\alpha} \frac{\partial f_i}{\partial x_\alpha}  = \frac{f_i^{\rm eq }(\rho,\bm u, \theta_0) -f_i}{\tau} + F_i ,
\label{nonidealbe}
\end{equation}
where the forcing term $F_i$ is given by
\begin{equation}
F_i = \frac{g^{\rm nid}_\alpha (c_{i\alpha}-u_\alpha)  }{\theta_0}f_i^{\rm eq}(\rho,\bm u, \theta_0).
\end{equation}
The nonideal contributions are captured in $g^{\rm nid}_\alpha$,
\begin{equation}
g^{\rm nid}_\alpha = -\frac{1}{\rho} \partial_\beta P_{\alpha\beta}^{\rm nid}, \quad  P_{\alpha\beta}^{\rm nid} = P_{\alpha\beta}-\rho\theta_0 \delta_{\alpha\beta}.
\label{pressure_formulation}
\end{equation}
The above form of $g^{\rm nid}_\alpha$, known as the pressure formulation, is sufficient to incorporate the nonideal interactions and the Korteweg's stress tensor in the lattice Boltzmann model \citep{LEE200516,leefischer,chikatamarla2015entropic}. Alternatively, by exploiting the Gibbs-Duhem relation one can write the chemical potential formulation as \citep{leefischer,suryanarayanan_singh_ansumali_2013}
\begin{equation}
g^{\rm nid}_\alpha = - \partial_\alpha \mu^{\rm nid} ,
\label{chemicalpotential_formulation}
\end{equation}
where $ \mu^{\rm nid} = \mu_0^{\rm nid} - \kappa \partial^2\rho.$

The Eq.\eqref{nonidealbe} is integrated along the characteristics using the trapezoid rule to obtain the discrete (in space and time) evolution of populations as
\begin{align}
\begin{split}
\tilde{f_i}(\bm x +\bm c_i \Delta t, t+\Delta t) = \tilde{f_i}(\bm x, t) + \alpha\beta \big[{f_i}^{\rm eq}(\rho,{\bm u},\theta_0) \\- \tilde{f_i} (\bm x, t) \big] 
+ \left( 1-\frac{\alpha\beta}{2} \right)\Delta t F_i,
\end{split}
\label{multiphaseEvolution}
\end{align}
where $\tilde{f_i}$ is a transformation of the populations $f_i$ defined as
\begin{equation}
\tilde{f_i} = f_i - \frac{\Delta t}{2\tau}[{f_i}^{\rm eq}(\rho,{\bm u},\theta_0) - {f_i} ] - \frac{\Delta t}{2}F_i,
\end{equation}
$\beta=\Delta t/(2\tau+\Delta t)$ and $\alpha=2$ for the standard LBM. For the entropic LBM the parameter $\alpha$ needs to be computed such that the dynamics obeys the $H$ theorem. The macroscopic variables are calculated as
\begin{equation}
\rho = \sum_{i} \tilde{f_i}, \qquad u_\alpha = \frac{1}{\rho}\sum_{i} \tilde{f_i} c_{i\alpha} + \frac{\Delta t}{2} g^{\rm nid}_{\alpha}.
\end{equation}

It is worth noticing that the Eq.\eqref{multiphaseEvolution} does not conserve momentum. The change in momentum at each site during a time step is 
obtained by multiplying Eq.\eqref{multiphaseEvolution} with $c_{i\alpha}$ and summing over all directions as
\begin{equation}
\rho u_\alpha(t+\Delta t) - \rho u_\alpha(t) = \Delta t \rho g^{\rm nid}_\alpha. 
\label{momentumchange}
\end{equation}

In fact, this change in momentum is due to the nonideal nature of the fluid and leads to two stable phases separated by an interface.
However, the global momentum of the system should be exactly conserved provided no net momentum exchange occurs at the boundary \citep{shanchen}.
This is an important feature of the discrete dynamics, satisfied by the pressure formulation [Eq.\eqref{pressure_formulation}] but not by the chemical potential formulation [Eq.\eqref{chemicalpotential_formulation}].
It is known that the chemical potential formulation is more accurate than the pressure formulation
and leads to smaller spurious currents \citep{leefischer,suryanarayanan_singh_ansumali_2013}.
However, the thermodynamic consistency requires that the global momentum should stay preserved. 
This will be elaborated in the forthcoming sections, where we propose an alternate formulation that is similar to the chemical potential formulation but preserves global momentum. 

\subsection{Discretization scheme}
As discussed earlier, for a continuous system the Gibbs-Duhem relation is trivially satisfied.
However, for a discrete system it gets violated, i.e.,
\begin{equation}
\tilde \partial_\alpha P_{\alpha\beta} \neq \rho \tilde \partial_\beta \mu,
\end{equation}
where $\tilde \partial_\alpha$ represents the discrete derivative operator \citep{wagner2006}.
This is the reason why the two formulations, namely the pressure and the chemical potential show different accuracy, stability, and spurious currents. 

The chemical potential formulation is more accurate and has smaller spurious currents as compared to the pressure formulation \citep{leefischer}.
However, one of its drawback is that the global momentum is not conserved. This is further explained in what follows:
In the pressure formulation in 1D, the second-order discrete derivative $\tilde \partial^{(2)}_\alpha$ at the $n^{\rm th}$ grid point is written as
\begin{equation}
\tilde \partial^{(2)}_\alpha P_{\alpha\beta} = \frac{1}{2\Delta x_\alpha} \left[ P_{\alpha\beta}(n+1) - P_{\alpha\beta}(n-1) \right].
\label{gradp2}
\end{equation}
With appropriate boundary conditions (say periodic), one can sum over the entire domain to show that
\begin{equation}
\sum_{n=1}^N \left[ P_{\alpha\beta}(n+1) - P_{\alpha\beta}(n-1) \right] = 0.
\label{gradpmomcons}
\end{equation}
The local change of momentum from Eq.\eqref{momentumchange} is $\Delta t \rho g^{\rm nid}_\alpha$. 
The global momentum conservation upon using Eqs.\eqref{pressure_formulation},\eqref{gradp2},\eqref{gradpmomcons} follows as
\begin{align}
\begin{split}
\sum_{n=1}^N &\Delta t \rho g^{\rm nid}_\alpha 
= -\sum_{n=1}^N \Delta t \tilde \partial^{(2)}_\alpha P^{\rm nid}_{\alpha\beta} 
\\&= -\frac{\Delta t}{2\Delta x_\alpha} \sum_{n=1}^N \left[ P^{\rm nid}_{\alpha\beta}(n+1) - P^{\rm nid}_{\alpha\beta}(n-1) \right] = 0.
\end{split}
\end{align} 
However, for the chemical potential formulation the net global momentum is
\begin{equation}
\sum_{n=1}^N \Delta t \rho g^{\rm nid}_\alpha =  -\frac{\Delta t}{2\Delta x_\alpha} \sum_{n=1}^N \rho(n) \left[ \mu^{\rm nid}({n+1}) - \mu^{\rm nid}({n-1}) \right] ,
\label{chempotbad}
\end{equation} 
which is non-zero. 

Fundamentally, this lack of momentum conservation is emerging due to the violation of the Leibniz rule.
For this analysis we ignore the interfacial terms which are added separately.
The bulk nonideal pressure using the thermodynamic relations is written as 
\begin{equation}
p^{\rm nid} = \mu_0^{\rm nid} \rho - {\cal F}^{\rm nid}.
\end{equation}
Taking the discrete derivative of the above equation one obtains
\begin{equation}
\tilde \partial_\alpha p^{\rm nid} = \left\{ \mu_0^{\rm nid}\tilde \partial_\alpha \rho \right\}+ \rho \tilde \partial_\alpha \mu^{\rm nid} - \tilde \partial_\alpha {\cal F}^{\rm nid},
\label{leibiznew}
\end{equation}
where the left hand side is the pressure formulation which conserves the global momentum. The term in curly braces on the right hand side 
is the chemical potential formulation which does not conserve the global momentum because the other two terms (although they cancel in the continuous case) are ignored in the discrete chemical potential
formulation. It is interesting to note that if one defines the discrete derivative as
\begin{equation}
\tilde \partial_\alpha (AB) = A \tilde \partial_\alpha B + B \tilde \partial_\alpha A ,
\end{equation}
$A,B$ being arbitrary functions, the Leibniz rule as well as the global momentum conservation holds.

Therefore, the global momentum conserving new formulation using Eq.\eqref{leibiznew} is written as
\begin{equation}
\rho g^{\rm nid}_\alpha = -\left [\rho \tilde \partial^{(2)}_\alpha \mu_0^{\rm nid} + \mu_0^{\rm nid} \tilde \partial^{(2)}_\alpha \rho - \tilde \partial^{(2)}_\alpha {\cal F}^{\rm nid} + \kappa \tilde \partial^{(2)}_\beta I_{\alpha\beta} \right],
\end{equation}
where the interfacial stresses $I_{\alpha\beta}$ are given by
\begin{equation}
I_{\alpha\beta} = \left[ - \frac{1}{2}\partial_\gamma \rho \, \partial_\gamma \rho - \rho \, \partial^2 \rho \right]\delta_{\alpha\beta}  + \partial_\alpha \rho \, \partial_\beta \rho.
\end{equation}
Ideally, one would prefer to work with the fourth-order discrete derivatives [Eq.\eqref{gradient4}] but they lead to violation of the global momentum conservation. One can, however, use the fourth order discrete derivative for $\tilde \partial_\alpha {\cal F}^{\rm nid}$, for which we use a convex combination of the second and fourth order discrete derivative (see Appendix \ref{appendixA} for details). This brings us to the final form of $g^{\rm nid}_\alpha$ 
\begin{align}
\begin{split}
\rho g^{\rm nid}_\alpha = -\bigg[\rho \tilde \partial^{(2)}_\alpha \mu_0^{\rm nid} + \mu_0^{\rm nid} \tilde \partial^{(2)}_\alpha \rho - \eta \tilde \partial^{(2)}_\alpha {\cal F}^{\rm nid} \\- (1-\eta)\tilde \partial^{(4)}_\alpha {\cal F}^{\rm nid} + \kappa \tilde \partial^{(2)}_\beta I_{\alpha\beta} \bigg].
\end{split}
\label{gradientFinal}
\end{align}
The above route to calculating the discrete derivative is thermodynamically consistent, preserves the global momentum, 
and has a parameter $\eta \in [0,1]$ that can be fine tuned to improve the accuracy, but for this work we restrict ourselves to $\eta=1/2$.

\begin{figure}\centering
 \includegraphics[width=0.4\textwidth]{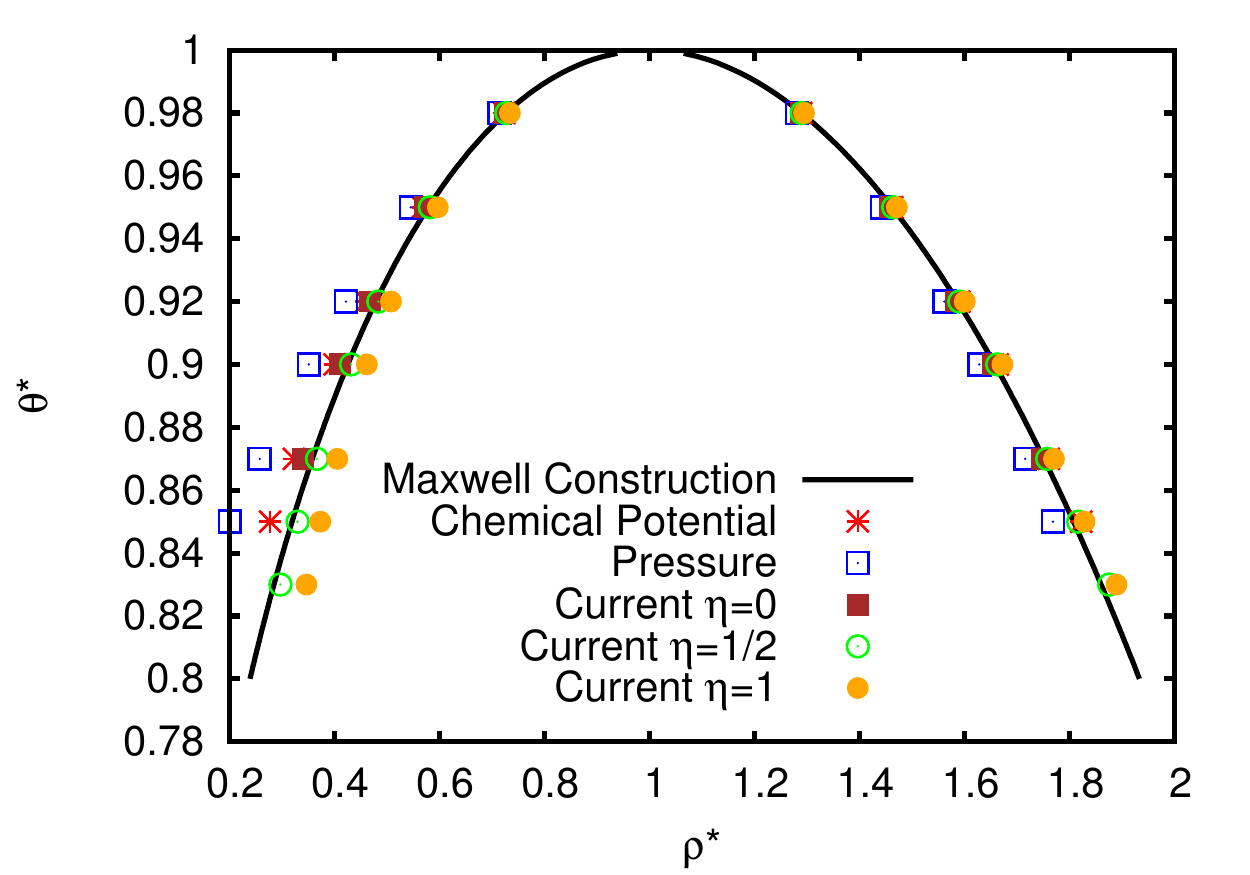}
 \includegraphics[width=0.4\textwidth]{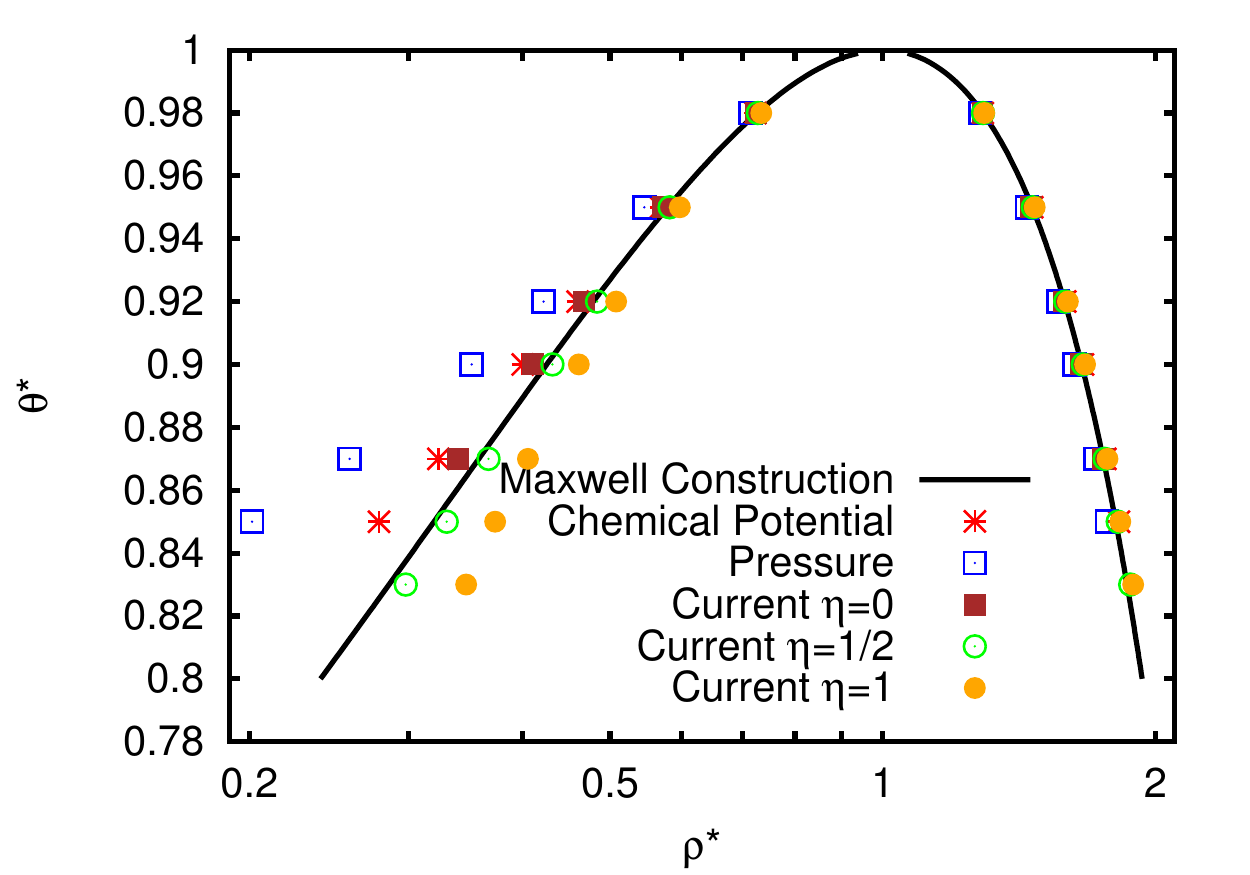}
\caption{Liquid vapour densities of the van der Waals fluid for various formulations at different reduced temperature $\theta^*=\theta/\theta_c$. The same plot is represented on the linear scale (left) and the logarithmic scale (right) to emphasize the error in density of the gas phase.}
 \label{vdwDensity}
\end{figure}

\begin{figure}\centering
 \includegraphics[width=0.4\textwidth]{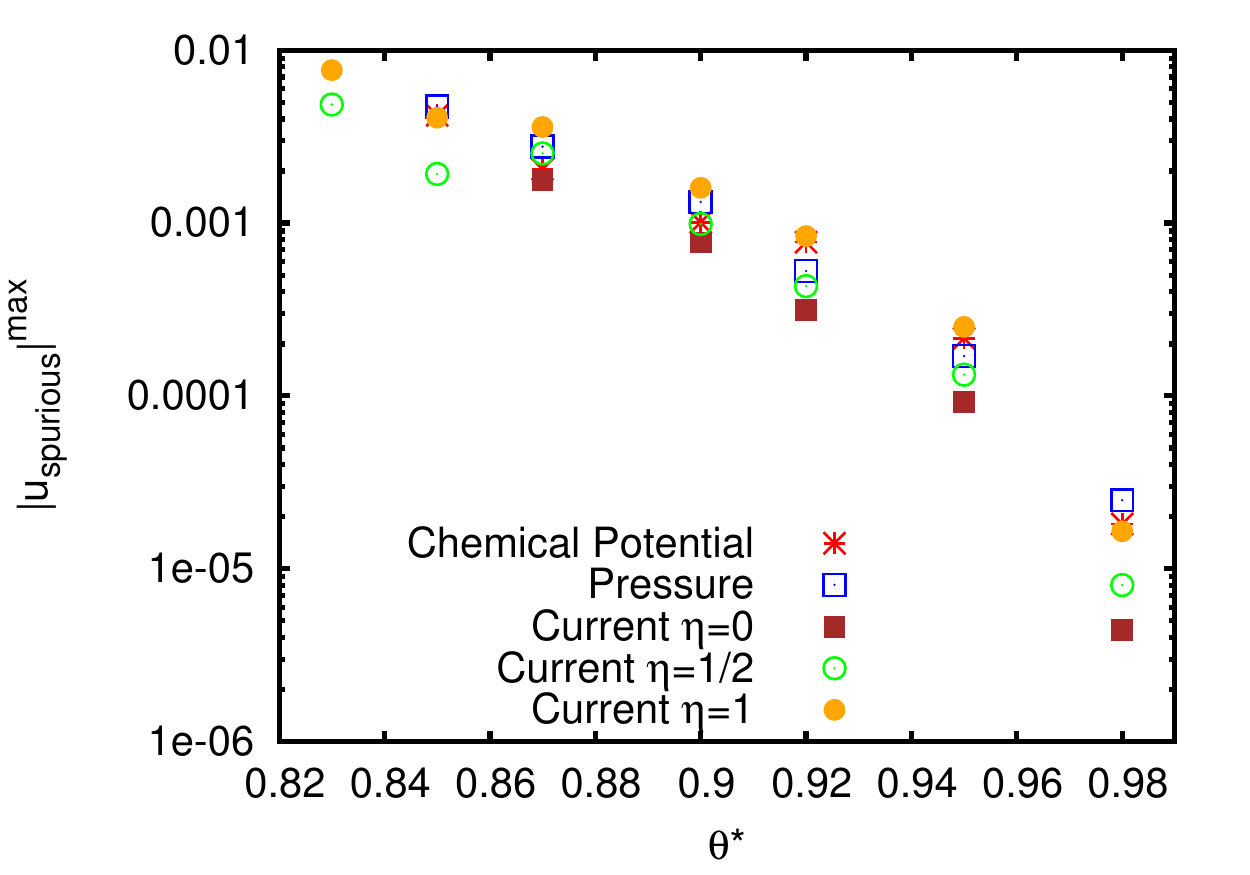}
\caption{Spurious currents of the van der Waals fluid for various formulations at different reduced temperature $\theta^*=\theta/\theta_c$.}
 \label{spuriousvdw}
\end{figure}

To demonstrate the accuracy of the present model, we simulate a one dimensional interface of a van der Waals fluid on a grid of size $192 \times 4 \times 4$ using the RD3Q41 model. 
Figure \ref{vdwDensity} shows the densities from the pressure formulation, chemical potential formulation, and the current scheme for three $\eta$ values.
It is evident that the current scheme predicts the values of phase density that matches with the theoretical Maxwell's construction values.
From Fig. \ref{spuriousvdw} and Table \ref{scheme_vs_umax}, it can be seen the maximum magnitude of the spurious current is also reduced upon using the current formulation. 
Here, it should be pointed out that although
$\eta=0$ gives smaller spurious currents than $\eta=1/2$, it succumbs to numerical instabilities at a higher $\theta/\theta_c=0.87$.

\begin{table}
\begin{center}
\begin{tabular}{ccc} \hline \hline
$g^{\rm nid}_\alpha$  & PR [Eq.\eqref{cspreos}]                                  & CS     \\  \hline
Pressure              & $\quad 6.376\times 10^{-3}$                & $\quad 4.317\times 10^{-3}$  \\ 
Chemical potential    & $\quad 5.139\times 10^{-3}$                & $\quad 4.281\times 10^{-3}$  \\ 
Current, $\eta=1.0$   & $\quad 4.765\times 10^{-3}$                & $\quad 5.565\times 10^{-3}$  \\ 
Current, $\eta=0.5$   & $\quad 2.790\times 10^{-3}$                & $\quad 3.706\times 10^{-3}$  \\ \hline  \hline
 \end{tabular}
 \caption{Discretization scheme and the maximum magnitude of spurious current for Peng-Robinson (PR) and   Carnahan-Starling (CS) EoS on a grid of size $80\times 80\times 4$ at $\theta/\theta_c=0.9$ using the RD3Q41 model.}
  \label{scheme_vs_umax}
\end{center}
\end{table}

\subsection{Sound propagation in a nonideal gas}
In this section, we validate the proposed model for sound propagation in an isothermal nonideal gas.
For a gas following an ideal equation of state $p = \rho \theta_0$ the sound speed  is fixed at $\sqrt{\theta_0}$ where $\theta_0$ is a reference temperature.
However, in real gases, the speed of sound becomes dependent on the phase density too as pressure is a non-trivial function of phase density.
To confirm this, we introduce a density perturbation at constant temperature in a nonideal fluid with Peng-Robinson type equation of state given by 
\begin{equation}
p = \rho\theta_0\frac{1+\eta+\eta^2-\eta^3}{(1-\eta)^3} -\frac{a \rho^2}{1+2\rho b -\rho^2 b^2},
\label{cspreos}
\end{equation}  
where $a=1.851427622\theta_c/\rho_c$, $b=0.353748714/\rho_c$ are van der Waals like critical parameters and $\eta=\rho b/4$. 
Here, $\theta_c$ and $\rho_c$ are critical temperature and critical density, respectively.

We compute the speed of sound at various values of $\theta^*=\theta_0/\theta_c$ and their corresponding equilibrium phase densities ($\rho_{\rm{ph}}$) 
by initializing a 1D density fluctuation of the form
\begin{equation}
 \rho(x,t=0) =  \rho_{\rm{ph}}  \left( 1.0 + \epsilon \cos(\pi x) \right),
\end{equation}
in domain of size $[-\pi,\pi]$ in $x$-direction. 
The domain has two lattice points in the $y$ and $z$ directions and periodic boundary conditions are applied in all three directions.
The wave is expected to reach the same configuration as the initial condition after one wave period ($t_p$).
We track the l2-norm of the density fluctuation computed using the present state and initial state, which is minimum when the waves are in-phase after completing one cycle.
The domain size ($L$) and the time period at which l2-norm is minimum are used to compute the speed of sound ($c_s = L/t_p$).

\begin{figure}
\centering
 \includegraphics[scale=0.63]{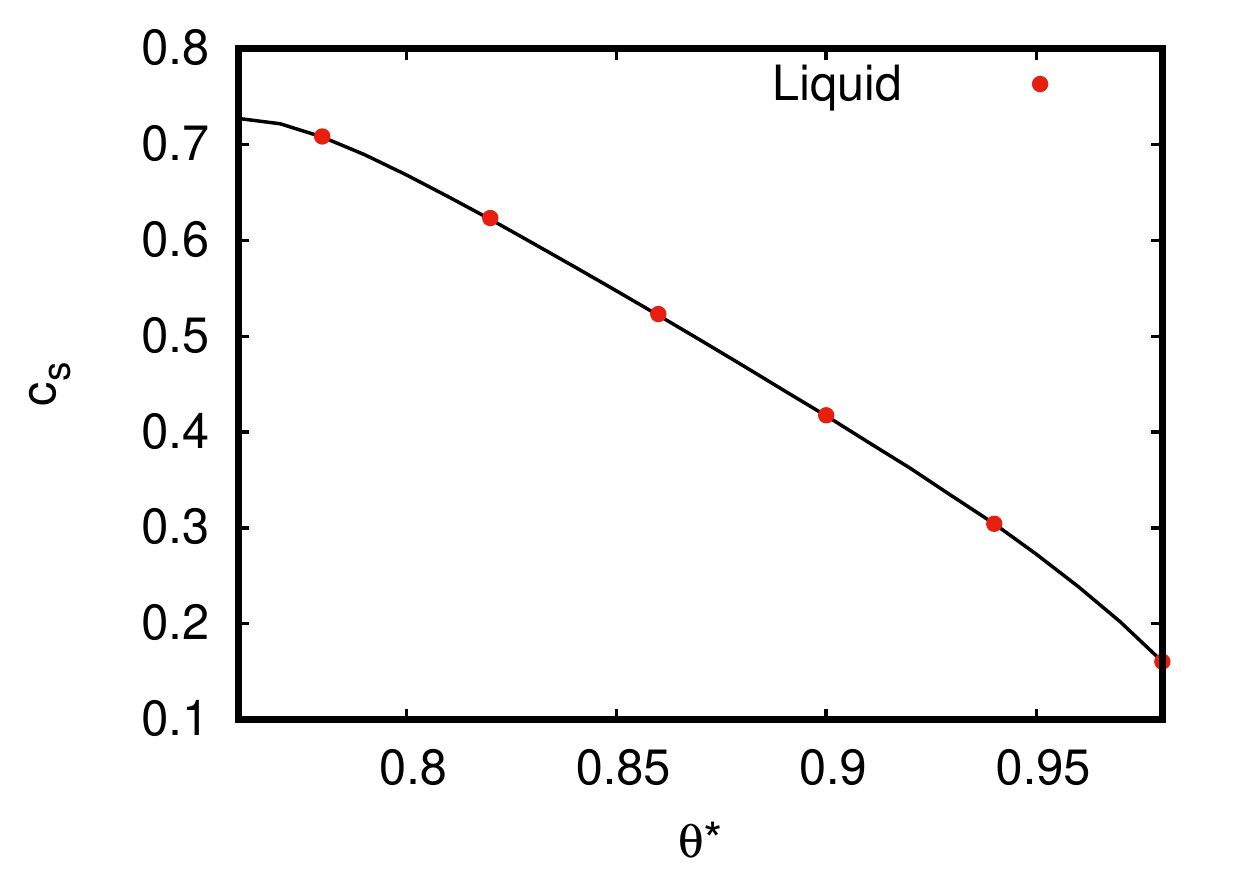}
 \includegraphics[scale=0.63]{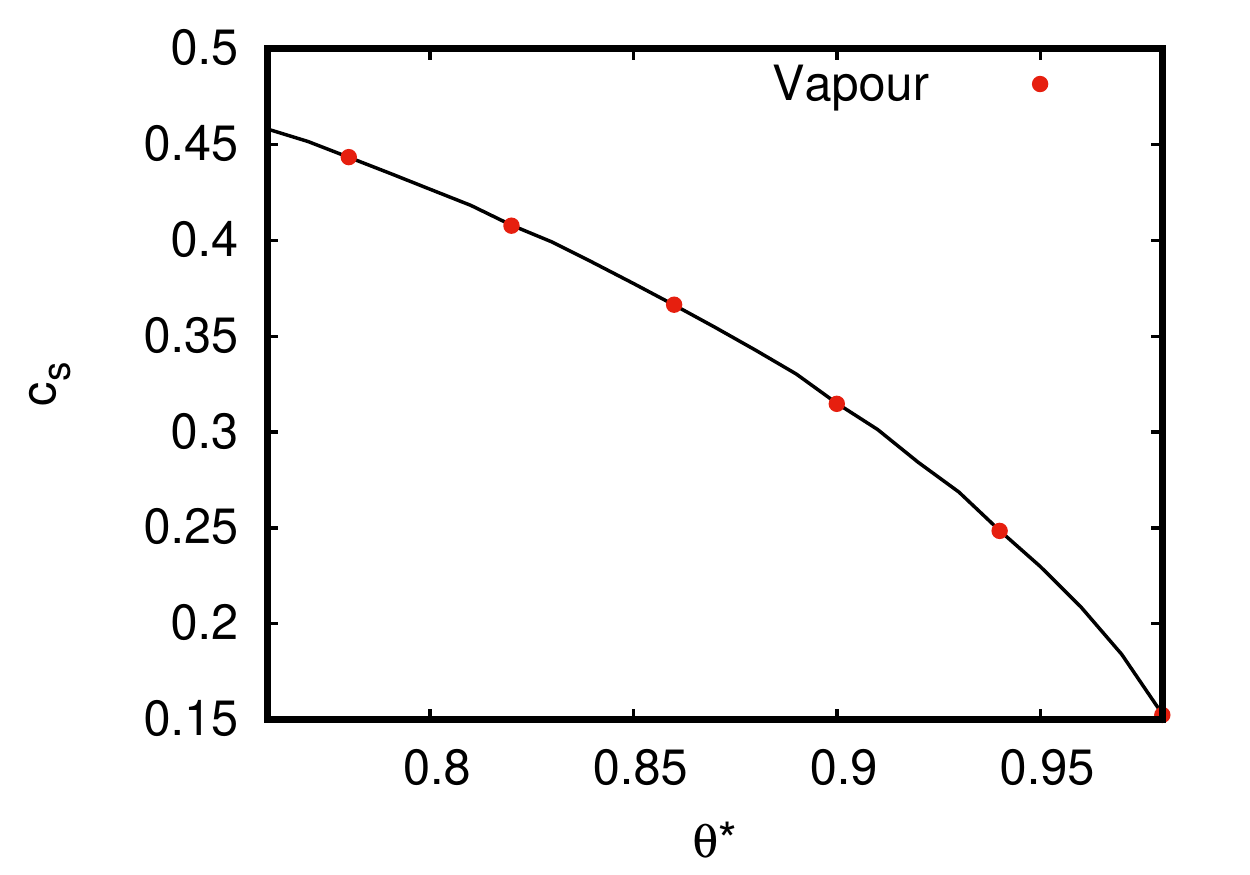}
\caption{Speed of sound in a nonideal gas for liquid and vapour phases -- theoretical prediction (solid line) and simulation (points).}
 \label{fig:soundSpeed_NI}
\end{figure}

We observe a good agreement with the theoretical prediction of the sound speed in both the phases for a broad range of $\theta^*$ as shown in Fig. \ref{fig:soundSpeed_NI}. 

\subsection{Quenching of a liquid-gas system}

\begin{figure}\centering
 \includegraphics[width=0.4\textwidth]{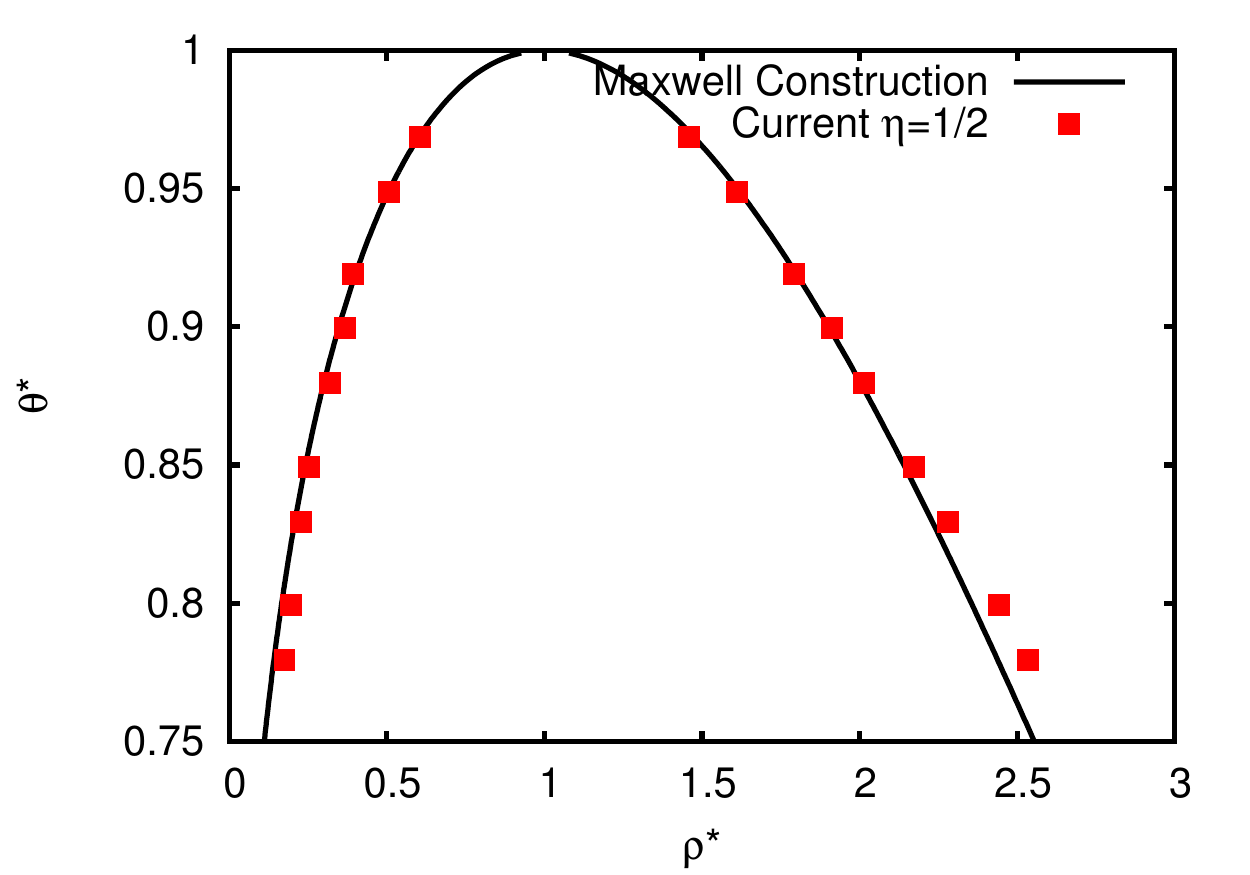}\\
 \includegraphics[width=0.4\textwidth]{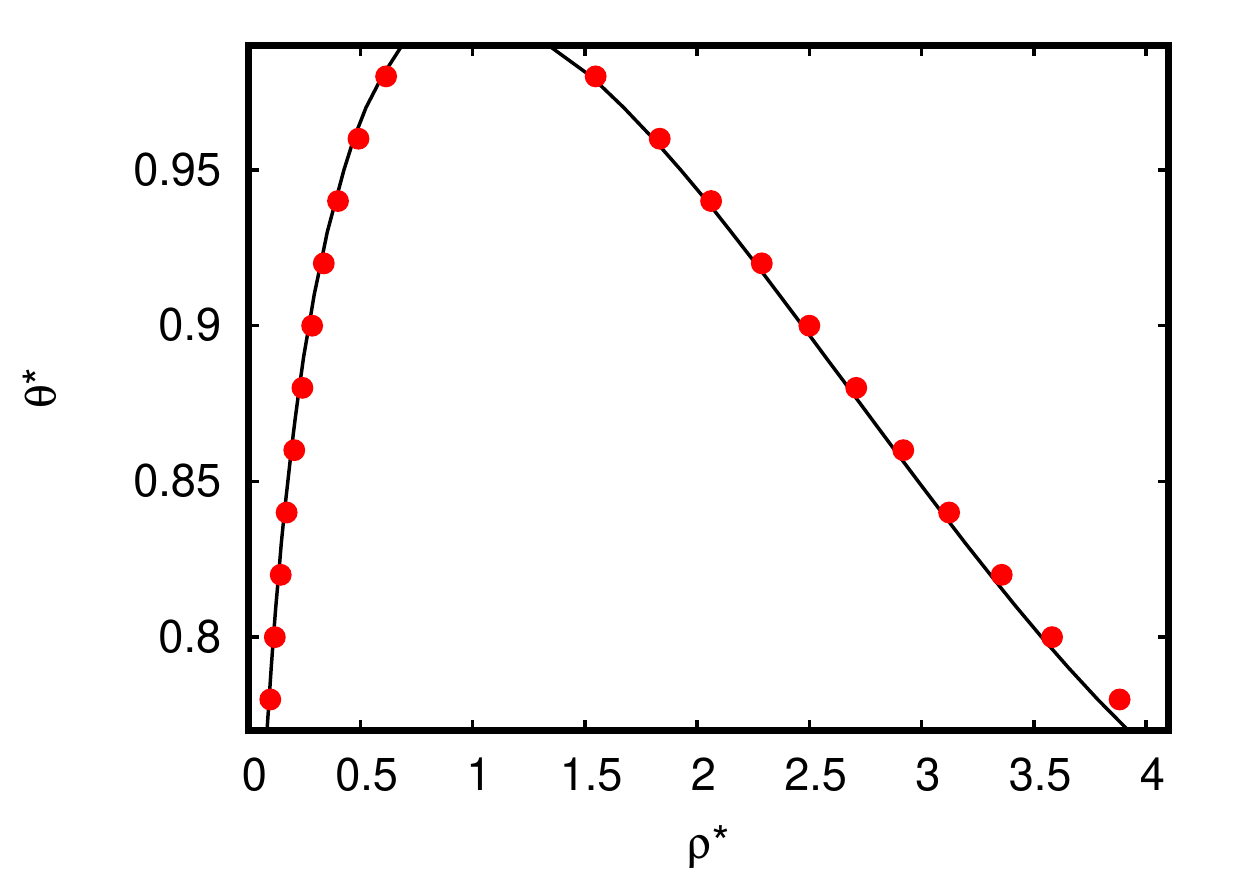}
\caption{Equilibrium liquid and gas density $\rho/\rho_c=\rho^*$ obtained from $RD3Q41$ model for Carnahan Starling EoS (top) and Peng-Robinson type EoS (bottom) compared against their respective Maxwell equal area construction at various $\theta/\theta_c=\theta^*$.}
 \label{csprquench}
\end{figure}

Quenching of a liquid-gas system exhibits phase separation and has been widely accepted as a test for thermodynamic 
consistency and stability of a multiphase lattice Boltzmann model \cite{wagner2007interface}. 
We use the Peng-Robinson EoS as defined in Eq.(\ref{cspreos}), and the Carnahan-Starling EoS. 
For $\theta<\theta_c$ the system shows phase separations wherein the initial stages tiny bubbles of the liquid phase
surrounded by the ambient gaseous phase are formed.
As time progresses, the bubbles merge to form a stable lamellar film, a cylinder or a spherical droplet.  
Figure \ref{csprquench} shows the liquid and gas density obtained from RD3Q41 model compared against the theoretical 
values obtained from Maxwell's equal area construction. The model accurately recovers liquid-gas bulk densities, hence is 
confirmed to be thermodynamically consistent.


\subsection{Droplet Collision} \label{dropletcollision}
\begin{figure}
\centering
  {\includegraphics[trim={0.5cm 0cm 0cm 0cm},clip,scale=0.8]{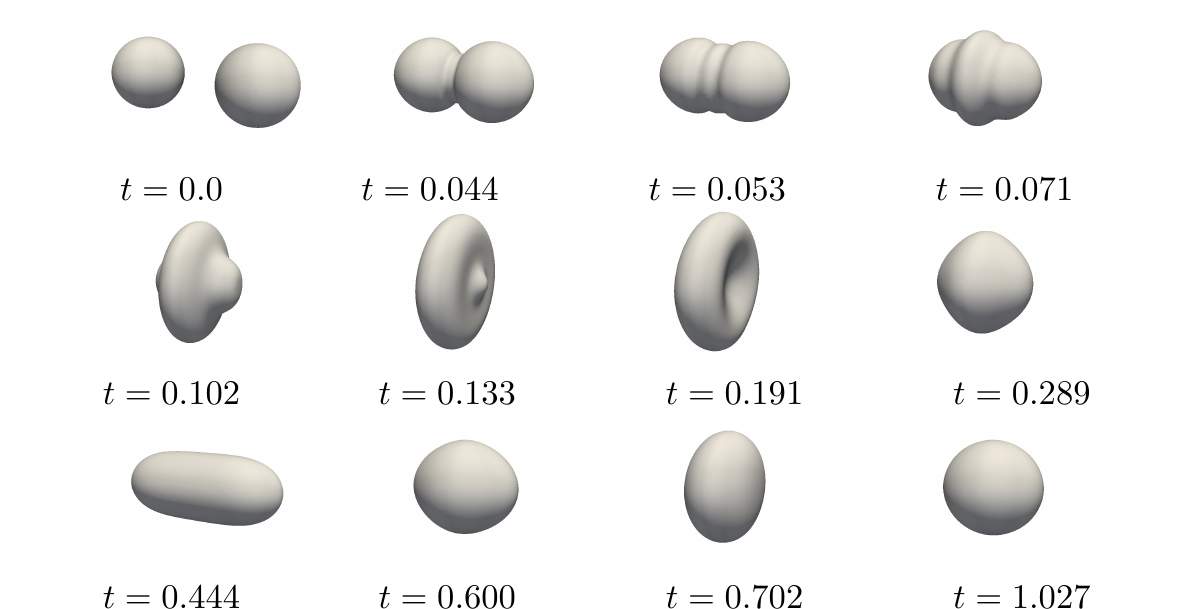} }
  \caption{Head-on collision between two droplets at Re $=297.03$ and We$=19.47$.}
  \label{collision}
\end{figure}

We employ the proposed multiphase model and the Carnahan-Starling EoS \cite{carnahan1969equation} to simulate binary 
droplet collision as it offers interesting outcomes depending upon the control parameters 
\cite{chikatamarla2015entropic}. 
Three particular outcomes observed upon collision at various Reynolds number, Weber number and the impact parameter are 
coalescence, stretching separation, and reflexive separation \cite{inamuro2004lattice}. 

    The setup consists of a rectangular box of $160  \times 200  \times 200 $ lattice units with two liquid phase 
    droplets of diameter $D_0=45$ units and interface width $5$ units located in the ambient of gas phase with a 
    distance of $30$ units between their centres. At $t=0$, they are imparted a relative velocity $U_0 = 0.2$ towards each other. 
    The liquid phase having a density of $\rho_{\rm liq}/\rho_c = 2.412$ is in equilibrium 
    with the gas phase of density $\rho_{\rm gas}/\rho_c=0.185$ at $\theta/\theta_c = 0.8$. The chosen value of 
    $\kappa=0.001a \Delta x^2$ corresponds to a surface tension $\sigma=0.223$ when $\Delta x = 1$. The kinematic viscosity of the liquid phase 
    $\nu_{\rm liq}=0.0303$, hence, Reynolds number (Re) and the Weber number (We) can be calculated as
\begin{equation}
\rm Re = \frac{U_0 D_0}{\nu_{\rm liq}}=297.03 ,\quad We = \frac{\rho_{\rm liq} U_0^2 D_0}{\sigma}=19.47.
\end{equation}

For calculating $\alpha$ required for the entropic formulation of LBM, we rewrite Eq.\eqref{multiphaseEvolution} as
\begin{align}
\tilde{f_i}(\bm x+\bm c_i \Delta t, t+\Delta t) = \hat{f_i}(\bm x, t) +& \alpha\beta [\hat{f_i}^{\rm eq}(\rho,{\bm 
u},\theta_0)  - \hat{f_i}(\bm x, t) ], 
\end{align}
where we have the transformations $\hat f_i = \tilde{f_i} + \Delta t F_i$ and $\hat f_i^{\rm eq} = f_i^{\rm eq} + 
(\Delta t/2) F_i$ with $\Delta t = 1$. These transformations permit us to define $x_i = \hat f_i^{\rm eq}/\hat f_i-1$ and compute the path 
length $\alpha$
in the spirit of Ref. \citep{atif2017,atifDetailed}. Figure \ref{collision} shows head-on collision between two droplets in 
thermodynamic equilibrium with the ambient.

\section{Turbulent Flows}\label{section:turbulent}
Higher-order isothermal lattice Boltzmann methods are known to be stable and have been proposed as an alternative to direct numerical simulations (DNS) for fluid turbulence~\citep{chikatamarla2010lattice}.
It is shown that energy conserving LB models are numerically stable than their non-energy conserving counterparts \citep{singh2013energy}. 
The proposed RD3Q41 model, which is an energy conserving higher-order LB model, takes advantage of these two features and is a viable 
alternative for DNS simulations of turbulent flows. 
In this section, we present simulations of turbulent flows for three cases - a fully periodic test case of decaying turbulence, flow in a rectangular channel
and flow over a sphere.

\subsection{Kida-Peltz flow}
Kida-Peltz flow is a periodic flow with highly-symmetric initial conditions and is a
good test case for the computational study of high Reynolds number (Re) flows~\citep{kida1985three}. The initial conditions for the flow are 
\begin{align}
\begin{split}
  u_x(x,y,z) &= U_0\ \sin x\ (\cos(3y)\ \cos z - \cos  y\ \cos(3z)), \\
  u_y(x,y,z) &= U_0\ \sin y\ (\cos(3z)\ \cos x - \cos  z\ \cos(3x)), \\
  u_z(x,y,z) &= U_0\ \sin z\ (\cos(3x)\ \cos y - \cos  x\ \cos(3y)),
\end{split}
\end{align}
with $ x,y,z \in [0,2\pi] $. 
The simulations were performed at a Reynolds number $5000$ defined based on the domain length and velocity $U_0$. 
In this flow, enstrophy($\Omega$) increases very sharply in the initial time steps and reaches a maximum value and then 
decays with time. Mean enstrophy is calculated from the symmetric velocity gradient tensor\cite{frisch1995turbulence}
\begin{equation}
S_{\alpha \beta} = \frac{2}{\rho \theta (2\tau + \Delta t)}  \sum_i \left(f_i - f_i^{eq}\right) c_{i \alpha}c_{i 
\beta}, 
\end{equation}
where $\tau = \nu / c_s^2$. 

\begin{figure}
\centering
 \includegraphics[scale=0.63]{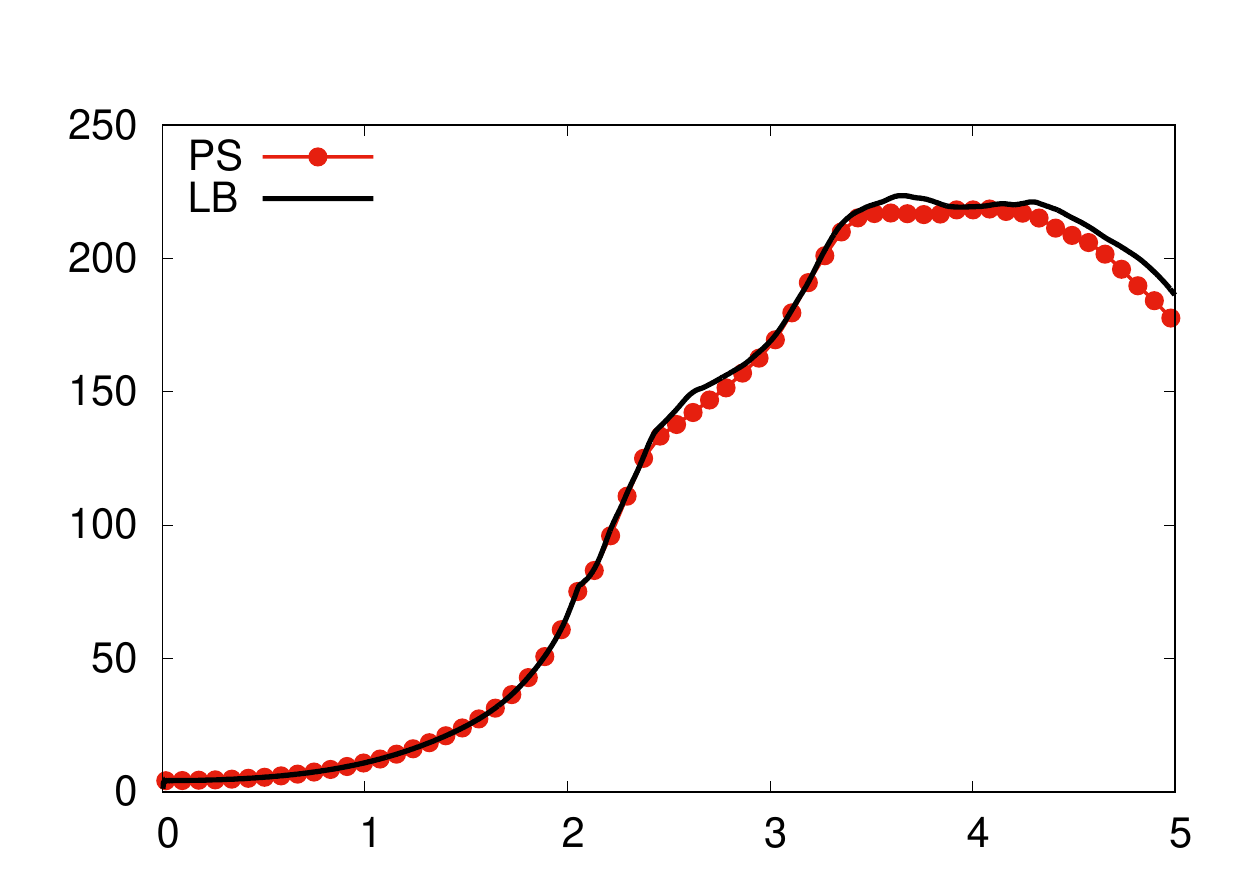}
 \caption{Comparison of evolution of mean enstrophy with time (line--LBM, points--PS).}
 \label{fig:Kida}
\end{figure}

We demonstrate the efficiency of the proposed RD3Q41 model by comparing the evolution of mean enstrophy of 
a Kida-Peltz flow setup with that of pseudo-spectral(PS) method in Fig. \ref{fig:Kida}. 

\subsection{Turbulent channel flow}
In this section, the proposed RD3Q41 model is benchmarked for the classic wall-bounded turbulent flow in a rectangular channel.
We perform simulations at a friction Reynolds number $(Re_\tau) \approx 180$.
The friction Reynolds number is defined based on the wall shear velocity ($u_\tau$) and the channel half-width ($\delta$).
The simulations were performed a domain of size $12\delta \times 2\delta \times 6\delta$, and the channel half-width was chosen to be $96$ grid points.
This corresponds to a nondimensional grid spacing $y^+= yu_\tau / \nu$  $ \approx 2$.
This resolution is sufficient to resolve the Kolmogorov length scale~\citep{bespalko2012analysis}.
The flow is driven by a body-force, and periodic boundary conditions are applied in streamwise and spanwise directions.
Turbulence in the domain is triggered by adding a divergence-free noise to the initial conditions~\citep{bridson2007curl}.

\begin{figure}
 \centering
 \includegraphics[scale=0.6]{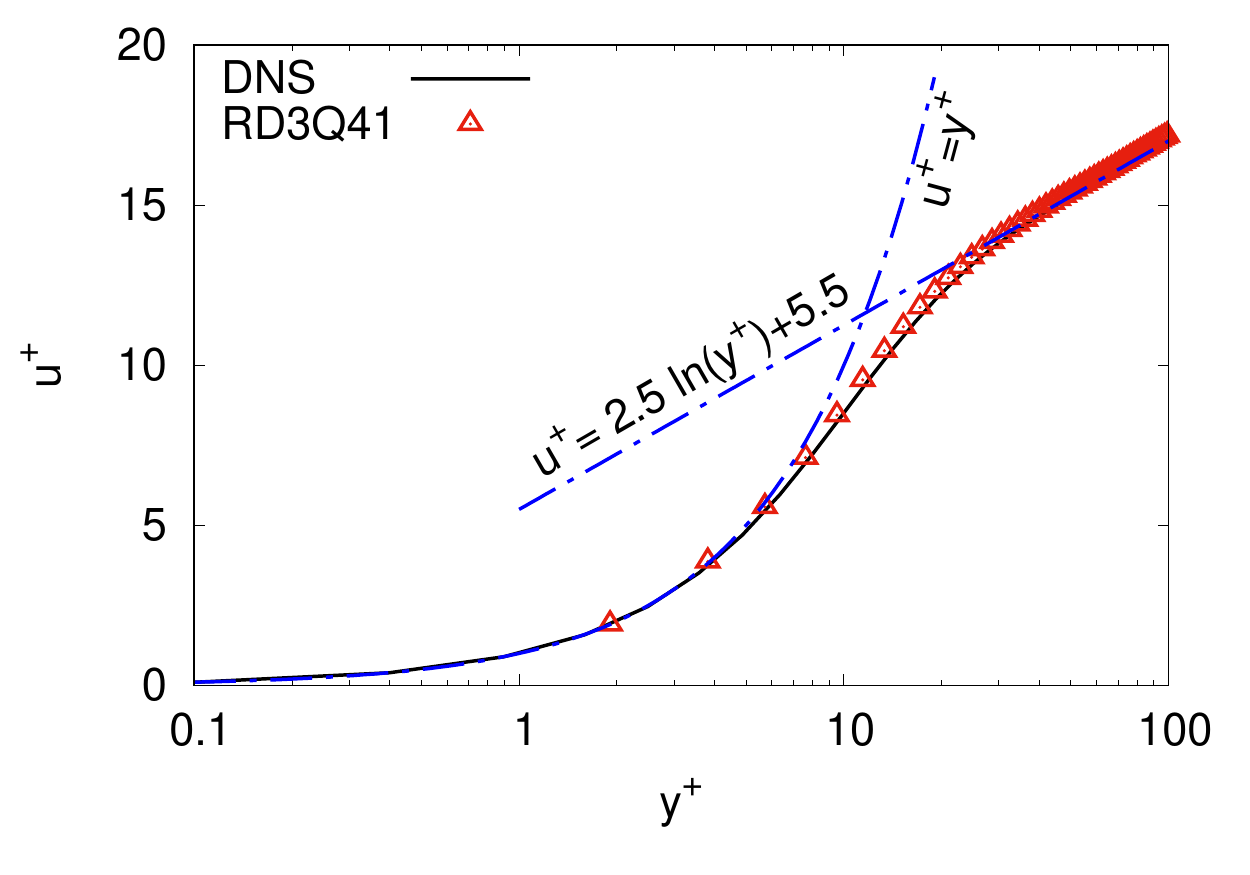}
 \caption{Mean velocity profiles from our simulation (points) and DNS results (line) from~\citep{moser1999direct}.}
 \label{fig:wall}
\end{figure}

\begin{figure}
 \centering
 \includegraphics[scale=0.6]{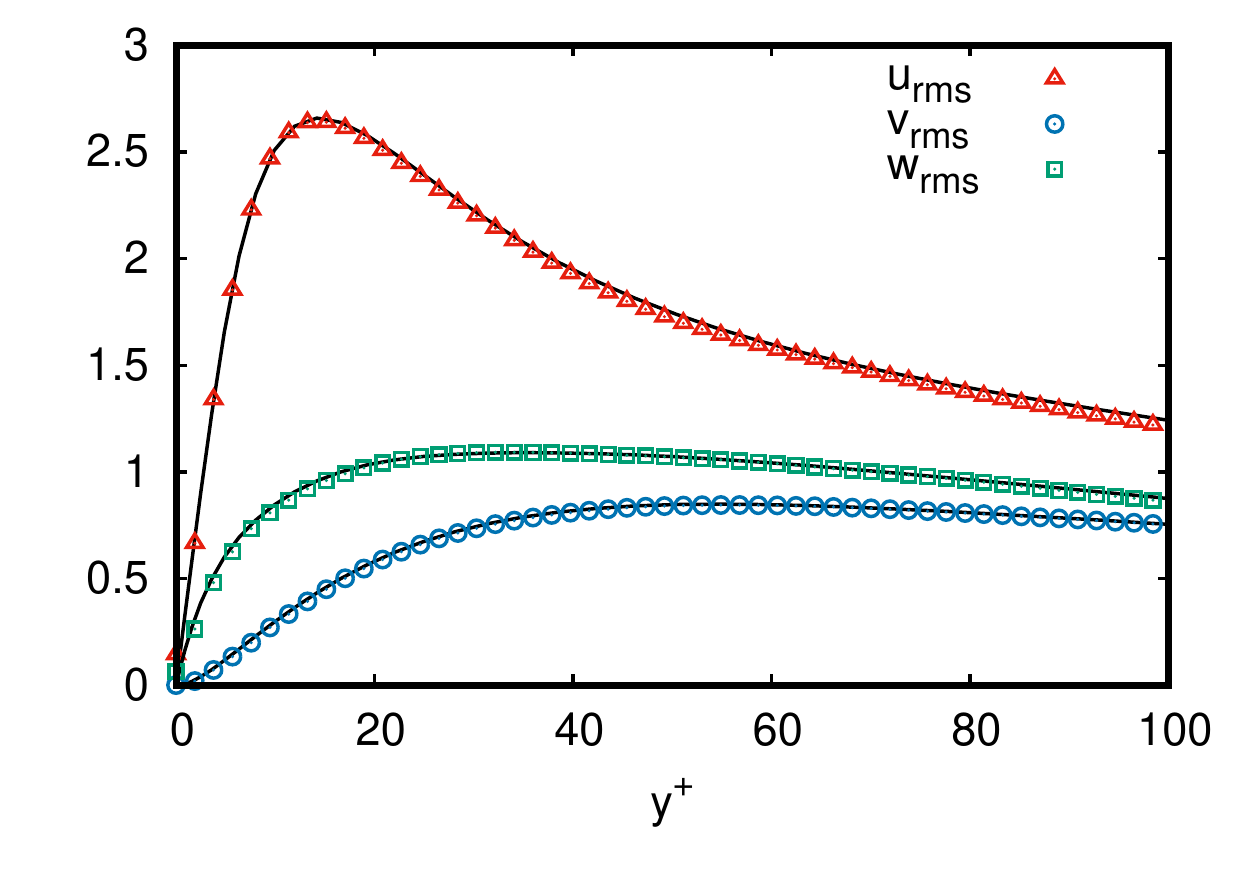}
 \caption{RMS velocity profiles  from our simulation (points) and DNS results (line) from~\citep{moser1999direct}.}
 \label{fig:RMS}
\end{figure}

At sufficiently high Reynolds numbers in channel flows, the variation of the mean velocity in wall coordinates ($u^+$) is known to follow
the law of the wall~\citep{pope2001turbulent}.
The mean velocity ($u^+$) scales linearly with wall coordinate $y^+$ in viscous layer and in the log-law region scales as $u^+ = (1/\kappa) ln(y^+) + B $
where $\kappa = 0.42$ and $B=5.5$  are constants. 
Fig. \ref{fig:wall} shows that the mean flow of velocity in wall coordinates from our simulations follows the law of the
wall and is in excellent agreement with that reported in Ref.~\citep{moser1999direct}.
RMS velocity profiles also show an excellent agreement, as shown in Fig. \ref{fig:RMS}.

The other mean flow properties like skin friction coefficient ($C_f$), the ratio of bulk mean velocity ($U_m$) and centreline
velocity ($U_c$) to wall shear velocity ($u_\tau$) show good agreement with Kim et al.~\citep{kim1987turbulence} and
are reported in Table \ref{table:Mean}. This shows that the proposed model is well suited to simulate wall-bounded turbulent flows.

\begin{table}
\begin{center}
\begin{tabular}{ccccccc} \hline \hline
                              & \ \ & $C_{f_0} $ & $C_{f}$ & $U_m/u_{\tau}$ & $U_c/u_{\tau}$ & $U_c/U_m$  \\ \hline
  RD3Q41                      & \ \ & $ 5.95 \times 10^{-3}$ & $ 8.23 \times 10^{-3}$ & 15.58 & 18.33 & 1.17  \\ 
  Ref.\cite{kim1987turbulence}& \ \ & $6.04 \times 10^{-3}$ & $8.18 \times 10^{-3}$ & 15.63 & 18.20 & 1.16 \\   \hline \hline
 \end{tabular}
 \caption{Mean flow properties from present RD3Q41 simulation and from Ref.~\citep{kim1987turbulence} at $Re_\tau = 180$.}
  \label{table:Mean}
\end{center}
\end{table}

\subsection{Flow past a sphere}
\begin{figure}
 \includegraphics[scale=0.07]{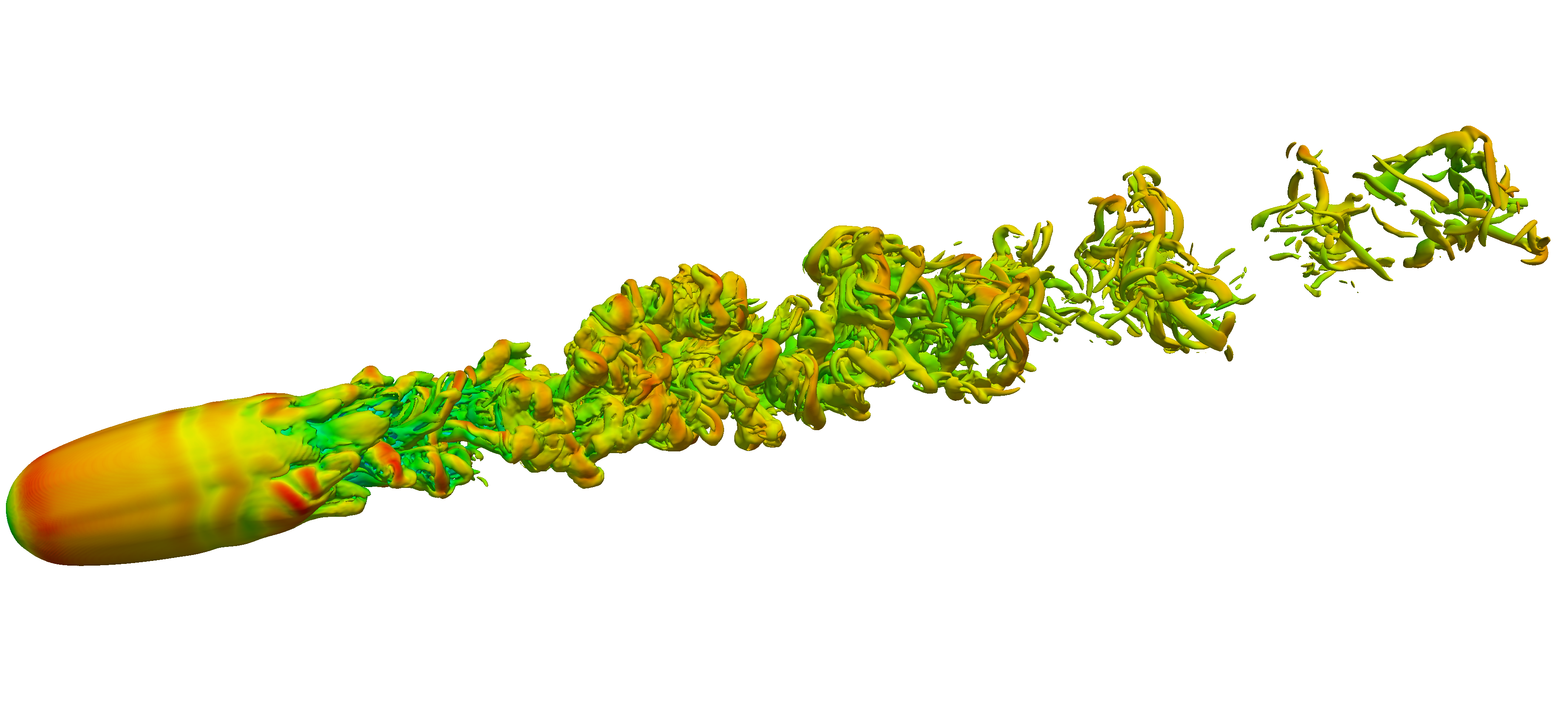}
 \caption{Isovorticity contours of flow over sphere at Reynolds number$=3700$.}
 \label{fig:sphere}
\end{figure}

Flow over a bluff bodies is a well studied problem that is of considerable academic and practical interest.
In this section, we simulate flow past a sphere at Reynolds number of $3700$.
The Reynolds number is defined based on the freestream velocity $U_0$ and the diameter of the sphere $D_0$.
A computational domain of $[-4.725D_0,15.525D_0]\times[-6.125D_0,6.125D_0]\times[-6.125D_0,6.125D_0]$ is used with $80$ grid points per diameter of the
sphere and the centre of sphere located at the origin.
The inlet and outlet boundary conditions are based on Ref.~\citep{chikatamarla2006grad}.
A diffused bounce-back boundary condition described in Ref.~\citep{krithivasan2014diffused} is implemented on the surface of the sphere.

\begin{table}
\begin{center}
\begin{tabular}{lcccc} \hline \hline
                                            & $C_d$  & Base  $C_p $  & $L_R$    & $\phi_s$      \\  \hline
   DNS~\citep{rodriguez2013flow}            & 0.394  & -0.207   & 2.28     & 89.4               \\ 
   LES~\citep{yun2006vortical}              & 0.355  &  -0.194  & 2.622    & 90                 \\ 
   Experiments\cite{kim1988observations}    &  -     &  -0.224  &  -       & -                  \\ 
   RD3Q41                                   &  0.427 &  -0.211  & 2.241    & 90                 \\  \hline \hline
   \end{tabular}                                                                                
 \caption{Drag coefficient ($C_d$), average base pressure coefficient ($C_p$), recirculation length ($L_R$), separation
angle ($\phi_s$).}
  \label{table:sphereData}
\end{center}
\end{table}

\begin{figure}
 \centering
 \includegraphics[scale=0.6]{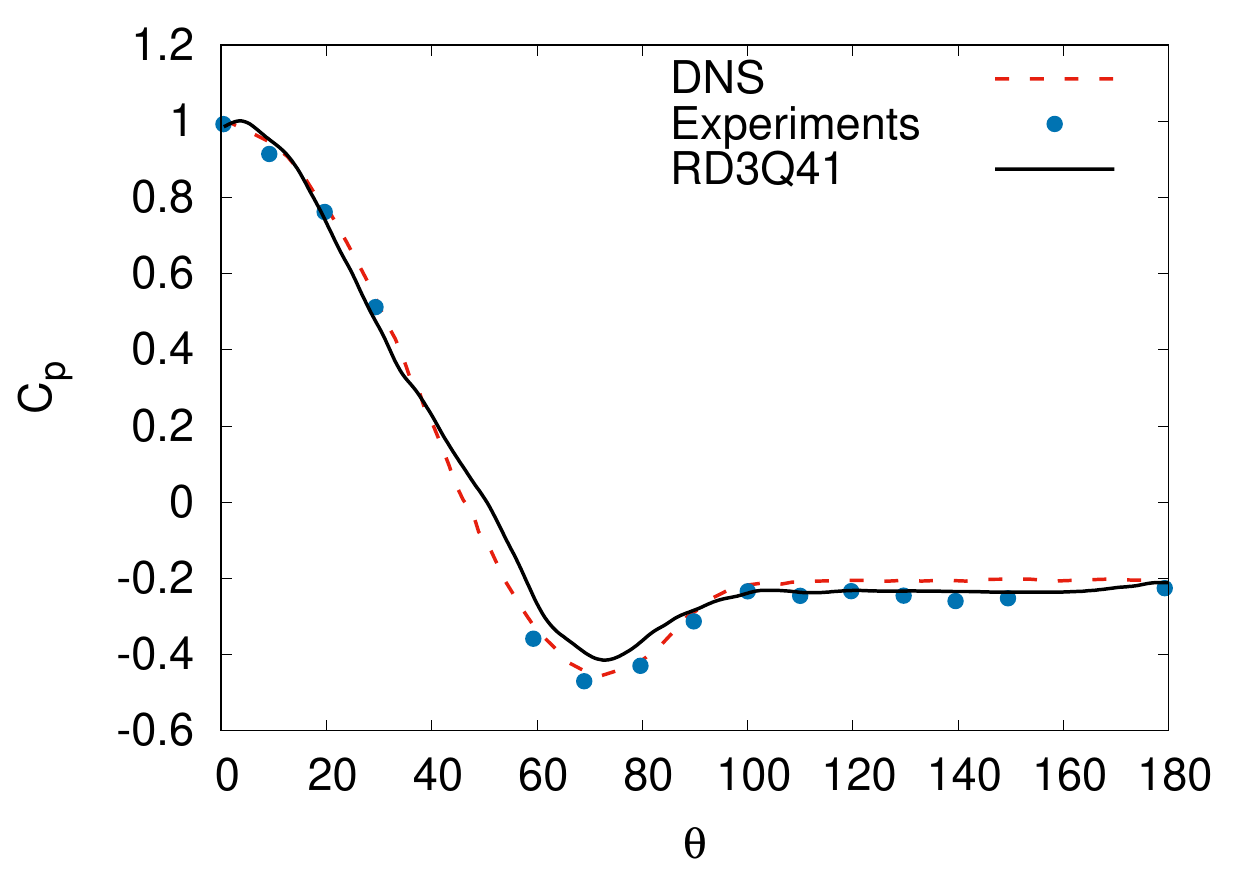}
 \caption{$C_p$ distribution on the surface of the sphere compared with with the experimental~\citep{kim1988observations} and DNS~\citep{rodriguez2013flow} studies.}
 \label{fig:sphereCP}
\end{figure}

We compare the flow variables such as the drag coefficient, base coefficient pressure, separation angle, and mean
recirculation length with the available numerical and experimental data in Table \ref{table:sphereData}. The drag coefficient $C_d$ is defined as
\begin{equation}
C_d = \frac{F_D}{\frac{1}{2} \rho U_0^2 A},
\end{equation}
where $F_D$ is the drag force and $A = \pi D_0^2/4$. The recirculation length is defined as the distance between the rear end of the sphere and the location where the
velocity in the streamwise direction changes its sign.
The angular distribution of the pressure coefficient $C_p$  defined as
\begin{equation}
 C_p = \frac{P - P_0}{\frac{1}{2} \rho U^2_0}
\end{equation}
is plotted over the surface of the sphere in Fig. \ref{fig:sphereCP}. $C_p$ profiles from an experimental study, and a DNS study are also shown for reference. 
It is to be noted that capturing the $C_p$ distribution requires a very high resolution of the sphere
\cite{dorschner2016grid} while with the current model, we find a reasonable agreement with only $80$ points per diameter.

\begin{figure*}
  \includegraphics[trim={0cm    0cm 0.8cm 0cm},clip,scale=0.55]{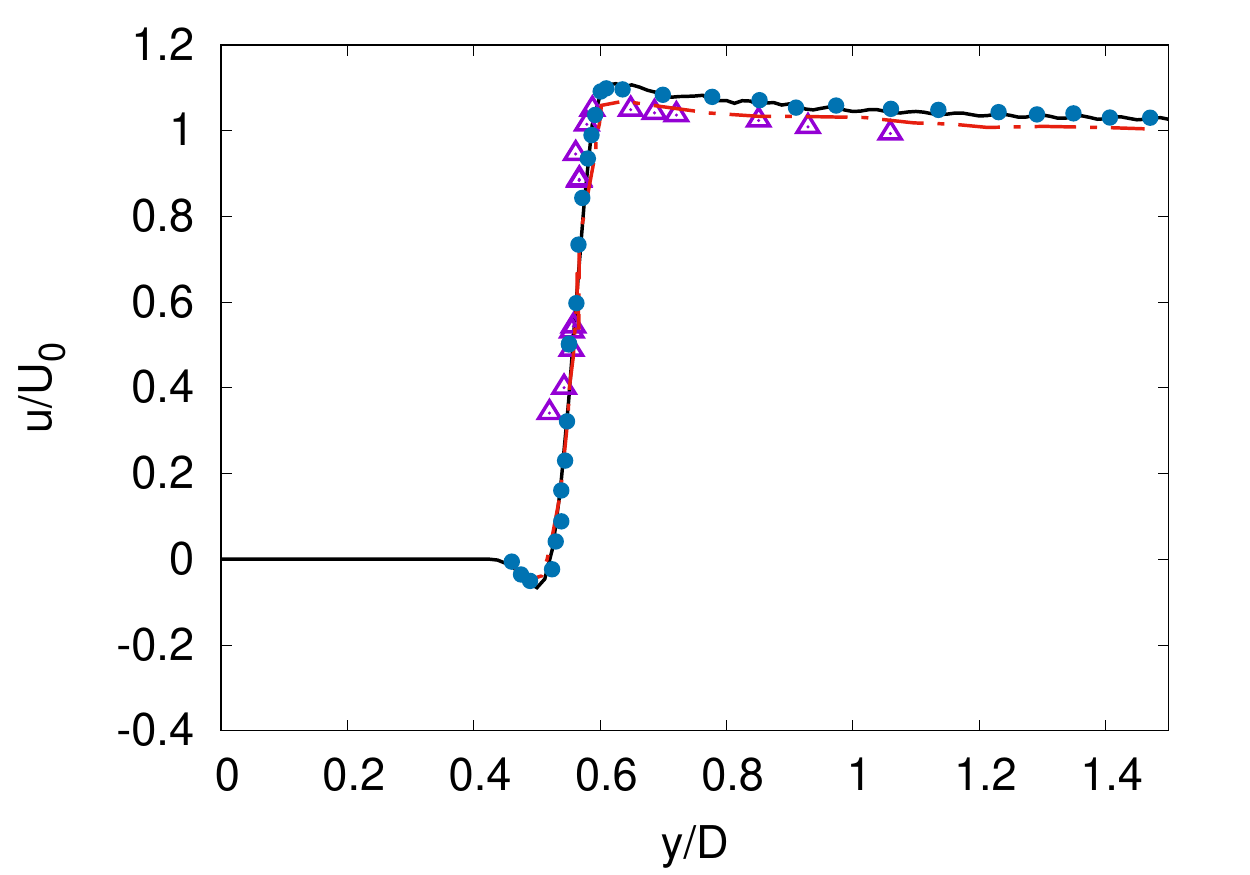}
  \includegraphics[trim={2.23cm 0cm 0.8cm 0cm},clip,scale=0.55]{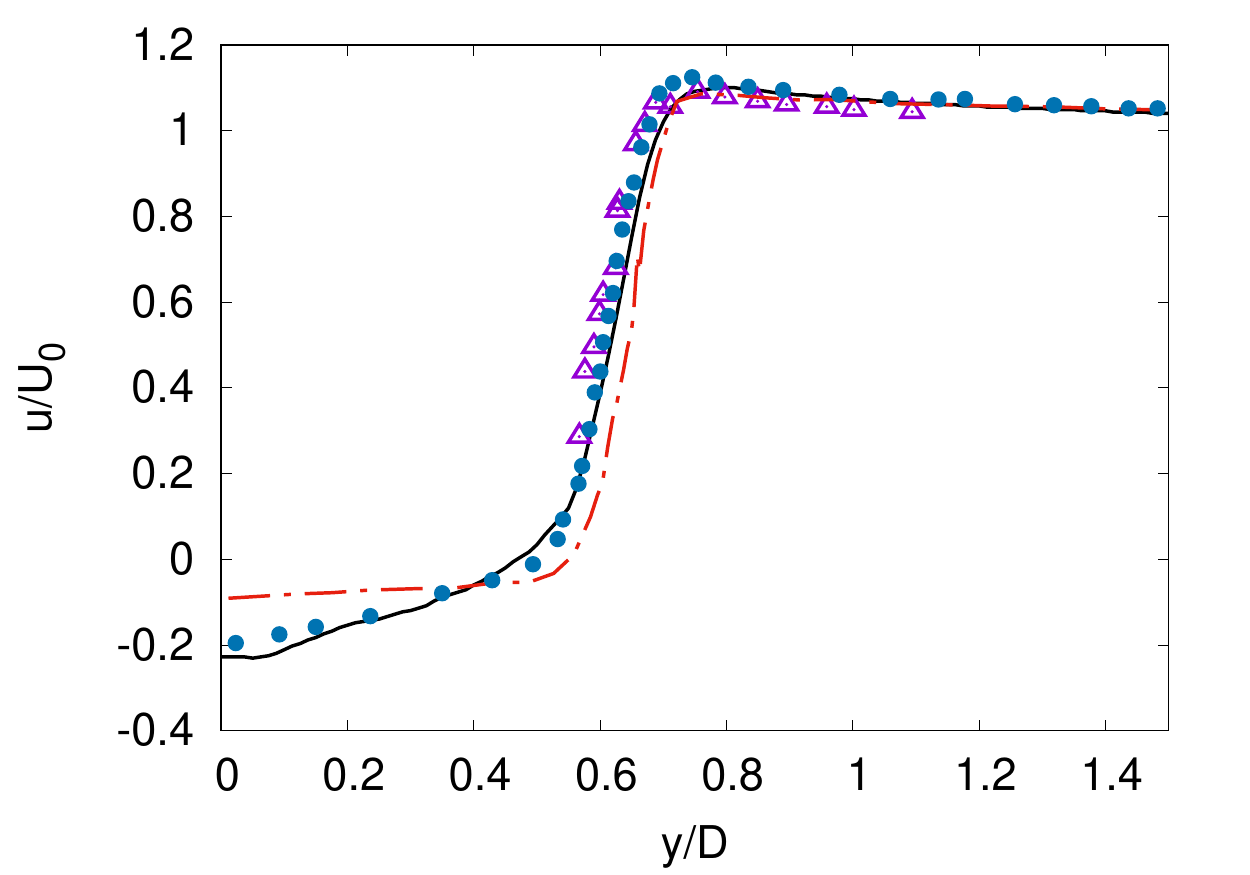}
  \includegraphics[trim={2.23cm 0cm 0.8cm 0cm},clip,scale=0.55]{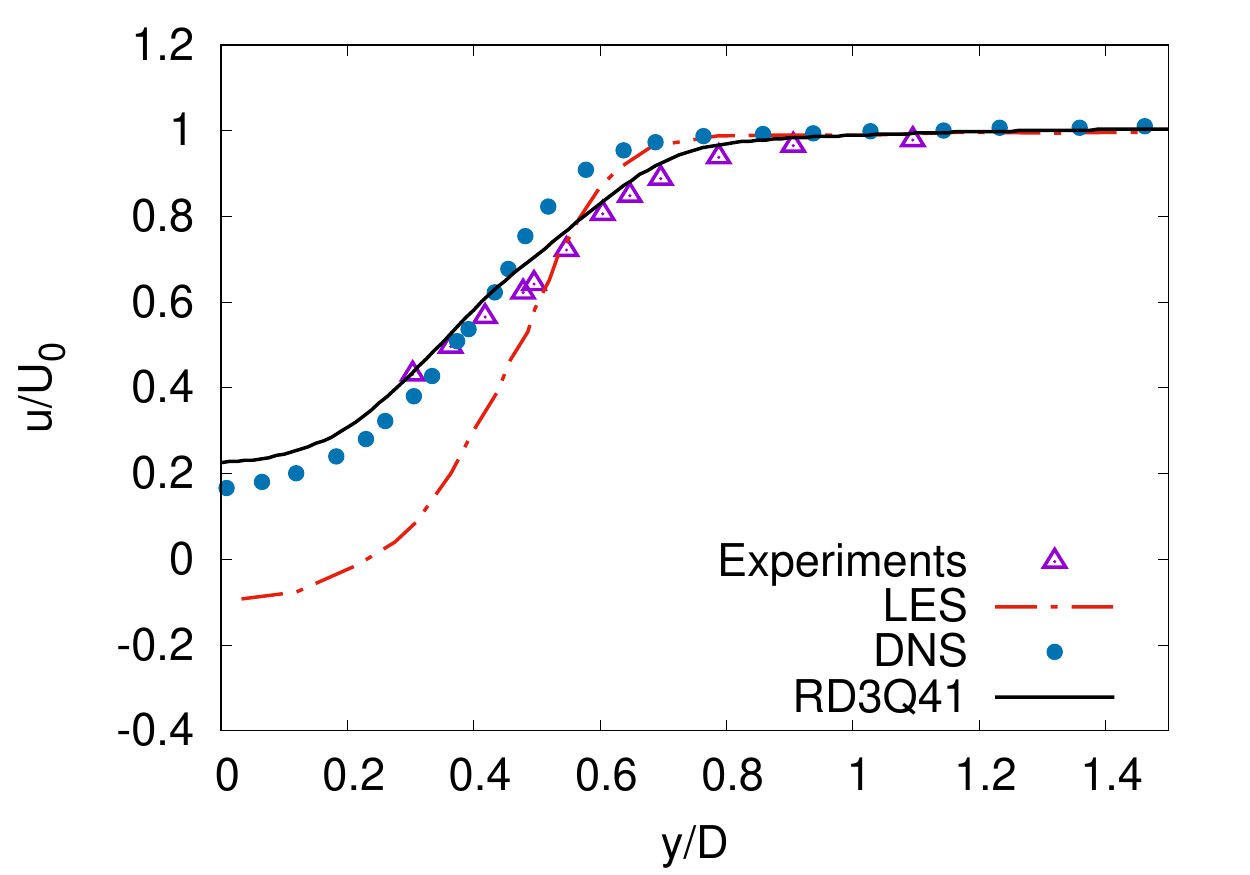}  
  \caption{Averaged profile of the normalised velocity at three different locations ($x/D = 0.2, 1.6, 3.0$)  
  in the wake of the sphere from the present study (solid line)
  compared with DNS~\citep{rodriguez2013flow}, LES~\citep{yun2006vortical} and experimental~\citep{kim1988observations} studies.}
  \label{fig:meanStreamwiseWake1}
\end{figure*}

The averaged profile of the velocity in the streamwise direction normalized with the freestream velocity at three different locations ($x/D_0 = 0.2, 1.6, 3.0$)  in the wake of the sphere are shown in Fig. \ref{fig:meanStreamwiseWake1}.
The  averaged profiles of the flow show a good agreement with the DNS~\citep{rodriguez2013flow}, LES~\citep{yun2006vortical} and experimental~\citep{kim1988observations} studies.

\section{Outlook}
In this paper, we have presented an energy conserving
lattice Boltzmann model which is suitable for compressible hydrodynamics, aeroacoustics and thermoacoustic problems. 
It recovers the pressure dynamics and the isentropic sound speed in addition to the effects of viscous heating, heat conduction
with a high degree of accuracy. 
The theoretical requirements and the methodology to construct this model have been discussed,
and the test cases confirm its stability for a wide range of parameters.
With improved accuracy in the velocity space and better representation of the curved surfaces, this model 
raises the prospect of direct simulations of turbulent flows involving objects.
This has been shown via simulations of Kida-Peltz flow, channel flow and flow past a sphere.
The specific heat ratio and Prandtl number of this model are fixed. This restriction on the Prandtl number can be addressed by a number of collision kernels~\citep{holway1966new,Shakhov1968,levermore1996moment,ansumali2007quasi}. 
The interactions between turbulence and acoustics along with methods to fix the specific heat ratio to 
that of polyatomic gases is subject of further investigation~\citep{nie2008thermal}.   

\acknowledgments{The authors would like to thank Soumyadeep Bhattacharya for his help in generating Fig. \ref{fig::BCC_SC}. }

\appendix
\section{Evaluating derivatives on a lattice} 
\label{appendixA}
In this section, we briefly review the way to evaluate the discrete gradients and Laplacian in an isotropic manner. 
The discrete derivative  $\tilde{\partial}$ is defined as
\begin{equation}
 \tilde{\partial}_\alpha \psi = \frac{1}{\hat \theta_0 \Delta t} \sum_{i} \hat w_i {c}_{i\alpha} \psi(\bm{x}+\bm{c}_{i}\Delta t).
\end{equation}
Here, $\hat w_i$ are the weights of the lattice with discrete velocities $c_{i\alpha}$ and $\psi$ is the function whose derivative needs to be calculated.
The above definition can be understood as follows: we consider the Taylor expansion of 
${\cal Q}_\alpha =  \sum_i \hat w_i c_{i\alpha} \psi(\bm{x}+\bm{c}_i\Delta t) /(\hat \theta_0 \Delta t) $ \citep{thampi2013isotropic,ramadugu2013lattice},
\begin{align}
\begin{split} 
{\cal Q}_\alpha &\equiv  \frac{1}{\hat \theta_0 \Delta t} \sum_i \hat w_i c_{i\alpha} \psi(\bm{x}+\bm{c}_i\Delta t) 
 = \frac{1}{\hat \theta_0 \Delta t} \bigg[ \psi(\bm{x})\sum_i \hat w_i c_{i\alpha} \\&+\Delta t \frac{\partial \psi}{\partial x_\beta}\sum_i \hat w_i c_{i\alpha} c_{i\beta} 
 + \frac{\Delta t^2}{2}\frac{\partial^2 \psi}{\partial x_\beta \partial x_\gamma} \sum_i \hat w_i c_{i\alpha} c_{i\beta} c_{i\gamma} 
 \\&+ \frac{\Delta t^3}{6}\frac{\partial^3 \psi}{\partial x_\beta \partial x_\gamma \partial x_\kappa} \sum_i \hat w_i c_{i\alpha} c_{i\beta} c_{i\gamma} c_{i\kappa} 
+ \mathcal{O}(\Delta t^4) \bigg].
 \label{TaylorExp}
\end{split}
\end{align}
The odd-order moments of weight are zero due to symmetry of the underlying lattice and further simplification of ${\cal Q}_\alpha$ leads to
\begin{align}
\begin{split} 
{\cal Q}_\alpha &\equiv \frac{1}{\hat \theta_0 \Delta t} \sum_i \hat w_i c_{i\alpha} \psi(\bm{x}+\bm{c}_i\Delta t) 
 =  \frac{\partial \psi}{\partial x_\alpha}
\\& + \frac{\Delta t^2 \hat\theta_0 }{2} \frac{\partial}{\partial x_\alpha} \partial^2 \psi
 + \frac{3\Delta t^4 \hat\theta_0^2}{8} \frac{\partial}{\partial x_\alpha} \partial^4 \psi + \mathcal{O}(\Delta t^6),
 \label{TaylorExpSimple}
\end{split}
\end{align}
assuming the lattice is sufficiently isotropic. The emergence of the derivatives of $\psi$ is seen on the right hand side of the above equation.

The second-order gradient is hence
\begin{align}
\begin{split} 
 \tilde{\partial}_\alpha^{(2)} \psi = \frac{1}{\hat \theta_0 \Delta t} \sum_{i} \hat w_i {c}_{i\alpha} \psi(\bm{x}+\bm{c}_{i}\Delta t).
 \label{gradient2}
\end{split}
\end{align}
The fourth-order gradient is written as 
\begin{align}
\begin{split} 
 \tilde{\partial}_\alpha^{(4)} \psi & =  \frac{1}{\hat \theta_0 \Delta t } \sum_{i} \hat w_i {c}_{i\alpha} \psi(\bm{x}+\bm{c}_{i}\Delta t) 
  -\frac{\hat \theta_0}{2} (\Delta t)^2 {\nabla}_\alpha ({\nabla}^2 \psi).
 \label{gradient4}
\end{split}
\end{align}
The Laplacian can also be evaluated as 
\begin{align}
\begin{split} 
  \tilde{\partial}^2 \psi = \frac{2}{\Delta t^2 \hat \theta_0} \left[\sum_{i} \hat w_i \psi(\bm{x}+\bm{c}_i\Delta t) - \psi(\bm{x}) \right].
\end{split}
\end{align}
Here, it remains to justify the choice of stencil for calculating derivatives. From 
Eq.\eqref{TaylorExpSimple} it is evident that the discretization errors are proportional to $\hat \theta_0 =  \sum \hat w_i c_{ix}^2 $ 
of the chosen stencil. 
For the crystallographic grid, the $15$ velocity stencil comprising of the discrete velocities
\begin{equation}
c_i = \{(\pm 1,0,0),(0,\pm 1,0),(0,0,\pm 1),(\pm 0.5, \pm 0.5, \pm 0.5) \},
\end{equation}
is found to have the least $\hat \theta_0=1/6$ with $\hat w_{0} = 14/36,$ $\hat w_{\rm SC} = 1/36$ and $ \hat w_{\rm BCC} = 2/36$, therefore is the ideal choice for computing derivatives.

\bibliography{RD3Q41}

\begin{thebibliography}{102}%
\makeatletter
\providecommand \@ifxundefined [1]{%
 \@ifx{#1\undefined}
}%
\providecommand \@ifnum [1]{%
 \ifnum #1\expandafter \@firstoftwo
 \else \expandafter \@secondoftwo
 \fi
}%
\providecommand \@ifx [1]{%
 \ifx #1\expandafter \@firstoftwo
 \else \expandafter \@secondoftwo
 \fi
}%
\providecommand \natexlab [1]{#1}%
\providecommand \enquote  [1]{``#1''}%
\providecommand \bibnamefont  [1]{#1}%
\providecommand \bibfnamefont [1]{#1}%
\providecommand \citenamefont [1]{#1}%
\providecommand \href@noop [0]{\@secondoftwo}%
\providecommand \href [0]{\begingroup \@sanitize@url \@href}%
\providecommand \@href[1]{\@@startlink{#1}\@@href}%
\providecommand \@@href[1]{\endgroup#1\@@endlink}%
\providecommand \@sanitize@url [0]{\catcode `\\12\catcode `\$12\catcode
  `\&12\catcode `\#12\catcode `\^12\catcode `\_12\catcode `\%12\relax}%
\providecommand \@@startlink[1]{}%
\providecommand \@@endlink[0]{}%
\providecommand \url  [0]{\begingroup\@sanitize@url \@url }%
\providecommand \@url [1]{\endgroup\@href {#1}{\urlprefix }}%
\providecommand \urlprefix  [0]{URL }%
\providecommand \Eprint [0]{\href }%
\providecommand \doibase [0]{http://dx.doi.org/}%
\providecommand \selectlanguage [0]{\@gobble}%
\providecommand \bibinfo  [0]{\@secondoftwo}%
\providecommand \bibfield  [0]{\@secondoftwo}%
\providecommand \translation [1]{[#1]}%
\providecommand \BibitemOpen [0]{}%
\providecommand \bibitemStop [0]{}%
\providecommand \bibitemNoStop [0]{.\EOS\space}%
\providecommand \EOS [0]{\spacefactor3000\relax}%
\providecommand \BibitemShut  [1]{\csname bibitem#1\endcsname}%
\let\auto@bib@innerbib\@empty
\bibitem [{\citenamefont {Succi}(2001)}]{succi2001lattice}%
  \BibitemOpen
  \bibfield  {author} {\bibinfo {author} {\bibfnamefont {S.}~\bibnamefont
  {Succi}},\ }\href@noop {} {\emph {\bibinfo {title} {The lattice Boltzmann
  equation: for fluid dynamics and beyond}}}\ (\bibinfo  {publisher} {Oxford
  university press},\ \bibinfo {year} {2001})\BibitemShut {NoStop}%
\bibitem [{\citenamefont {Chen}\ and\ \citenamefont
  {Doolen}(1998)}]{chen1998lattice}%
  \BibitemOpen
  \bibfield  {author} {\bibinfo {author} {\bibfnamefont {S.}~\bibnamefont
  {Chen}}\ and\ \bibinfo {author} {\bibfnamefont {G.~D.}\ \bibnamefont
  {Doolen}},\ }\href@noop {} {\bibfield  {journal} {\bibinfo  {journal} {Annu.
  Rev. Fluid Mech.}\ }\textbf {\bibinfo {volume} {30}},\ \bibinfo {pages} {329}
  (\bibinfo {year} {1998})}\BibitemShut {NoStop}%
\bibitem [{\citenamefont {Aidun}\ and\ \citenamefont
  {Clausen}(2010)}]{aidun2010lattice}%
  \BibitemOpen
  \bibfield  {author} {\bibinfo {author} {\bibfnamefont {C.~K.}\ \bibnamefont
  {Aidun}}\ and\ \bibinfo {author} {\bibfnamefont {J.~R.}\ \bibnamefont
  {Clausen}},\ }\href@noop {} {\bibfield  {journal} {\bibinfo  {journal} {Annu.
  Rev. Fluid Mech.}\ }\textbf {\bibinfo {volume} {42}},\ \bibinfo {pages} {439}
  (\bibinfo {year} {2010})}\BibitemShut {NoStop}%
\bibitem [{\citenamefont {Ansumali}\ and\ \citenamefont
  {Karlin}(2005)}]{ansumali2005consistent}%
  \BibitemOpen
  \bibfield  {author} {\bibinfo {author} {\bibfnamefont {S.}~\bibnamefont
  {Ansumali}}\ and\ \bibinfo {author} {\bibfnamefont {I.~V.}\ \bibnamefont
  {Karlin}},\ }\href@noop {} {\bibfield  {journal} {\bibinfo  {journal} {Phys.
  Rev. Lett.}\ }\textbf {\bibinfo {volume} {95}},\ \bibinfo {pages} {260605}
  (\bibinfo {year} {2005})}\BibitemShut {NoStop}%
\bibitem [{\citenamefont {Dellar}(2001)}]{dellar2001bulk}%
  \BibitemOpen
  \bibfield  {author} {\bibinfo {author} {\bibfnamefont {P.~J.}\ \bibnamefont
  {Dellar}},\ }\href@noop {} {\bibfield  {journal} {\bibinfo  {journal} {Phys.
  Rev. E}\ }\textbf {\bibinfo {volume} {64}},\ \bibinfo {pages} {031203}
  (\bibinfo {year} {2001})}\BibitemShut {NoStop}%
\bibitem [{\citenamefont {Namburi}\ \emph {et~al.}(2016)\citenamefont
  {Namburi}, \citenamefont {Krithivasan},\ and\ \citenamefont
  {Ansumali}}]{namburi2016crystallographic}%
  \BibitemOpen
  \bibfield  {author} {\bibinfo {author} {\bibfnamefont {M.}~\bibnamefont
  {Namburi}}, \bibinfo {author} {\bibfnamefont {S.}~\bibnamefont
  {Krithivasan}}, \ and\ \bibinfo {author} {\bibfnamefont {S.}~\bibnamefont
  {Ansumali}},\ }\href@noop {} {\bibfield  {journal} {\bibinfo  {journal} {Sci.
  Rep.}\ }\textbf {\bibinfo {volume} {6}},\ \bibinfo {pages} {27172} (\bibinfo
  {year} {2016})}\BibitemShut {NoStop}%
\bibitem [{\citenamefont {Ladd}(1994)}]{ladd1994numerical}%
  \BibitemOpen
  \bibfield  {author} {\bibinfo {author} {\bibfnamefont {A.~J.}\ \bibnamefont
  {Ladd}},\ }\href@noop {} {\bibfield  {journal} {\bibinfo  {journal} {J. Fluid
  Mech.}\ }\textbf {\bibinfo {volume} {271}},\ \bibinfo {pages} {285} (\bibinfo
  {year} {1994})}\BibitemShut {NoStop}%
\bibitem [{\citenamefont {Ladd}\ and\ \citenamefont
  {Verberg}(2001)}]{ladd2001lattice}%
  \BibitemOpen
  \bibfield  {author} {\bibinfo {author} {\bibfnamefont {A.}~\bibnamefont
  {Ladd}}\ and\ \bibinfo {author} {\bibfnamefont {R.}~\bibnamefont {Verberg}},\
  }\href@noop {} {\bibfield  {journal} {\bibinfo  {journal} {J. Stat. Phys.}\
  }\textbf {\bibinfo {volume} {104}},\ \bibinfo {pages} {1191} (\bibinfo {year}
  {2001})}\BibitemShut {NoStop}%
\bibitem [{\citenamefont {Ladd}(1993)}]{ladd1993short}%
  \BibitemOpen
  \bibfield  {author} {\bibinfo {author} {\bibfnamefont {A.~J.}\ \bibnamefont
  {Ladd}},\ }\href@noop {} {\bibfield  {journal} {\bibinfo  {journal} {Phys.
  Rev. Lett.}\ }\textbf {\bibinfo {volume} {70}},\ \bibinfo {pages} {1339}
  (\bibinfo {year} {1993})}\BibitemShut {NoStop}%
\bibitem [{\citenamefont {Buick}\ \emph {et~al.}(1998)\citenamefont {Buick},
  \citenamefont {Greated},\ and\ \citenamefont {Campbell}}]{buick1998lattice}%
  \BibitemOpen
  \bibfield  {author} {\bibinfo {author} {\bibfnamefont {J.}~\bibnamefont
  {Buick}}, \bibinfo {author} {\bibfnamefont {C.}~\bibnamefont {Greated}}, \
  and\ \bibinfo {author} {\bibfnamefont {D.}~\bibnamefont {Campbell}},\
  }\href@noop {} {\bibfield  {journal} {\bibinfo  {journal} {Europhys. Lett.}\
  }\textbf {\bibinfo {volume} {43}},\ \bibinfo {pages} {235} (\bibinfo {year}
  {1998})}\BibitemShut {NoStop}%
\bibitem [{\citenamefont {Crouse}\ \emph {et~al.}(2006)\citenamefont {Crouse},
  \citenamefont {Freed}, \citenamefont {Balasubramanian}, \citenamefont
  {Senthooran}, \citenamefont {Lew},\ and\ \citenamefont
  {Mongeau}}]{crouse2006fundamental}%
  \BibitemOpen
  \bibfield  {author} {\bibinfo {author} {\bibfnamefont {B.}~\bibnamefont
  {Crouse}}, \bibinfo {author} {\bibfnamefont {D.}~\bibnamefont {Freed}},
  \bibinfo {author} {\bibfnamefont {G.}~\bibnamefont {Balasubramanian}},
  \bibinfo {author} {\bibfnamefont {S.}~\bibnamefont {Senthooran}}, \bibinfo
  {author} {\bibfnamefont {P.~T.}\ \bibnamefont {Lew}}, \ and\ \bibinfo
  {author} {\bibfnamefont {L.}~\bibnamefont {Mongeau}},\ }in\ \href@noop {}
  {\emph {\bibinfo {booktitle} {12th AIAA/CEAS Aeroacoustics Conference (27th
  AIAA Aeroacoustics Conference)}}}\ (\bibinfo {year} {2006})\ p.\ \bibinfo
  {pages} {2571}\BibitemShut {NoStop}%
\bibitem [{\citenamefont {Mari{\'e}}\ \emph {et~al.}(2009)\citenamefont
  {Mari{\'e}}, \citenamefont {Ricot},\ and\ \citenamefont
  {Sagaut}}]{marie2009comparison}%
  \BibitemOpen
  \bibfield  {author} {\bibinfo {author} {\bibfnamefont {S.}~\bibnamefont
  {Mari{\'e}}}, \bibinfo {author} {\bibfnamefont {D.}~\bibnamefont {Ricot}}, \
  and\ \bibinfo {author} {\bibfnamefont {P.}~\bibnamefont {Sagaut}},\
  }\href@noop {} {\bibfield  {journal} {\bibinfo  {journal} {J. Comput. Phys.}\
  }\textbf {\bibinfo {volume} {228}},\ \bibinfo {pages} {1056} (\bibinfo {year}
  {2009})}\BibitemShut {NoStop}%
\bibitem [{\citenamefont {Li}\ and\ \citenamefont
  {Shan}(2011)}]{li2011lattice}%
  \BibitemOpen
  \bibfield  {author} {\bibinfo {author} {\bibfnamefont {Y.}~\bibnamefont
  {Li}}\ and\ \bibinfo {author} {\bibfnamefont {X.}~\bibnamefont {Shan}},\
  }\href@noop {} {\bibfield  {journal} {\bibinfo  {journal} {Philosophical
  Transactions of the Royal Society A: Mathematical, Physical and Engineering
  Sciences}\ }\textbf {\bibinfo {volume} {369}},\ \bibinfo {pages} {2371}
  (\bibinfo {year} {2011})}\BibitemShut {NoStop}%
\bibitem [{\citenamefont {Viggen}(2013)}]{viggen1}%
  \BibitemOpen
  \bibfield  {author} {\bibinfo {author} {\bibfnamefont {E.~M.}\ \bibnamefont
  {Viggen}},\ }\href {\doibase 10.1103/PhysRevE.87.023306} {\bibfield
  {journal} {\bibinfo  {journal} {Phys. Rev. E}\ }\textbf {\bibinfo {volume}
  {87}},\ \bibinfo {pages} {023306} (\bibinfo {year} {2013})}\BibitemShut
  {NoStop}%
\bibitem [{\citenamefont {Viggen}(2014)}]{viggen2}%
  \BibitemOpen
  \bibfield  {author} {\bibinfo {author} {\bibfnamefont {E.~M.}\ \bibnamefont
  {Viggen}},\ }\href {\doibase 10.1103/PhysRevE.90.013310} {\bibfield
  {journal} {\bibinfo  {journal} {Phys. Rev. E}\ }\textbf {\bibinfo {volume}
  {90}},\ \bibinfo {pages} {013310} (\bibinfo {year} {2014})}\BibitemShut
  {NoStop}%
\bibitem [{\citenamefont {Landau}\ and\ \citenamefont
  {Lifshitz}(1959)}]{landau1959course}%
  \BibitemOpen
  \bibfield  {author} {\bibinfo {author} {\bibfnamefont {L.~D.}\ \bibnamefont
  {Landau}}\ and\ \bibinfo {author} {\bibfnamefont {E.}~\bibnamefont
  {Lifshitz}},\ }\href@noop {} {\emph {\bibinfo {title} {Course of Theoretical
  Physics Vol. 6 Fluid Mechanics}}}\ (\bibinfo  {publisher} {Pergamon Press},\
  \bibinfo {year} {1959})\BibitemShut {NoStop}%
\bibitem [{\citenamefont {Chaikin}\ and\ \citenamefont
  {Lubensky}(2000)}]{chaikin2000principles}%
  \BibitemOpen
  \bibfield  {author} {\bibinfo {author} {\bibfnamefont {P.~M.}\ \bibnamefont
  {Chaikin}}\ and\ \bibinfo {author} {\bibfnamefont {T.~C.}\ \bibnamefont
  {Lubensky}},\ }\href@noop {} {\emph {\bibinfo {title} {Principles of
  condensed matter physics}}},\ Vol.~\bibinfo {volume} {1}\ (\bibinfo
  {publisher} {Cambridge university press Cambridge},\ \bibinfo {year}
  {2000})\BibitemShut {NoStop}%
\bibitem [{\citenamefont {Callen}(1998)}]{callen1998thermodynamics}%
  \BibitemOpen
  \bibfield  {author} {\bibinfo {author} {\bibfnamefont {H.~B.}\ \bibnamefont
  {Callen}},\ }\href@noop {} {\enquote {\bibinfo {title} {Thermodynamics and an
  introduction to thermostatistics},}\ } (\bibinfo {year} {1998})\BibitemShut
  {NoStop}%
\bibitem [{\citenamefont {Leal}(2007)}]{leal2007advanced}%
  \BibitemOpen
  \bibfield  {author} {\bibinfo {author} {\bibfnamefont {L.~G.}\ \bibnamefont
  {Leal}},\ }\href@noop {} {\emph {\bibinfo {title} {Advanced transport
  phenomena: fluid mechanics and convective transport processes}}},\
  Vol.~\bibinfo {volume} {7}\ (\bibinfo  {publisher} {Cambridge University
  Press},\ \bibinfo {year} {2007})\BibitemShut {NoStop}%
\bibitem [{\citenamefont {Singh}\ \emph {et~al.}(2013)\citenamefont {Singh},
  \citenamefont {Krithivasan}, \citenamefont {Karlin}, \citenamefont {Succi},\
  and\ \citenamefont {Ansumali}}]{singh2013energy}%
  \BibitemOpen
  \bibfield  {author} {\bibinfo {author} {\bibfnamefont {S.}~\bibnamefont
  {Singh}}, \bibinfo {author} {\bibfnamefont {S.}~\bibnamefont {Krithivasan}},
  \bibinfo {author} {\bibfnamefont {I.~V.}\ \bibnamefont {Karlin}}, \bibinfo
  {author} {\bibfnamefont {S.}~\bibnamefont {Succi}}, \ and\ \bibinfo {author}
  {\bibfnamefont {S.}~\bibnamefont {Ansumali}},\ }\href@noop {} {\bibfield
  {journal} {\bibinfo  {journal} {Commun. Comput. Phys.}\ }\textbf {\bibinfo
  {volume} {13}},\ \bibinfo {pages} {603} (\bibinfo {year} {2013})}\BibitemShut
  {NoStop}%
\bibitem [{\citenamefont {Frapolli}\ \emph {et~al.}(2014)\citenamefont
  {Frapolli}, \citenamefont {Chikatamarla},\ and\ \citenamefont
  {Karlin}}]{frapolli2014multispeed}%
  \BibitemOpen
  \bibfield  {author} {\bibinfo {author} {\bibfnamefont {N.}~\bibnamefont
  {Frapolli}}, \bibinfo {author} {\bibfnamefont {S.}~\bibnamefont
  {Chikatamarla}}, \ and\ \bibinfo {author} {\bibfnamefont {I.}~\bibnamefont
  {Karlin}},\ }\href@noop {} {\bibfield  {journal} {\bibinfo  {journal} {Phys.
  Rev. E}\ }\textbf {\bibinfo {volume} {90}},\ \bibinfo {pages} {043306}
  (\bibinfo {year} {2014})}\BibitemShut {NoStop}%
\bibitem [{\citenamefont {Atif}\ \emph {et~al.}(2018)\citenamefont {Atif},
  \citenamefont {Namburi},\ and\ \citenamefont {Ansumali}}]{atif2018higher}%
  \BibitemOpen
  \bibfield  {author} {\bibinfo {author} {\bibfnamefont {M.}~\bibnamefont
  {Atif}}, \bibinfo {author} {\bibfnamefont {M.}~\bibnamefont {Namburi}}, \
  and\ \bibinfo {author} {\bibfnamefont {S.}~\bibnamefont {Ansumali}},\
  }\href@noop {} {\bibfield  {journal} {\bibinfo  {journal} {Phys. Rev. E}\
  }\textbf {\bibinfo {volume} {98}},\ \bibinfo {pages} {053311} (\bibinfo
  {year} {2018})}\BibitemShut {NoStop}%
\bibitem [{\citenamefont {Yudistiawan}\ \emph
  {et~al.}(2010{\natexlab{a}})\citenamefont {Yudistiawan}, \citenamefont
  {Kwak}, \citenamefont {Patil},\ and\ \citenamefont {Ansumali}}]{Wahyu2010}%
  \BibitemOpen
  \bibfield  {author} {\bibinfo {author} {\bibfnamefont {W.~P.}\ \bibnamefont
  {Yudistiawan}}, \bibinfo {author} {\bibfnamefont {S.~K.}\ \bibnamefont
  {Kwak}}, \bibinfo {author} {\bibfnamefont {D.~V.}\ \bibnamefont {Patil}}, \
  and\ \bibinfo {author} {\bibfnamefont {S.}~\bibnamefont {Ansumali}},\
  }\href@noop {} {\bibfield  {journal} {\bibinfo  {journal} {Phys. Rev. E}\
  }\textbf {\bibinfo {volume} {82}},\ \bibinfo {pages} {046701} (\bibinfo
  {year} {2010}{\natexlab{a}})}\BibitemShut {NoStop}%
\bibitem [{\citenamefont {Bhatnagar}\ \emph {et~al.}(1954)\citenamefont
  {Bhatnagar}, \citenamefont {Gross},\ and\ \citenamefont
  {Krook}}]{bhatnagar1954model}%
  \BibitemOpen
  \bibfield  {author} {\bibinfo {author} {\bibfnamefont {P.~L.}\ \bibnamefont
  {Bhatnagar}}, \bibinfo {author} {\bibfnamefont {E.~P.}\ \bibnamefont
  {Gross}}, \ and\ \bibinfo {author} {\bibfnamefont {M.}~\bibnamefont
  {Krook}},\ }\href@noop {} {\bibfield  {journal} {\bibinfo  {journal} {Phys.
  Rev.}\ }\textbf {\bibinfo {volume} {94}},\ \bibinfo {pages} {511} (\bibinfo
  {year} {1954})}\BibitemShut {NoStop}%
\bibitem [{\citenamefont {Ansumali}\ \emph {et~al.}(2003)\citenamefont
  {Ansumali}, \citenamefont {Karlin},\ and\ \citenamefont
  {{\"O}ttinger}}]{ansumali2003minimal}%
  \BibitemOpen
  \bibfield  {author} {\bibinfo {author} {\bibfnamefont {S.}~\bibnamefont
  {Ansumali}}, \bibinfo {author} {\bibfnamefont {I.~V.}\ \bibnamefont
  {Karlin}}, \ and\ \bibinfo {author} {\bibfnamefont {H.~C.}\ \bibnamefont
  {{\"O}ttinger}},\ }\href@noop {} {\bibfield  {journal} {\bibinfo  {journal}
  {Europhys. Lett.}\ }\textbf {\bibinfo {volume} {63}},\ \bibinfo {pages} {798}
  (\bibinfo {year} {2003})}\BibitemShut {NoStop}%
\bibitem [{\citenamefont {Wagner}(1998)}]{wagner1998h}%
  \BibitemOpen
  \bibfield  {author} {\bibinfo {author} {\bibfnamefont {A.~J.}\ \bibnamefont
  {Wagner}},\ }\href@noop {} {\bibfield  {journal} {\bibinfo  {journal}
  {Europhys. Lett.}\ }\textbf {\bibinfo {volume} {44}},\ \bibinfo {pages} {144}
  (\bibinfo {year} {1998})}\BibitemShut {NoStop}%
\bibitem [{\citenamefont {Chen}\ \emph {et~al.}(1992)\citenamefont {Chen},
  \citenamefont {Chen},\ and\ \citenamefont {Matthaeus}}]{chen1992recovery}%
  \BibitemOpen
  \bibfield  {author} {\bibinfo {author} {\bibfnamefont {H.}~\bibnamefont
  {Chen}}, \bibinfo {author} {\bibfnamefont {S.}~\bibnamefont {Chen}}, \ and\
  \bibinfo {author} {\bibfnamefont {W.~H.}\ \bibnamefont {Matthaeus}},\
  }\href@noop {} {\bibfield  {journal} {\bibinfo  {journal} {Phys. Rev. A}\
  }\textbf {\bibinfo {volume} {45}},\ \bibinfo {pages} {R5339} (\bibinfo {year}
  {1992})}\BibitemShut {NoStop}%
\bibitem [{\citenamefont {Qian}\ \emph {et~al.}(1992)\citenamefont {Qian},
  \citenamefont {d'Humi{\`e}res},\ and\ \citenamefont
  {Lallemand}}]{qian1992lattice}%
  \BibitemOpen
  \bibfield  {author} {\bibinfo {author} {\bibfnamefont {Y.}~\bibnamefont
  {Qian}}, \bibinfo {author} {\bibfnamefont {D.}~\bibnamefont
  {d'Humi{\`e}res}}, \ and\ \bibinfo {author} {\bibfnamefont {P.}~\bibnamefont
  {Lallemand}},\ }\href@noop {} {\bibfield  {journal} {\bibinfo  {journal}
  {Europhys. Lett.}\ }\textbf {\bibinfo {volume} {17}},\ \bibinfo {pages} {479}
  (\bibinfo {year} {1992})}\BibitemShut {NoStop}%
\bibitem [{\citenamefont {Benzi}\ \emph {et~al.}(1992)\citenamefont {Benzi},
  \citenamefont {Succi},\ and\ \citenamefont {Vergassola}}]{benzi1992lattice}%
  \BibitemOpen
  \bibfield  {author} {\bibinfo {author} {\bibfnamefont {R.}~\bibnamefont
  {Benzi}}, \bibinfo {author} {\bibfnamefont {S.}~\bibnamefont {Succi}}, \ and\
  \bibinfo {author} {\bibfnamefont {M.}~\bibnamefont {Vergassola}},\
  }\href@noop {} {\bibfield  {journal} {\bibinfo  {journal} {Phys. Rep.}\
  }\textbf {\bibinfo {volume} {222}},\ \bibinfo {pages} {145} (\bibinfo {year}
  {1992})}\BibitemShut {NoStop}%
\bibitem [{\citenamefont {Shan}\ and\ \citenamefont
  {He}(1998)}]{shan1998discretization}%
  \BibitemOpen
  \bibfield  {author} {\bibinfo {author} {\bibfnamefont {X.}~\bibnamefont
  {Shan}}\ and\ \bibinfo {author} {\bibfnamefont {X.}~\bibnamefont {He}},\
  }\href@noop {} {\bibfield  {journal} {\bibinfo  {journal} {Phys. Rev. Lett.}\
  }\textbf {\bibinfo {volume} {80}},\ \bibinfo {pages} {65} (\bibinfo {year}
  {1998})}\BibitemShut {NoStop}%
\bibitem [{\citenamefont {Yudistiawan}\ \emph
  {et~al.}(2010{\natexlab{b}})\citenamefont {Yudistiawan}, \citenamefont
  {Kwak}, \citenamefont {Patil},\ and\ \citenamefont
  {Ansumali}}]{yudistiawan2010higher}%
  \BibitemOpen
  \bibfield  {author} {\bibinfo {author} {\bibfnamefont {W.~P.}\ \bibnamefont
  {Yudistiawan}}, \bibinfo {author} {\bibfnamefont {S.~K.}\ \bibnamefont
  {Kwak}}, \bibinfo {author} {\bibfnamefont {D.}~\bibnamefont {Patil}}, \ and\
  \bibinfo {author} {\bibfnamefont {S.}~\bibnamefont {Ansumali}},\ }\href@noop
  {} {\bibfield  {journal} {\bibinfo  {journal} {Phys. Rev. E}\ }\textbf
  {\bibinfo {volume} {82}},\ \bibinfo {pages} {046701} (\bibinfo {year}
  {2010}{\natexlab{b}})}\BibitemShut {NoStop}%
\bibitem [{\citenamefont {Chikatamarla}\ and\ \citenamefont
  {Karlin}(2009)}]{chikatamarla2009lattices}%
  \BibitemOpen
  \bibfield  {author} {\bibinfo {author} {\bibfnamefont {S.~S.}\ \bibnamefont
  {Chikatamarla}}\ and\ \bibinfo {author} {\bibfnamefont {I.~V.}\ \bibnamefont
  {Karlin}},\ }\href@noop {} {\bibfield  {journal} {\bibinfo  {journal} {Phys.
  Rev. E}\ }\textbf {\bibinfo {volume} {79}},\ \bibinfo {pages} {046701}
  (\bibinfo {year} {2009})}\BibitemShut {NoStop}%
\bibitem [{\citenamefont {Yudistiawan}\ \emph {et~al.}(2008)\citenamefont
  {Yudistiawan}, \citenamefont {Ansumali},\ and\ \citenamefont
  {Karlin}}]{yudistiawan2008hydrodynamics}%
  \BibitemOpen
  \bibfield  {author} {\bibinfo {author} {\bibfnamefont {W.~P.}\ \bibnamefont
  {Yudistiawan}}, \bibinfo {author} {\bibfnamefont {S.}~\bibnamefont
  {Ansumali}}, \ and\ \bibinfo {author} {\bibfnamefont {I.~V.}\ \bibnamefont
  {Karlin}},\ }\href@noop {} {\bibfield  {journal} {\bibinfo  {journal} {Phys.
  Rev. E}\ }\textbf {\bibinfo {volume} {78}},\ \bibinfo {pages} {016705}
  (\bibinfo {year} {2008})}\BibitemShut {NoStop}%
\bibitem [{\citenamefont {Chikatamarla}\ and\ \citenamefont
  {Karlin}(2006)}]{PhysRevLett.97.190601}%
  \BibitemOpen
  \bibfield  {author} {\bibinfo {author} {\bibfnamefont {S.~S.}\ \bibnamefont
  {Chikatamarla}}\ and\ \bibinfo {author} {\bibfnamefont {I.~V.}\ \bibnamefont
  {Karlin}},\ }\href {\doibase 10.1103/PhysRevLett.97.190601} {\bibfield
  {journal} {\bibinfo  {journal} {Phys. Rev. Lett.}\ }\textbf {\bibinfo
  {volume} {97}},\ \bibinfo {pages} {190601} (\bibinfo {year}
  {2006})}\BibitemShut {NoStop}%
\bibitem [{\citenamefont {Qian}\ and\ \citenamefont
  {Zhou}(1998)}]{qian1998complete}%
  \BibitemOpen
  \bibfield  {author} {\bibinfo {author} {\bibfnamefont {Y.-H.}\ \bibnamefont
  {Qian}}\ and\ \bibinfo {author} {\bibfnamefont {Y.}~\bibnamefont {Zhou}},\
  }\href@noop {} {\bibfield  {journal} {\bibinfo  {journal} {Europhys. Lett.}\
  }\textbf {\bibinfo {volume} {42}},\ \bibinfo {pages} {359} (\bibinfo {year}
  {1998})}\BibitemShut {NoStop}%
\bibitem [{\citenamefont {Shan}\ \emph {et~al.}(2006)\citenamefont {Shan},
  \citenamefont {Yuan},\ and\ \citenamefont {Chen}}]{shan2006kinetic}%
  \BibitemOpen
  \bibfield  {author} {\bibinfo {author} {\bibfnamefont {X.}~\bibnamefont
  {Shan}}, \bibinfo {author} {\bibfnamefont {X.-F.}\ \bibnamefont {Yuan}}, \
  and\ \bibinfo {author} {\bibfnamefont {H.}~\bibnamefont {Chen}},\ }\href@noop
  {} {\bibfield  {journal} {\bibinfo  {journal} {J. Fluid Mech.}\ }\textbf
  {\bibinfo {volume} {550}},\ \bibinfo {pages} {413} (\bibinfo {year}
  {2006})}\BibitemShut {NoStop}%
\bibitem [{\citenamefont {Chikatamarla}\ \emph {et~al.}(2010)\citenamefont
  {Chikatamarla}, \citenamefont {Frouzakis}, \citenamefont {Karlin},
  \citenamefont {Tomboulides},\ and\ \citenamefont
  {Boulouchos}}]{chikatamarla2010lattice}%
  \BibitemOpen
  \bibfield  {author} {\bibinfo {author} {\bibfnamefont {S.}~\bibnamefont
  {Chikatamarla}}, \bibinfo {author} {\bibfnamefont {C.}~\bibnamefont
  {Frouzakis}}, \bibinfo {author} {\bibfnamefont {I.}~\bibnamefont {Karlin}},
  \bibinfo {author} {\bibfnamefont {A.}~\bibnamefont {Tomboulides}}, \ and\
  \bibinfo {author} {\bibfnamefont {K.}~\bibnamefont {Boulouchos}},\
  }\href@noop {} {\bibfield  {journal} {\bibinfo  {journal} {J. Fluid Mech.}\
  }\textbf {\bibinfo {volume} {656}},\ \bibinfo {pages} {298} (\bibinfo {year}
  {2010})}\BibitemShut {NoStop}%
\bibitem [{\citenamefont {Reichl}(1999)}]{reichl1999modern}%
  \BibitemOpen
  \bibfield  {author} {\bibinfo {author} {\bibfnamefont {L.~E.}\ \bibnamefont
  {Reichl}},\ }\href@noop {} {\enquote {\bibinfo {title} {A modern course in
  statistical physics},}\ } (\bibinfo {year} {1999})\BibitemShut {NoStop}%
\bibitem [{\citenamefont {Liboff}(2003)}]{liboff2003kinetic}%
  \BibitemOpen
  \bibfield  {author} {\bibinfo {author} {\bibfnamefont {R.~L.}\ \bibnamefont
  {Liboff}},\ }\href@noop {} {\emph {\bibinfo {title} {Kinetic theory:
  classical, quantum, and relativistic descriptions}}}\ (\bibinfo  {publisher}
  {Springer Science \& Business Media},\ \bibinfo {year} {2003})\BibitemShut
  {NoStop}%
\bibitem [{\citenamefont {Mazloomi~M}\ \emph {et~al.}(2015)\citenamefont
  {Mazloomi~M}, \citenamefont {Chikatamarla},\ and\ \citenamefont
  {Karlin}}]{chikatamarla2015entropic}%
  \BibitemOpen
  \bibfield  {author} {\bibinfo {author} {\bibfnamefont {A.}~\bibnamefont
  {Mazloomi~M}}, \bibinfo {author} {\bibfnamefont {S.~S.}\ \bibnamefont
  {Chikatamarla}}, \ and\ \bibinfo {author} {\bibfnamefont {I.~V.}\
  \bibnamefont {Karlin}},\ }\href@noop {} {\bibfield  {journal} {\bibinfo
  {journal} {Phys. Rev. Lett.}\ }\textbf {\bibinfo {volume} {114}},\ \bibinfo
  {pages} {174502} (\bibinfo {year} {2015})}\BibitemShut {NoStop}%
\bibitem [{\citenamefont {Murthy}\ \emph {et~al.}(2018)\citenamefont {Murthy},
  \citenamefont {Kolluru}, \citenamefont {Kumaran},\ and\ \citenamefont
  {Ansumali}}]{murthy2018lattice}%
  \BibitemOpen
  \bibfield  {author} {\bibinfo {author} {\bibfnamefont {J.~S.~N.}\
  \bibnamefont {Murthy}}, \bibinfo {author} {\bibfnamefont {P.~K.}\
  \bibnamefont {Kolluru}}, \bibinfo {author} {\bibfnamefont {V.}~\bibnamefont
  {Kumaran}}, \ and\ \bibinfo {author} {\bibfnamefont {S.}~\bibnamefont
  {Ansumali}},\ }\href@noop {} {\bibfield  {journal} {\bibinfo  {journal}
  {Commun. Comput. Phys.}\ }\textbf {\bibinfo {volume} {23}},\ \bibinfo {pages}
  {1223} (\bibinfo {year} {2018})}\BibitemShut {NoStop}%
\bibitem [{\citenamefont {Alim}\ \emph {et~al.}(2009)\citenamefont {Alim},
  \citenamefont {Entezari},\ and\ \citenamefont {M{\"o}ller}}]{alim2009}%
  \BibitemOpen
  \bibfield  {author} {\bibinfo {author} {\bibfnamefont {U.~R.}\ \bibnamefont
  {Alim}}, \bibinfo {author} {\bibfnamefont {A.}~\bibnamefont {Entezari}}, \
  and\ \bibinfo {author} {\bibfnamefont {T.}~\bibnamefont {M{\"o}ller}},\
  }\href@noop {} {\bibfield  {journal} {\bibinfo  {journal} {‎IEEE Trans.
  Vis. Comput. Graph}\ }\textbf {\bibinfo {volume} {15}},\ \bibinfo {pages}
  {630} (\bibinfo {year} {2009})}\BibitemShut {NoStop}%
\bibitem [{\citenamefont {Ansumali}\ \emph
  {et~al.}(2007{\natexlab{a}})\citenamefont {Ansumali}, \citenamefont {Karlin},
  \citenamefont {Arcidiacono}, \citenamefont {Abbas},\ and\ \citenamefont
  {Prasianakis}}]{ansumali2007hydrodynamics}%
  \BibitemOpen
  \bibfield  {author} {\bibinfo {author} {\bibfnamefont {S.}~\bibnamefont
  {Ansumali}}, \bibinfo {author} {\bibfnamefont {I.}~\bibnamefont {Karlin}},
  \bibinfo {author} {\bibfnamefont {S.}~\bibnamefont {Arcidiacono}}, \bibinfo
  {author} {\bibfnamefont {A.}~\bibnamefont {Abbas}}, \ and\ \bibinfo {author}
  {\bibfnamefont {N.}~\bibnamefont {Prasianakis}},\ }\href@noop {} {\bibfield
  {journal} {\bibinfo  {journal} {Phys. Rev. Lett.}\ }\textbf {\bibinfo
  {volume} {98}},\ \bibinfo {pages} {124502} (\bibinfo {year}
  {2007}{\natexlab{a}})}\BibitemShut {NoStop}%
\bibitem [{\citenamefont {Karlin}\ \emph {et~al.}(1998)\citenamefont {Karlin},
  \citenamefont {Gorban}, \citenamefont {Succi},\ and\ \citenamefont
  {Boffi}}]{karlin1998maximum}%
  \BibitemOpen
  \bibfield  {author} {\bibinfo {author} {\bibfnamefont {I.~V.}\ \bibnamefont
  {Karlin}}, \bibinfo {author} {\bibfnamefont {A.~N.}\ \bibnamefont {Gorban}},
  \bibinfo {author} {\bibfnamefont {S.}~\bibnamefont {Succi}}, \ and\ \bibinfo
  {author} {\bibfnamefont {V.}~\bibnamefont {Boffi}},\ }\href@noop {}
  {\bibfield  {journal} {\bibinfo  {journal} {Phys. Rev. Lett.}\ }\textbf
  {\bibinfo {volume} {81}},\ \bibinfo {pages} {6} (\bibinfo {year}
  {1998})}\BibitemShut {NoStop}%
\bibitem [{\citenamefont {Boghosian}\ \emph {et~al.}(2003)\citenamefont
  {Boghosian}, \citenamefont {Love}, \citenamefont {Coveney}, \citenamefont
  {Karlin}, \citenamefont {Succi},\ and\ \citenamefont
  {Yepez}}]{boghosian2003galilean}%
  \BibitemOpen
  \bibfield  {author} {\bibinfo {author} {\bibfnamefont {B.~M.}\ \bibnamefont
  {Boghosian}}, \bibinfo {author} {\bibfnamefont {P.~J.}\ \bibnamefont {Love}},
  \bibinfo {author} {\bibfnamefont {P.~V.}\ \bibnamefont {Coveney}}, \bibinfo
  {author} {\bibfnamefont {I.~V.}\ \bibnamefont {Karlin}}, \bibinfo {author}
  {\bibfnamefont {S.}~\bibnamefont {Succi}}, \ and\ \bibinfo {author}
  {\bibfnamefont {J.}~\bibnamefont {Yepez}},\ }\href@noop {} {\bibfield
  {journal} {\bibinfo  {journal} {Phys. Rev. E}\ }\textbf {\bibinfo {volume}
  {68}},\ \bibinfo {pages} {025103} (\bibinfo {year} {2003})}\BibitemShut
  {NoStop}%
\bibitem [{\citenamefont {Chen}\ and\ \citenamefont
  {Teixeira}(2000)}]{chen2000h}%
  \BibitemOpen
  \bibfield  {author} {\bibinfo {author} {\bibfnamefont {H.}~\bibnamefont
  {Chen}}\ and\ \bibinfo {author} {\bibfnamefont {C.}~\bibnamefont
  {Teixeira}},\ }\href@noop {} {\bibfield  {journal} {\bibinfo  {journal}
  {Comp. Phys. Commun.}\ }\textbf {\bibinfo {volume} {129}},\ \bibinfo {pages}
  {21} (\bibinfo {year} {2000})}\BibitemShut {NoStop}%
\bibitem [{\citenamefont {Karlin}\ \emph {et~al.}(1999)\citenamefont {Karlin},
  \citenamefont {Ferrante},\ and\ \citenamefont
  {{\"O}ttinger}}]{karlin1999perfect}%
  \BibitemOpen
  \bibfield  {author} {\bibinfo {author} {\bibfnamefont {I.~V.}\ \bibnamefont
  {Karlin}}, \bibinfo {author} {\bibfnamefont {A.}~\bibnamefont {Ferrante}}, \
  and\ \bibinfo {author} {\bibfnamefont {H.~C.}\ \bibnamefont {{\"O}ttinger}},\
  }\href@noop {} {\bibfield  {journal} {\bibinfo  {journal} {Europhys. Lett.}\
  }\textbf {\bibinfo {volume} {47}},\ \bibinfo {pages} {182} (\bibinfo {year}
  {1999})}\BibitemShut {NoStop}%
\bibitem [{\citenamefont {Succi}\ \emph {et~al.}(2002)\citenamefont {Succi},
  \citenamefont {Karlin},\ and\ \citenamefont {Chen}}]{succi2002colloquium}%
  \BibitemOpen
  \bibfield  {author} {\bibinfo {author} {\bibfnamefont {S.}~\bibnamefont
  {Succi}}, \bibinfo {author} {\bibfnamefont {I.~V.}\ \bibnamefont {Karlin}}, \
  and\ \bibinfo {author} {\bibfnamefont {H.}~\bibnamefont {Chen}},\ }\href@noop
  {} {\bibfield  {journal} {\bibinfo  {journal} {Rev. Mod. Phys.}\ }\textbf
  {\bibinfo {volume} {74}},\ \bibinfo {pages} {1203} (\bibinfo {year}
  {2002})}\BibitemShut {NoStop}%
\bibitem [{\citenamefont {Chikatamarla}\ \emph
  {et~al.}(2006{\natexlab{a}})\citenamefont {Chikatamarla}, \citenamefont
  {Ansumali},\ and\ \citenamefont {Karlin}}]{chikatamarla2006entropic}%
  \BibitemOpen
  \bibfield  {author} {\bibinfo {author} {\bibfnamefont {S.}~\bibnamefont
  {Chikatamarla}}, \bibinfo {author} {\bibfnamefont {S.}~\bibnamefont
  {Ansumali}}, \ and\ \bibinfo {author} {\bibfnamefont {I.~V.}\ \bibnamefont
  {Karlin}},\ }\href@noop {} {\bibfield  {journal} {\bibinfo  {journal} {Phys.
  Rev. Lett.}\ }\textbf {\bibinfo {volume} {97}},\ \bibinfo {pages} {010201}
  (\bibinfo {year} {2006}{\natexlab{a}})}\BibitemShut {NoStop}%
\bibitem [{\citenamefont {Tam}\ and\ \citenamefont
  {Webb}(1993)}]{tam1993dispersion}%
  \BibitemOpen
  \bibfield  {author} {\bibinfo {author} {\bibfnamefont {C.~K.}\ \bibnamefont
  {Tam}}\ and\ \bibinfo {author} {\bibfnamefont {J.~C.}\ \bibnamefont {Webb}},\
  }\href@noop {} {\bibfield  {journal} {\bibinfo  {journal} {J. Comput. Phys.}\
  }\textbf {\bibinfo {volume} {107}},\ \bibinfo {pages} {262} (\bibinfo {year}
  {1993})}\BibitemShut {NoStop}%
\bibitem [{\citenamefont {Gendre}\ \emph {et~al.}(2017)\citenamefont {Gendre},
  \citenamefont {Ricot}, \citenamefont {Fritz},\ and\ \citenamefont
  {Sagaut}}]{gendre2017grid}%
  \BibitemOpen
  \bibfield  {author} {\bibinfo {author} {\bibfnamefont {F.}~\bibnamefont
  {Gendre}}, \bibinfo {author} {\bibfnamefont {D.}~\bibnamefont {Ricot}},
  \bibinfo {author} {\bibfnamefont {G.}~\bibnamefont {Fritz}}, \ and\ \bibinfo
  {author} {\bibfnamefont {P.}~\bibnamefont {Sagaut}},\ }\href@noop {}
  {\bibfield  {journal} {\bibinfo  {journal} {Phys. Rev. E}\ }\textbf {\bibinfo
  {volume} {96}},\ \bibinfo {pages} {023311} (\bibinfo {year}
  {2017})}\BibitemShut {NoStop}%
\bibitem [{\citenamefont {Abramowitz}\ and\ \citenamefont
  {Stegun}(1965)}]{abramowitz1965handbook}%
  \BibitemOpen
  \bibfield  {author} {\bibinfo {author} {\bibfnamefont {M.}~\bibnamefont
  {Abramowitz}}\ and\ \bibinfo {author} {\bibfnamefont {I.~A.}\ \bibnamefont
  {Stegun}},\ }\href@noop {} {\emph {\bibinfo {title} {Handbook of mathematical
  functions: with formulas, graphs, and mathematical tables}}},\ Vol.~\bibinfo
  {volume} {55}\ (\bibinfo  {publisher} {Courier Corporation},\ \bibinfo {year}
  {1965})\BibitemShut {NoStop}%
\bibitem [{\citenamefont {Bogey}\ and\ \citenamefont
  {Bailly}(2002)}]{bogey2002three}%
  \BibitemOpen
  \bibfield  {author} {\bibinfo {author} {\bibfnamefont {C.}~\bibnamefont
  {Bogey}}\ and\ \bibinfo {author} {\bibfnamefont {C.}~\bibnamefont {Bailly}},\
  }\href@noop {} {\bibfield  {journal} {\bibinfo  {journal} {Acta Acust united
  Ac.}\ }\textbf {\bibinfo {volume} {88}},\ \bibinfo {pages} {463} (\bibinfo
  {year} {2002})}\BibitemShut {NoStop}%
\bibitem [{\citenamefont {Hardin}\ \emph {et~al.}(1994)\citenamefont {Hardin},
  \citenamefont {Ristorcelli},\ and\ \citenamefont {Tam}}]{hardin1994icase}%
  \BibitemOpen
  \bibfield  {author} {\bibinfo {author} {\bibfnamefont {J.}~\bibnamefont
  {Hardin}}, \bibinfo {author} {\bibfnamefont {J.}~\bibnamefont {Ristorcelli}},
  \ and\ \bibinfo {author} {\bibfnamefont {C.}~\bibnamefont {Tam}},\
  }\href@noop {} {\emph {\bibinfo {title} {ICASE/LaRC workshop on benchmark
  problems in computational aeroacoustics}}}\ (\bibinfo {year}
  {1994})\BibitemShut {NoStop}%
\bibitem [{\citenamefont {Tam}\ and\ \citenamefont
  {Hardin}(1997)}]{tam1997second}%
  \BibitemOpen
  \bibfield  {author} {\bibinfo {author} {\bibfnamefont {C.~K.}\ \bibnamefont
  {Tam}}\ and\ \bibinfo {author} {\bibfnamefont {J.}~\bibnamefont {Hardin}},\
  }\href@noop {} {\emph {\bibinfo {title} {Second computational aeroacoustics
  (CAA) workshop on benchmark problems}}}\ (\bibinfo {year} {1997})\BibitemShut
  {NoStop}%
\bibitem [{\citenamefont {Tam}\ and\ \citenamefont
  {Hu}(2004)}]{tam2004optimized}%
  \BibitemOpen
  \bibfield  {author} {\bibinfo {author} {\bibfnamefont {C.}~\bibnamefont
  {Tam}}\ and\ \bibinfo {author} {\bibfnamefont {F.}~\bibnamefont {Hu}},\
  }\href@noop {} {\emph {\bibinfo {title} {10th AIAA/CEAS aeroacoustics
  conference}}}\ (\bibinfo {year} {2004})\ p.\ \bibinfo {pages}
  {2812}\BibitemShut {NoStop}%
\bibitem [{\citenamefont {Bird}\ \emph {et~al.}(2015)\citenamefont {Bird},
  \citenamefont {Stewart}, \citenamefont {Lightfoot},\ and\ \citenamefont
  {Klingenberg}}]{bird2015introductory}%
  \BibitemOpen
  \bibfield  {author} {\bibinfo {author} {\bibfnamefont {R.~B.}\ \bibnamefont
  {Bird}}, \bibinfo {author} {\bibfnamefont {W.~E.}\ \bibnamefont {Stewart}},
  \bibinfo {author} {\bibfnamefont {E.~N.}\ \bibnamefont {Lightfoot}}, \ and\
  \bibinfo {author} {\bibfnamefont {D.~J.}\ \bibnamefont {Klingenberg}},\
  }\href@noop {} {\emph {\bibinfo {title} {Introductory Transport
  Phenomena}}},\ Vol.~\bibinfo {volume} {1}\ (\bibinfo  {publisher} {Wiley New
  York},\ \bibinfo {year} {2015})\BibitemShut {NoStop}%
\bibitem [{\citenamefont {Ansumali}\ and\ \citenamefont
  {Karlin}(2002)}]{ansumali2002kinetic}%
  \BibitemOpen
  \bibfield  {author} {\bibinfo {author} {\bibfnamefont {S.}~\bibnamefont
  {Ansumali}}\ and\ \bibinfo {author} {\bibfnamefont {I.~V.}\ \bibnamefont
  {Karlin}},\ }\href@noop {} {\bibfield  {journal} {\bibinfo  {journal} {Phys.
  Rev. E}\ }\textbf {\bibinfo {volume} {66}},\ \bibinfo {pages} {026311}
  (\bibinfo {year} {2002})}\BibitemShut {NoStop}%
\bibitem [{\citenamefont {Larkin}(1967)}]{larkin1967heat}%
  \BibitemOpen
  \bibfield  {author} {\bibinfo {author} {\bibfnamefont {B.}~\bibnamefont
  {Larkin}},\ }in\ \href@noop {} {\emph {\bibinfo {booktitle} {Thermophysics of
  Spacecraft and Planetary Bodies}}}\ (\bibinfo  {publisher} {Elsevier},\
  \bibinfo {year} {1967})\ pp.\ \bibinfo {pages} {819--832}\BibitemShut
  {NoStop}%
\bibitem [{\citenamefont {Parang}\ and\ \citenamefont
  {Salah-Eddine}(1984)}]{parang1984thermoacoustic}%
  \BibitemOpen
  \bibfield  {author} {\bibinfo {author} {\bibfnamefont {M.}~\bibnamefont
  {Parang}}\ and\ \bibinfo {author} {\bibfnamefont {A.}~\bibnamefont
  {Salah-Eddine}},\ }\href@noop {} {\bibfield  {journal} {\bibinfo  {journal}
  {AIAA J.}\ }\textbf {\bibinfo {volume} {22}},\ \bibinfo {pages} {1020}
  (\bibinfo {year} {1984})}\BibitemShut {NoStop}%
\bibitem [{\citenamefont {Huang}\ and\ \citenamefont
  {Bau}(1997)}]{HUANG1997407}%
  \BibitemOpen
  \bibfield  {author} {\bibinfo {author} {\bibfnamefont {Y.}~\bibnamefont
  {Huang}}\ and\ \bibinfo {author} {\bibfnamefont {H.~H.}\ \bibnamefont
  {Bau}},\ }\href {\doibase https://doi.org/10.1016/0017-9310(96)00068-3}
  {\bibfield  {journal} {\bibinfo  {journal} {Int. J. Heat Mass Transf.}\
  }\textbf {\bibinfo {volume} {40}},\ \bibinfo {pages} {407 } (\bibinfo {year}
  {1997})}\BibitemShut {NoStop}%
\bibitem [{\citenamefont {Spradley}\ and\ \citenamefont
  {Churchill}(1975)}]{spradley_churchill_1975}%
  \BibitemOpen
  \bibfield  {author} {\bibinfo {author} {\bibfnamefont {L.~W.}\ \bibnamefont
  {Spradley}}\ and\ \bibinfo {author} {\bibfnamefont {S.~W.}\ \bibnamefont
  {Churchill}},\ }\href {\doibase 10.1017/S0022112075002303} {\bibfield
  {journal} {\bibinfo  {journal} {J. Fluid Mech.}\ }\textbf {\bibinfo {volume}
  {70}},\ \bibinfo {pages} {705–720} (\bibinfo {year} {1975})}\BibitemShut
  {NoStop}%
\bibitem [{\citenamefont {Gunstensen}\ \emph {et~al.}(1991)\citenamefont
  {Gunstensen}, \citenamefont {Rothman}, \citenamefont {Zaleski},\ and\
  \citenamefont {Zanetti}}]{colorfluid}%
  \BibitemOpen
  \bibfield  {author} {\bibinfo {author} {\bibfnamefont {A.~K.}\ \bibnamefont
  {Gunstensen}}, \bibinfo {author} {\bibfnamefont {D.~H.}\ \bibnamefont
  {Rothman}}, \bibinfo {author} {\bibfnamefont {S.}~\bibnamefont {Zaleski}}, \
  and\ \bibinfo {author} {\bibfnamefont {G.}~\bibnamefont {Zanetti}},\ }\href
  {\doibase 10.1103/PhysRevA.43.4320} {\bibfield  {journal} {\bibinfo
  {journal} {Phys. Rev. A}\ }\textbf {\bibinfo {volume} {43}},\ \bibinfo
  {pages} {4320} (\bibinfo {year} {1991})}\BibitemShut {NoStop}%
\bibitem [{\citenamefont {Shan}\ and\ \citenamefont {Chen}(1993)}]{shanchen}%
  \BibitemOpen
  \bibfield  {author} {\bibinfo {author} {\bibfnamefont {X.}~\bibnamefont
  {Shan}}\ and\ \bibinfo {author} {\bibfnamefont {H.}~\bibnamefont {Chen}},\
  }\href {\doibase 10.1103/PhysRevE.47.1815} {\bibfield  {journal} {\bibinfo
  {journal} {Phys. Rev. E}\ }\textbf {\bibinfo {volume} {47}},\ \bibinfo
  {pages} {1815} (\bibinfo {year} {1993})}\BibitemShut {NoStop}%
\bibitem [{\citenamefont {Swift}\ \emph {et~al.}(1995)\citenamefont {Swift},
  \citenamefont {Osborn},\ and\ \citenamefont {Yeomans}}]{freeenergy}%
  \BibitemOpen
  \bibfield  {author} {\bibinfo {author} {\bibfnamefont {M.~R.}\ \bibnamefont
  {Swift}}, \bibinfo {author} {\bibfnamefont {W.~R.}\ \bibnamefont {Osborn}}, \
  and\ \bibinfo {author} {\bibfnamefont {J.~M.}\ \bibnamefont {Yeomans}},\
  }\href {\doibase 10.1103/PhysRevLett.75.830} {\bibfield  {journal} {\bibinfo
  {journal} {Phys. Rev. Lett.}\ }\textbf {\bibinfo {volume} {75}},\ \bibinfo
  {pages} {830} (\bibinfo {year} {1995})}\BibitemShut {NoStop}%
\bibitem [{\citenamefont {He}\ \emph {et~al.}(1998)\citenamefont {He},
  \citenamefont {Shan},\ and\ \citenamefont {Doolen}}]{hechendoolen}%
  \BibitemOpen
  \bibfield  {author} {\bibinfo {author} {\bibfnamefont {X.}~\bibnamefont
  {He}}, \bibinfo {author} {\bibfnamefont {X.}~\bibnamefont {Shan}}, \ and\
  \bibinfo {author} {\bibfnamefont {G.~D.}\ \bibnamefont {Doolen}},\ }\href
  {\doibase 10.1103/PhysRevE.57.R13} {\bibfield  {journal} {\bibinfo  {journal}
  {Phys. Rev. E}\ }\textbf {\bibinfo {volume} {57}},\ \bibinfo {pages} {R13}
  (\bibinfo {year} {1998})}\BibitemShut {NoStop}%
\bibitem [{\citenamefont {Ansumali}(2011)}]{ansumalihardsphere}%
  \BibitemOpen
  \bibfield  {author} {\bibinfo {author} {\bibfnamefont {S.}~\bibnamefont
  {Ansumali}},\ }\href {\doibase 10.4208/cicp.301009.240910s} {\bibfield
  {journal} {\bibinfo  {journal} {Commun. Comput. Phys.}\ }\textbf {\bibinfo
  {volume} {9}},\ \bibinfo {pages} {056703} (\bibinfo {year}
  {2011})}\BibitemShut {NoStop}%
\bibitem [{\citenamefont {He}\ and\ \citenamefont
  {Doolen}(2002)}]{he2002thermodynamic}%
  \BibitemOpen
  \bibfield  {author} {\bibinfo {author} {\bibfnamefont {X.}~\bibnamefont
  {He}}\ and\ \bibinfo {author} {\bibfnamefont {G.~D.}\ \bibnamefont
  {Doolen}},\ }\href@noop {} {\bibfield  {journal} {\bibinfo  {journal} {J.
  Stat. Phys.}\ }\textbf {\bibinfo {volume} {107}},\ \bibinfo {pages} {309}
  (\bibinfo {year} {2002})}\BibitemShut {NoStop}%
\bibitem [{\citenamefont {Lee}\ and\ \citenamefont
  {Fischer}(2006)}]{leefischer}%
  \BibitemOpen
  \bibfield  {author} {\bibinfo {author} {\bibfnamefont {T.}~\bibnamefont
  {Lee}}\ and\ \bibinfo {author} {\bibfnamefont {P.~F.}\ \bibnamefont
  {Fischer}},\ }\href {\doibase 10.1103/PhysRevE.74.046709} {\bibfield
  {journal} {\bibinfo  {journal} {Phys. Rev. E}\ }\textbf {\bibinfo {volume}
  {74}},\ \bibinfo {pages} {046709} (\bibinfo {year} {2006})}\BibitemShut
  {NoStop}%
\bibitem [{\citenamefont {Shan}(2006)}]{shan2006analysis}%
  \BibitemOpen
  \bibfield  {author} {\bibinfo {author} {\bibfnamefont {X.}~\bibnamefont
  {Shan}},\ }\href {\doibase 10.1103/PhysRevE.73.047701} {\bibfield  {journal}
  {\bibinfo  {journal} {Phys. Rev. E}\ }\textbf {\bibinfo {volume} {73}},\
  \bibinfo {pages} {047701} (\bibinfo {year} {2006})}\BibitemShut {NoStop}%
\bibitem [{\citenamefont {Wagner}(2006)}]{wagner2006}%
  \BibitemOpen
  \bibfield  {author} {\bibinfo {author} {\bibfnamefont {A.~J.}\ \bibnamefont
  {Wagner}},\ }\href {\doibase 10.1103/PhysRevE.74.056703} {\bibfield
  {journal} {\bibinfo  {journal} {Phys. Rev. E}\ }\textbf {\bibinfo {volume}
  {74}},\ \bibinfo {pages} {056703} (\bibinfo {year} {2006})}\BibitemShut
  {NoStop}%
\bibitem [{\citenamefont {van~der Waals}(1979)}]{van1979thermodynamic}%
  \BibitemOpen
  \bibfield  {author} {\bibinfo {author} {\bibfnamefont {J.~D.}\ \bibnamefont
  {van~der Waals}},\ }\href@noop {} {\bibfield  {journal} {\bibinfo  {journal}
  {J. Stat. Phys.}\ }\textbf {\bibinfo {volume} {20}},\ \bibinfo {pages} {200}
  (\bibinfo {year} {1979})}\BibitemShut {NoStop}%
\bibitem [{\citenamefont {Rowlinson}\ and\ \citenamefont
  {Widom}(1982)}]{rowlinson}%
  \BibitemOpen
  \bibfield  {author} {\bibinfo {author} {\bibfnamefont {J.}~\bibnamefont
  {Rowlinson}}\ and\ \bibinfo {author} {\bibfnamefont {B.}~\bibnamefont
  {Widom}},\ }\href@noop {} {\enquote {\bibinfo {title} {Molecular theory of
  capillarity},}\ } (\bibinfo {year} {1982})\BibitemShut {NoStop}%
\bibitem [{\citenamefont {Lee}\ and\ \citenamefont {Lin}(2005)}]{LEE200516}%
  \BibitemOpen
  \bibfield  {author} {\bibinfo {author} {\bibfnamefont {T.}~\bibnamefont
  {Lee}}\ and\ \bibinfo {author} {\bibfnamefont {C.-L.}\ \bibnamefont {Lin}},\
  }\href {\doibase https://doi.org/10.1016/j.jcp.2004.12.001} {\bibfield
  {journal} {\bibinfo  {journal} {J. Comput. Phys.}\ }\textbf {\bibinfo
  {volume} {206}},\ \bibinfo {pages} {16 } (\bibinfo {year}
  {2005})}\BibitemShut {NoStop}%
\bibitem [{\citenamefont {Suryanarayanan}\ \emph {et~al.}(2013)\citenamefont
  {Suryanarayanan}, \citenamefont {Singh},\ and\ \citenamefont
  {Ansumali}}]{suryanarayanan_singh_ansumali_2013}%
  \BibitemOpen
  \bibfield  {author} {\bibinfo {author} {\bibfnamefont {S.}~\bibnamefont
  {Suryanarayanan}}, \bibinfo {author} {\bibfnamefont {S.}~\bibnamefont
  {Singh}}, \ and\ \bibinfo {author} {\bibfnamefont {S.}~\bibnamefont
  {Ansumali}},\ }\href {\doibase 10.4208/cicp.401011.220212s} {\bibfield
  {journal} {\bibinfo  {journal} {Commun. Comput. Phys.}\ }\textbf {\bibinfo
  {volume} {13}},\ \bibinfo {pages} {629–648} (\bibinfo {year}
  {2013})}\BibitemShut {NoStop}%
\bibitem [{\citenamefont {Korteweg}(1901)}]{korteweg}%
  \BibitemOpen
  \bibfield  {author} {\bibinfo {author} {\bibfnamefont {D.~J.}\ \bibnamefont
  {Korteweg}},\ }\href@noop {} {\bibfield  {journal} {\bibinfo  {journal}
  {Arch. Neerl. Sci. Exactes Nat. Ser. II}\ }\textbf {\bibinfo {volume} {6}},\
  \bibinfo {pages} {1} (\bibinfo {year} {1901})}\BibitemShut {NoStop}%
\bibitem [{\citenamefont {Evans}(1979)}]{evans1979}%
  \BibitemOpen
  \bibfield  {author} {\bibinfo {author} {\bibfnamefont {R.}~\bibnamefont
  {Evans}},\ }\href {\doibase 10.1080/00018737900101365} {\bibfield  {journal}
  {\bibinfo  {journal} {Adv. Phys.}\ }\textbf {\bibinfo {volume} {28}},\
  \bibinfo {pages} {143} (\bibinfo {year} {1979})}\BibitemShut {NoStop}%
\bibitem [{\citenamefont {Wagner}\ and\ \citenamefont
  {Pooley}(2007)}]{wagner2007interface}%
  \BibitemOpen
  \bibfield  {author} {\bibinfo {author} {\bibfnamefont {A.}~\bibnamefont
  {Wagner}}\ and\ \bibinfo {author} {\bibfnamefont {C.}~\bibnamefont
  {Pooley}},\ }\href@noop {} {\bibfield  {journal} {\bibinfo  {journal} {Phys.
  Rev. E}\ }\textbf {\bibinfo {volume} {76}},\ \bibinfo {pages} {045702}
  (\bibinfo {year} {2007})}\BibitemShut {NoStop}%
\bibitem [{\citenamefont {Carnahan}\ and\ \citenamefont
  {Starling}(1969)}]{carnahan1969equation}%
  \BibitemOpen
  \bibfield  {author} {\bibinfo {author} {\bibfnamefont {N.~F.}\ \bibnamefont
  {Carnahan}}\ and\ \bibinfo {author} {\bibfnamefont {K.~E.}\ \bibnamefont
  {Starling}},\ }\href@noop {} {\bibfield  {journal} {\bibinfo  {journal} {J.
  Chem. Phys.}\ }\textbf {\bibinfo {volume} {51}},\ \bibinfo {pages} {635}
  (\bibinfo {year} {1969})}\BibitemShut {NoStop}%
\bibitem [{\citenamefont {Inamuro}\ \emph {et~al.}(2004)\citenamefont
  {Inamuro}, \citenamefont {Tajima},\ and\ \citenamefont
  {Ogino}}]{inamuro2004lattice}%
  \BibitemOpen
  \bibfield  {author} {\bibinfo {author} {\bibfnamefont {T.}~\bibnamefont
  {Inamuro}}, \bibinfo {author} {\bibfnamefont {S.}~\bibnamefont {Tajima}}, \
  and\ \bibinfo {author} {\bibfnamefont {F.}~\bibnamefont {Ogino}},\
  }\href@noop {} {\bibfield  {journal} {\bibinfo  {journal} {Int. J. Heat Mass
  Transf.}\ }\textbf {\bibinfo {volume} {47}},\ \bibinfo {pages} {4649}
  (\bibinfo {year} {2004})}\BibitemShut {NoStop}%
\bibitem [{\citenamefont {Atif}\ \emph {et~al.}(2017)\citenamefont {Atif},
  \citenamefont {Kolluru}, \citenamefont {Thantanapally},\ and\ \citenamefont
  {Ansumali}}]{atif2017}%
  \BibitemOpen
  \bibfield  {author} {\bibinfo {author} {\bibfnamefont {M.}~\bibnamefont
  {Atif}}, \bibinfo {author} {\bibfnamefont {P.~K.}\ \bibnamefont {Kolluru}},
  \bibinfo {author} {\bibfnamefont {C.}~\bibnamefont {Thantanapally}}, \ and\
  \bibinfo {author} {\bibfnamefont {S.}~\bibnamefont {Ansumali}},\ }\href
  {\doibase 10.1103/PhysRevLett.119.240602} {\bibfield  {journal} {\bibinfo
  {journal} {Phys. Rev. Lett.}\ }\textbf {\bibinfo {volume} {119}},\ \bibinfo
  {pages} {240602} (\bibinfo {year} {2017})}\BibitemShut {NoStop}%
\bibitem [{\citenamefont {Atif}\ \emph {et~al.}()\citenamefont {Atif},
  \citenamefont {Kolluru}, \citenamefont {Thantanapally},\ and\ \citenamefont
  {Ansumali}}]{atifDetailed}%
  \BibitemOpen
  \bibfield  {author} {\bibinfo {author} {\bibfnamefont {M.}~\bibnamefont
  {Atif}}, \bibinfo {author} {\bibfnamefont {P.~K.}\ \bibnamefont {Kolluru}},
  \bibinfo {author} {\bibfnamefont {C.}~\bibnamefont {Thantanapally}}, \ and\
  \bibinfo {author} {\bibfnamefont {S.}~\bibnamefont {Ansumali}},\ }\href@noop
  {} {\bibinfo  {journal} {(unpublished)}\ }\BibitemShut {NoStop}%
\bibitem [{\citenamefont {Kida}(1985)}]{kida1985three}%
  \BibitemOpen
\bibfield  {journal} {  }\bibfield  {author} {\bibinfo {author} {\bibfnamefont
  {S.}~\bibnamefont {Kida}},\ }\href@noop {} {\bibfield  {journal} {\bibinfo
  {journal} {‎J. Phys. Soc. Jpn}\ }\textbf {\bibinfo {volume} {54}},\
  \bibinfo {pages} {2132} (\bibinfo {year} {1985})}\BibitemShut {NoStop}%
\bibitem [{\citenamefont {Frisch}(1995)}]{frisch1995turbulence}%
  \BibitemOpen
  \bibfield  {author} {\bibinfo {author} {\bibfnamefont {U.}~\bibnamefont
  {Frisch}},\ }\href@noop {} {\emph {\bibinfo {title} {Turbulence: the legacy
  of AN Kolmogorov}}}\ (\bibinfo  {publisher} {Cambridge university press},\
  \bibinfo {year} {1995})\BibitemShut {NoStop}%
\bibitem [{\citenamefont {Bespalko}\ \emph {et~al.}(2012)\citenamefont
  {Bespalko}, \citenamefont {Pollard},\ and\ \citenamefont
  {Uddin}}]{bespalko2012analysis}%
  \BibitemOpen
  \bibfield  {author} {\bibinfo {author} {\bibfnamefont {D.}~\bibnamefont
  {Bespalko}}, \bibinfo {author} {\bibfnamefont {A.}~\bibnamefont {Pollard}}, \
  and\ \bibinfo {author} {\bibfnamefont {M.}~\bibnamefont {Uddin}},\
  }\href@noop {} {\bibfield  {journal} {\bibinfo  {journal} {Comput. Fluids}\
  }\textbf {\bibinfo {volume} {54}},\ \bibinfo {pages} {143} (\bibinfo {year}
  {2012})}\BibitemShut {NoStop}%
\bibitem [{\citenamefont {Bridson}\ \emph {et~al.}(2007)\citenamefont
  {Bridson}, \citenamefont {Houriham},\ and\ \citenamefont
  {Nordenstam}}]{bridson2007curl}%
  \BibitemOpen
  \bibfield  {author} {\bibinfo {author} {\bibfnamefont {R.}~\bibnamefont
  {Bridson}}, \bibinfo {author} {\bibfnamefont {J.}~\bibnamefont {Houriham}}, \
  and\ \bibinfo {author} {\bibfnamefont {M.}~\bibnamefont {Nordenstam}},\
  }\href@noop {} {\bibfield  {journal} {\bibinfo  {journal} {ACM Trans. Graph}\
  }\textbf {\bibinfo {volume} {26}},\ \bibinfo {pages} {46} (\bibinfo {year}
  {2007})}\BibitemShut {NoStop}%
\bibitem [{\citenamefont {Pope}(2001)}]{pope2001turbulent}%
  \BibitemOpen
  \bibfield  {author} {\bibinfo {author} {\bibfnamefont {S.~B.}\ \bibnamefont
  {Pope}},\ }\href@noop {} {\enquote {\bibinfo {title} {Turbulent flows},}\ }
  (\bibinfo {year} {2001})\BibitemShut {NoStop}%
\bibitem [{\citenamefont {Moser}\ \emph {et~al.}(1999)\citenamefont {Moser},
  \citenamefont {Kim},\ and\ \citenamefont {Mansour}}]{moser1999direct}%
  \BibitemOpen
  \bibfield  {author} {\bibinfo {author} {\bibfnamefont {R.~D.}\ \bibnamefont
  {Moser}}, \bibinfo {author} {\bibfnamefont {J.}~\bibnamefont {Kim}}, \ and\
  \bibinfo {author} {\bibfnamefont {N.~N.}\ \bibnamefont {Mansour}},\
  }\href@noop {} {\bibfield  {journal} {\bibinfo  {journal} {Phys. Fluids}\
  }\textbf {\bibinfo {volume} {11}},\ \bibinfo {pages} {943} (\bibinfo {year}
  {1999})}\BibitemShut {NoStop}%
\bibitem [{\citenamefont {Kim}\ \emph {et~al.}(1987)\citenamefont {Kim},
  \citenamefont {Moin},\ and\ \citenamefont {Moser}}]{kim1987turbulence}%
  \BibitemOpen
  \bibfield  {author} {\bibinfo {author} {\bibfnamefont {J.}~\bibnamefont
  {Kim}}, \bibinfo {author} {\bibfnamefont {P.}~\bibnamefont {Moin}}, \ and\
  \bibinfo {author} {\bibfnamefont {R.}~\bibnamefont {Moser}},\ }\href@noop {}
  {\bibfield  {journal} {\bibinfo  {journal} {J. Fluid Mech.}\ }\textbf
  {\bibinfo {volume} {177}},\ \bibinfo {pages} {133} (\bibinfo {year}
  {1987})}\BibitemShut {NoStop}%
\bibitem [{\citenamefont {Chikatamarla}\ \emph
  {et~al.}(2006{\natexlab{b}})\citenamefont {Chikatamarla}, \citenamefont
  {Ansumali},\ and\ \citenamefont {Karlin}}]{chikatamarla2006grad}%
  \BibitemOpen
  \bibfield  {author} {\bibinfo {author} {\bibfnamefont {S.}~\bibnamefont
  {Chikatamarla}}, \bibinfo {author} {\bibfnamefont {S.}~\bibnamefont
  {Ansumali}}, \ and\ \bibinfo {author} {\bibfnamefont {I.}~\bibnamefont
  {Karlin}},\ }\href@noop {} {\bibfield  {journal} {\bibinfo  {journal}
  {Europhys. Lett.}\ }\textbf {\bibinfo {volume} {74}},\ \bibinfo {pages} {215}
  (\bibinfo {year} {2006}{\natexlab{b}})}\BibitemShut {NoStop}%
\bibitem [{\citenamefont {Krithivasan}\ \emph {et~al.}(2014)\citenamefont
  {Krithivasan}, \citenamefont {Wahal},\ and\ \citenamefont
  {Ansumali}}]{krithivasan2014diffused}%
  \BibitemOpen
  \bibfield  {author} {\bibinfo {author} {\bibfnamefont {S.}~\bibnamefont
  {Krithivasan}}, \bibinfo {author} {\bibfnamefont {S.}~\bibnamefont {Wahal}},
  \ and\ \bibinfo {author} {\bibfnamefont {S.}~\bibnamefont {Ansumali}},\
  }\href@noop {} {\bibfield  {journal} {\bibinfo  {journal} {Phys. Rev. E}\
  }\textbf {\bibinfo {volume} {89}},\ \bibinfo {pages} {033313} (\bibinfo
  {year} {2014})}\BibitemShut {NoStop}%
\bibitem [{\citenamefont {Rodr{\'\i}guez}\ \emph {et~al.}(2013)\citenamefont
  {Rodr{\'\i}guez}, \citenamefont {Lehmkuhl}, \citenamefont {Borrell},\ and\
  \citenamefont {Oliva}}]{rodriguez2013flow}%
  \BibitemOpen
  \bibfield  {author} {\bibinfo {author} {\bibfnamefont {I.}~\bibnamefont
  {Rodr{\'\i}guez}}, \bibinfo {author} {\bibfnamefont {O.}~\bibnamefont
  {Lehmkuhl}}, \bibinfo {author} {\bibfnamefont {R.}~\bibnamefont {Borrell}}, \
  and\ \bibinfo {author} {\bibfnamefont {A.}~\bibnamefont {Oliva}},\
  }\href@noop {} {\bibfield  {journal} {\bibinfo  {journal} {Comput. Fluids}\
  }\textbf {\bibinfo {volume} {80}},\ \bibinfo {pages} {233} (\bibinfo {year}
  {2013})}\BibitemShut {NoStop}%
\bibitem [{\citenamefont {Yun}\ \emph {et~al.}(2006)\citenamefont {Yun},
  \citenamefont {Kim},\ and\ \citenamefont {Choi}}]{yun2006vortical}%
  \BibitemOpen
  \bibfield  {author} {\bibinfo {author} {\bibfnamefont {G.}~\bibnamefont
  {Yun}}, \bibinfo {author} {\bibfnamefont {D.}~\bibnamefont {Kim}}, \ and\
  \bibinfo {author} {\bibfnamefont {H.}~\bibnamefont {Choi}},\ }\href@noop {}
  {\bibfield  {journal} {\bibinfo  {journal} {Phys. Fluids}\ }\textbf {\bibinfo
  {volume} {18}},\ \bibinfo {pages} {015102} (\bibinfo {year}
  {2006})}\BibitemShut {NoStop}%
\bibitem [{\citenamefont {Kim}\ and\ \citenamefont
  {Durbin}(1988)}]{kim1988observations}%
  \BibitemOpen
  \bibfield  {author} {\bibinfo {author} {\bibfnamefont {H.}~\bibnamefont
  {Kim}}\ and\ \bibinfo {author} {\bibfnamefont {P.}~\bibnamefont {Durbin}},\
  }\href@noop {} {\bibfield  {journal} {\bibinfo  {journal} {Phys. Fluids}\
  }\textbf {\bibinfo {volume} {31}},\ \bibinfo {pages} {3260} (\bibinfo {year}
  {1988})}\BibitemShut {NoStop}%
\bibitem [{\citenamefont {Dorschner}\ \emph {et~al.}(2016)\citenamefont
  {Dorschner}, \citenamefont {Frapolli}, \citenamefont {Chikatamarla},\ and\
  \citenamefont {Karlin}}]{dorschner2016grid}%
  \BibitemOpen
  \bibfield  {author} {\bibinfo {author} {\bibfnamefont {B.}~\bibnamefont
  {Dorschner}}, \bibinfo {author} {\bibfnamefont {N.}~\bibnamefont {Frapolli}},
  \bibinfo {author} {\bibfnamefont {S.~S.}\ \bibnamefont {Chikatamarla}}, \
  and\ \bibinfo {author} {\bibfnamefont {I.~V.}\ \bibnamefont {Karlin}},\
  }\href@noop {} {\bibfield  {journal} {\bibinfo  {journal} {Phys. Rev. E}\
  }\textbf {\bibinfo {volume} {94}},\ \bibinfo {pages} {053311} (\bibinfo
  {year} {2016})}\BibitemShut {NoStop}%
\bibitem [{\citenamefont {Holway~Jr}(1966)}]{holway1966new}%
  \BibitemOpen
  \bibfield  {author} {\bibinfo {author} {\bibfnamefont {L.~H.}\ \bibnamefont
  {Holway~Jr}},\ }\href@noop {} {\bibfield  {journal} {\bibinfo  {journal}
  {Phys. Fluids}\ }\textbf {\bibinfo {volume} {9}},\ \bibinfo {pages} {1658}
  (\bibinfo {year} {1966})}\BibitemShut {NoStop}%
\bibitem [{\citenamefont {Shakhov}(1968)}]{Shakhov1968}%
  \BibitemOpen
  \bibfield  {author} {\bibinfo {author} {\bibfnamefont {E.~M.}\ \bibnamefont
  {Shakhov}},\ }\href@noop {} {\bibfield  {journal} {\bibinfo  {journal} {Fluid
  Dyn.}\ }\textbf {\bibinfo {volume} {3}},\ \bibinfo {pages} {95} (\bibinfo
  {year} {1968})}\BibitemShut {NoStop}%
\bibitem [{\citenamefont {Levermore}(1996)}]{levermore1996moment}%
  \BibitemOpen
  \bibfield  {author} {\bibinfo {author} {\bibfnamefont {C.~D.}\ \bibnamefont
  {Levermore}},\ }\href@noop {} {\bibfield  {journal} {\bibinfo  {journal} {J.
  Stat. Phys.}\ }\textbf {\bibinfo {volume} {83}},\ \bibinfo {pages} {1021}
  (\bibinfo {year} {1996})}\BibitemShut {NoStop}%
\bibitem [{\citenamefont {Ansumali}\ \emph
  {et~al.}(2007{\natexlab{b}})\citenamefont {Ansumali}, \citenamefont
  {Arcidiacono}, \citenamefont {Chikatamarla}, \citenamefont {Prasianakis},
  \citenamefont {Gorban},\ and\ \citenamefont {Karlin}}]{ansumali2007quasi}%
  \BibitemOpen
  \bibfield  {author} {\bibinfo {author} {\bibfnamefont {S.}~\bibnamefont
  {Ansumali}}, \bibinfo {author} {\bibfnamefont {S.}~\bibnamefont
  {Arcidiacono}}, \bibinfo {author} {\bibfnamefont {S.}~\bibnamefont
  {Chikatamarla}}, \bibinfo {author} {\bibfnamefont {N.}~\bibnamefont
  {Prasianakis}}, \bibinfo {author} {\bibfnamefont {A.}~\bibnamefont {Gorban}},
  \ and\ \bibinfo {author} {\bibfnamefont {I.}~\bibnamefont {Karlin}},\
  }\href@noop {} {\bibfield  {journal} {\bibinfo  {journal} {Eur. Phys. J. B}\
  }\textbf {\bibinfo {volume} {56}},\ \bibinfo {pages} {135} (\bibinfo {year}
  {2007}{\natexlab{b}})}\BibitemShut {NoStop}%
\bibitem [{\citenamefont {Nie}\ \emph {et~al.}(2008)\citenamefont {Nie},
  \citenamefont {Shan},\ and\ \citenamefont {Chen}}]{nie2008thermal}%
  \BibitemOpen
  \bibfield  {author} {\bibinfo {author} {\bibfnamefont {X.}~\bibnamefont
  {Nie}}, \bibinfo {author} {\bibfnamefont {X.}~\bibnamefont {Shan}}, \ and\
  \bibinfo {author} {\bibfnamefont {H.}~\bibnamefont {Chen}},\ }\href@noop {}
  {\bibfield  {journal} {\bibinfo  {journal} {Phys. Rev. E}\ }\textbf {\bibinfo
  {volume} {77}},\ \bibinfo {pages} {035701} (\bibinfo {year}
  {2008})}\BibitemShut {NoStop}%
\bibitem [{\citenamefont {Thampi}\ \emph {et~al.}(2013)\citenamefont {Thampi},
  \citenamefont {Ansumali}, \citenamefont {Adhikari},\ and\ \citenamefont
  {Succi}}]{thampi2013isotropic}%
  \BibitemOpen
  \bibfield  {author} {\bibinfo {author} {\bibfnamefont {S.~P.}\ \bibnamefont
  {Thampi}}, \bibinfo {author} {\bibfnamefont {S.}~\bibnamefont {Ansumali}},
  \bibinfo {author} {\bibfnamefont {R.}~\bibnamefont {Adhikari}}, \ and\
  \bibinfo {author} {\bibfnamefont {S.}~\bibnamefont {Succi}},\ }\href@noop {}
  {\bibfield  {journal} {\bibinfo  {journal} {J. Comput. Phys.}\ }\textbf
  {\bibinfo {volume} {234}},\ \bibinfo {pages} {1} (\bibinfo {year}
  {2013})}\BibitemShut {NoStop}%
\bibitem [{\citenamefont {Ramadugu}\ \emph {et~al.}(2013)\citenamefont
  {Ramadugu}, \citenamefont {Thampi}, \citenamefont {Adhikari}, \citenamefont
  {Succi},\ and\ \citenamefont {Ansumali}}]{ramadugu2013lattice}%
  \BibitemOpen
  \bibfield  {author} {\bibinfo {author} {\bibfnamefont {R.}~\bibnamefont
  {Ramadugu}}, \bibinfo {author} {\bibfnamefont {S.~P.}\ \bibnamefont
  {Thampi}}, \bibinfo {author} {\bibfnamefont {R.}~\bibnamefont {Adhikari}},
  \bibinfo {author} {\bibfnamefont {S.}~\bibnamefont {Succi}}, \ and\ \bibinfo
  {author} {\bibfnamefont {S.}~\bibnamefont {Ansumali}},\ }\href@noop {}
  {\bibfield  {journal} {\bibinfo  {journal} {Europhys. Lett.}\ }\textbf
  {\bibinfo {volume} {101}},\ \bibinfo {pages} {50006} (\bibinfo {year}
  {2013})}\BibitemShut {NoStop}%
\end{thebibliography}%

\end{document}